\documentclass[a4paper,12pt]{article}
\def\mydate{June 25, 2020}

\usepackage[dvipdfmx]{graphicx}
\usepackage{graphicx}

\usepackage{amsmath}        
\usepackage{amssymb}        
\usepackage{cite}
\usepackage{cancel}
\usepackage{fancybox}
\usepackage{color}
\usepackage{pifont}
\usepackage{colortbl}
\usepackage{multicol}
\usepackage{multirow}
\usepackage{lscape}
\usepackage[all,knot]{xy}
\usepackage{longtable}
\usepackage{braket}
\usepackage{bm}

\makeatletter
 
 \@addtoreset{equation}{section}
\makeatother

\renewcommand{\Re}{\mathop{\mbox{Re}}} % Real part "Re"
\newcommand{\SM}{\text{\rm SM}}

\def\ignore#1{{}}

\def\go{\rightarrow}

\def\SM{{\rm SM}}
\def\KK{{\rm KK}}

\def\la{\langle}
\def\ra{\rangle}

\def\mybig{\displaystyle \strut }

\def\myfrac#1#2{\frac{\mybig #1}{\mybig #2}}

\def\mymat#1#2{\begin{matrix}#1 \cr \noalign{\kern -2pt} #2\end{matrix}}

\def\mynoalign{\noalign{\kern 4pt}}
\def\mysnoalign{\noalign{\kern 3pt}}

\begin{document}
\allowdisplaybreaks

%\leftline{\mydate}
\thispagestyle{empty}

%%%%% PREPRINT NUMBERS %%%%%%

\leftline{\mydate  \hfill OU-HET-1060}
\rightline{KYUSHU-HET-214}

\vskip3.0cm

%%%%%%%%%%%%%%%% TITLE %%%%%%%%%%%%%%%
\baselineskip=30pt plus 1pt minus 1pt

\begin{center}

{\LARGE \bf Fermion pair production}

{\LARGE \bf at $e^- e^+$ linear collider experiments}

{\LARGE \bf  in GUT inspired gauge-Higgs unification}

\end{center}

%%%%%%%%%%%%% AUTHORS %%%%%%%%%%%%%

%\vspace{.0cm}
\baselineskip=22pt plus 1pt minus 1pt

\vskip 2.0cm

\begin{center}
{\bf Shuichiro Funatsu$^1$, Hisaki Hatanaka$^2$, Yutaka Hosotani$^3$,}

{\bf Yuta Orikasa$^4$ and Naoki Yamatsu$^5$}

\baselineskip=17pt plus 1pt minus 1pt

\vskip 10pt
{\small \it $^1$Institute of Particle Physics and Key Laboratory of Quark and Lepton 
Physics (MOE), Central China Normal University, Wuhan, Hubei 430079, China} \\
{\small \it $^2$Osaka, Osaka 536-0014, Japan} \\
{\small \it $^3$Department of Physics, Osaka University, 
Toyonaka, Osaka 560-0043, Japan} \\
{\small \it $^4$Institute of Experimental and Applied Physics, Czech Technical University in Prague,} \\
{\small \it Husova 240/5, 110 00 Prague 1, Czech Republic} \\
{\small \it $^5$Department of Physics, Kyushu University, Fukuoka 819-0395, Japan} \\

\end{center}

%%%%%%%%%%%% Abstract %%%%%%%%%%%%

\vskip 2.0cm

\begin{abstract}
Fermion pair production at  $e^-e^+$ linear collider experiments with polarized 
$e^-$ and $e^+$ beams is examined 
in the GUT inspired $SO(5)\times U(1)\times SU(3)$ gauge-Higgs  unification.
There arises large parity violation in the couplings of leptons and quarks to 
Kaluza-Klein (KK) excited neutral vector bosons $Z'$s, which leads to distinctive 
polarization dependence in cross sections, forward-backward asymmetries,
left-right asymmetries, and left-right forward-backward asymmetries
in various processes.  Those effects are detectable even for the KK mass scale up to about 15 TeV 
at future $e^-e^+$ linear collider experiments with energies 250$\,$GeV to 1$\,$TeV.
\end{abstract}

\newpage

\baselineskip=20pt plus 1pt minus 1pt
\parskip=0pt

\section{Introduction}

The standard model (SM) in particle physics has been established at low
energies. However, it is not yet clear that the observed Higgs boson has
exactly the same properties as those in the SM. It is necessary to
determine the Higgs couplings to quarks, leptons, SM gauge bosons, and
the Higgs self-couplings with better accuracy in future experiments. 

There remain uneasy points in the Higgs boson sector in the SM. 
While the dynamics of the SM gauge bosons, the photon, $W$ and $Z$
bosons and gluons is governed by the gauge principle,
dynamics of the Higgs boson in the SM is not.  
Higgs couplings of quarks and leptons as well as  
Higgs self-couplings are not regulated by any principle. At the quantum
level, there arise huge corrections to the Higgs boson mass, which have
to be canceled and tuned by hand to obtain the observed 125 GeV
mass. One way to achieve the stabilization of the Higgs boson mass
against quantum corrections is to identify the Higgs boson with
the zero mode of the fifth dimensional component of the gauge potential 
\cite{Hosotani:1983xw,Hosotani:1988bm,Davies:1987ei,Davies:1988wt,Hatanaka:1998yp,Hatanaka:1999sx}.
This scenario is referred to as gauge-Higgs unification (GHU).

In GHU  the Higgs field appears as a fluctuation mode of the
Aharonov-Bohm  (AB) phase $\theta_H$ in the fifth dimension. The
$SU(3)_C \times SO(5) \times U(1)_X$ gauge theory in the Randall-Sundrum
(RS) warped space has been proposed in 
Refs.~\cite{Agashe:2004rs,Medina:2007hz,Hosotani:2008tx,Funatsu:2013ni,Funatsu:2014fda,Funatsu:2016uvi,Funatsu:2019xwr,Funatsu:2019fry,Funatsu:2020znj}. 
It gives nearly the same phenomenology at low energies as the SM
\cite{Funatsu:2013ni,Funatsu:2014fda,Funatsu:2015xba,Funatsu:2016uvi}. 
Deviations of the gauge couplings of quarks and leptons from the SM
values are less than 0.1\% for $\theta_H \simeq 0.1$. 
Higgs couplings of quarks, leptons, $W$ and $Z$ bosons are approximately 
the SM values times $\cos \theta_H$; the deviation is about 1{\%}.
In one type of the models 
the Kaluza-Klein (KK) mass scale turns out about $m_\KK \simeq 8\,$TeV  for
$\theta_H \simeq 0.1$.
KK excited states contribute in intermediate states of
the two $\gamma$ decay of the Higgs boson. Their contribution is finite
and very small. The signal strengths of various Higgs decay modes are
approximately $\cos^2 \theta_H$ times the SM values.  The branching
fractions of those decay modes are approximately the same as in the SM.

GHU predicts $Z'$ bosons, which are the  KK modes of $\gamma$, $Z$, and $Z_R$.
They are mixed vector bosons of $U(1)_X$,  $U(1)_L \subset SU(2)_L $, and  $U(1)_R \subset SU(2)_R$
where $SU(2)_L\times SU(2)_R \subset SO(5)$.
In the model with quark-lepton multiplets introduced in the vector representation of $SO(5)$,
which is referred to as the {\it A-model} below, 
masses of $Z'$ bosons are in the 6\,TeV--9\,TeV range for $\theta_H=0.11$--0.07.  
They have broad widths and can be produced at 14$\,$TeV Large Hadron Collider (LHC). 
The current non-observation of $Z'$ signals  puts the limit $\theta_H \lesssim 0.11$. 
Distinct signals of the gauge-Higgs unification can be found in $e^- e^+$ collisions
\cite{Funatsu:2017nfm,Yoon:2018xud, Yoon:2018vsc, Funatsu:2019ujy,Hosotani:2019cnn}.
Large parity violation appears in the couplings of quarks and leptons to
KK gauge bosons, particularly to the $Z'$ bosons.  In the A-model,
right-handed quarks and charged leptons have rather large couplings to
$Z'$ bosons. The interference effects of $Z'$ bosons can be clearly observed at
250$\,$GeV $e^- e^+$ International Linear Collider (ILC)
\cite{Bilokin:2017lco,Richard:2018zhl,Irles:2019xny,Irles:2020gjh,Fujii:2017vwa,Aihara:2019gcq,Bambade:2019fyw}.
In the process $e^-e^+ \go \mu^-\mu^+$, the deviation from the SM
amounts to $-4${\%} with the electron beam polarized in the right-handed
mode by 80{\%} $(P_{e^-}=0.8)$ for $\theta_H\simeq 0.09$, whereas there
appears negligible deviation with the electron beam polarized in the
left-handed mode by 80{\%} $(P_{e^-}=-0.8)$. 
In the forward-backward asymmetry $A_{FB} (e^-e^+\to \mu^-\mu^+)$
the deviation from the SM becomes $-2${\%} for $P_{e^-} = 0.8$.
These deviations can be seen at 250$\,$GeV ILC even 
with 250$\,$fb$^{-1}$ data
\cite{Bilokin:2017lco,Richard:2018zhl,Irles:2019xny,Irles:2020gjh,Fujii:2017vwa,Aihara:2019gcq,Bambade:2019fyw}.
We note that the ILC designs
80\% polarization of the electron beam and 30\% polarization of the
positron beam according to the ILC Technical Design Report 
\cite{Behnke:2013xla,Baer:2013cma,Adolphsen:2013jya,Adolphsen:2013kya,Behnke:2013lya}.
The significance of polarized positrons and electrons for several new physics
searches at the ILC is summarized in Ref.~\cite{MoortgatPick:2005cw}.

Recently, an alternative gauge-Higgs unification model with quark-lepton
multiplets introduced in the spinor, vector, and singlet representations
of $SO(5)$, which is referred to as the {\it B-model} below, has been
proposed \cite{Funatsu:2019xwr}.
The B-model can be embedded in the
$SO(11)$ gauge-Higgs grand unification
\cite{Hosotani:2015hoa,Yamatsu:2015rge,Furui:2016owe,Hosotani:2017krs,Hosotani:2017edv,Hosotani:2017ghg,Englert:2019xhz,Englert:2020eep},
where the SM gauge group and quark-lepton content are incorporated into grand unified theory (GUT)
\cite{Georgi:1974sy,Inoue:1977qd,Fritzsch:1974nn,Gursey:1975ki,Slansky:1981yr,Yamatsu:2015gut}
in higher dimensional framework
\cite{Kawamura:1999nj,Kawamura:2000ir,Kawamura:2000ev,Burdman:2002se,Lim:2007jv,Kojima:2011ad,Kojima:2016fvv,Yamatsu:2017sgu,Yamatsu:2017ssg,Yamatsu:2018fsg,Maru:2019lit}.

In this paper, we evaluate cross sections,
forward-backward asymmetries \cite{Schrempp:1987zy,Kennedy:1988rt},
left-right asymmetries \cite{Schrempp:1987zy,Kennedy:1988rt,Abe:1994wx,Abe:1996nj},
and left-right forward-backward asymmetries
\cite{Blondel:1987gp,Kennedy:1988rt,Abe:1994bj,Abe:1994bm,Abe:1995yh}
in the processes  $e^-e^+\to f\bar{f}$
($f\bar{f}= \mu^-\mu^+, c\bar{c}, b\bar{b}, t\bar{t})$ in the GUT inspired GHU, the B-model.
The quantities in the process $e^- e^+ \to \tau^-  \tau^+ $ are almost the same as in the process 
$e^- e^+ \to \mu^-  \mu^+ $, as the couplings of $\tau^\pm$ to $Z'$s are nearly the same as those of $\mu^\pm$.
For the process $e^- e^+ \to e^-  e^+ $ there is an additional contribution from the BhaBha 
scattering\cite{Abe:1994sh,MoortgatPick:2005cw,Schael:2013ita,Bardin:2017mdd,Borodulin:2017pwh,Richard:2018zhl}, 
the analysis of which is given separately.
We shall find a significant difference between predictions from the SM and those from the B-model 
at  $e^-e^+$ linear collider experiments with polarized beams.

$Z'$ bosons appear in many  models beyond the SM and various physical consequences have been
examined\cite{Langacker:2008yv,Deguchi:2019tvp,Fujii:2019zll}. In most cases couplings of $Z'$ bosons to 
quarks and leptons are  comparable to those of $Z$ boson.
The situation is quite different in GHU.  As was shown in the A-model in Ref.\ \cite{Funatsu:2017nfm} and
is shown below in the B-model, either left-handed or right-handed components of quarks and leptons
have rather large couplings to $Z'$ bosons, particularly to  the first KK modes of $\gamma$, $Z$ and $Z_R$.
We shall see that substantial deviations from the SM can be seen in cross sections and 
other quantities in $e^- e^+ \to f  \bar f$ processes at ILC even though those $Z'$ bosons may be as heavy as 
10$\,$TeV.

There are similarities between composite Higgs models
\cite{Agashe:2004rs,Cacciapaglia:2020kgq,Contino:2010rs,Bellazzini:2014yua,Giudice:2007fh}
 and GHU models.  The Higgs boson
appears as a pseudo-Nambu-Goldstone boson in composite Higgs models whereas 
it appears as an AB phase in the fifth dimension in GHU models.  The Higgs
boson has a character of a phase
in both models and the couplings of the Higgs boson exhibit qualitatively similar behavior.
$Z'$ bosons appear KK modes of neutral gauge bosons in GHU models whose couplings to quarks and leptons
are unambiguously determined once the models are specified.
Analogues of $Z'$ bosons 
in the composite Higgs model are composite vector
bosons\cite{Bellazzini:2012tv}. 
It is interesting to explore implications of those composite vector bosons in $e^- e^+$ collisions.

The paper is organized as follows.  
In Sec.~\ref{Sec:GHU_model}, the model is introduced.
In Sec.~\ref{Sec:Observables}, we quickly review the definition of 
observables such as cross sections, forward-backward asymmetries, left-right
asymmetries, and left-right forward-backward asymmetries. 
In Sec.~\ref{Sec:Reults}, we evaluate cross sections and other observables in 
$e^- e^+ \to f \bar f$ with $f \bar f = \mu^-\mu^+$, $c\bar{c}$, $b\bar{b}$, and $t\bar{t}$.
Section~\ref{Sec:Summary} is devoted to summary and discussions.
{Useful formulas for decay widths are given in Appendix A.

\section{Model}
\label{Sec:GHU_model}

The GUT inspired $SU(3)_C \times SO(5)\times U(1)_X$ GHU model has been
introduced in Ref.~\cite{Funatsu:2019xwr} and further investigated in
Refs.~\cite{Funatsu:2019fry,Funatsu:2020znj}. 
It is defined in the RS warped space with metric given by
\begin{align}
 ds^2= g_{MN} dx^M dx^N =e^{-2\sigma(y)} \eta_{\mu\nu}dx^\mu dx^\nu+dy^2,
\end{align} 
where $M,N=0,1,2,3,5$, $\mu,\nu=0,1,2,3$, $y=x^5$,
$\eta_{\mu\nu}=\mbox{diag}(-1,+1,+1,+1)$,
$\sigma(y)=\sigma(y+ 2L)=\sigma(-y)$,
and $\sigma(y)=ky$ for $0 \le y \le L$.
In terms of the conformal coordinate $z=e^{ky}$
($1\leq z\leq z_L=e^{kL}$) in the region $0 \leq y \leq L$ 
\begin{align}
ds^2= \frac{1}{z^2}
\bigg(\eta_{\mu\nu}dx^{\mu} dx^{\nu} + \frac{dz^2}{k^2}\bigg).
\end{align} 
The bulk region $0<y<L$ ($1<z<z_L$) is anti-de Sitter (AdS) spacetime 
with a cosmological constant $\Lambda=-6k^2$, which is sandwiched by the
UV brane at $y=0$ ($z=1$) and the IR brane at $y=L$ ($z=z_L$).  
The KK mass scale is $m_{\rm KK}=\pi k/(z_L-1) \simeq \pi kz_L^{-1}$
for $z_L\gg 1$.

Let us denote gauge fields of $SU(3)_C$,  $SO(5)$, and $U(1)_X$ by
$A_M^{SU(3)_C}$, $A_M^{SO(5)}$, and $A_M^{U(1)_X}$, respectively.
The orbifold boundary conditions (BCs) are given by
\begin{align}
&\begin{pmatrix} A_\mu \cr  A_{y} \end{pmatrix} (x,y_j-y) =
P_{j} \begin{pmatrix} A_\mu \cr  - A_{y} \end{pmatrix} (x,y_j+y)P_{j}^{-1}
\label{Eq:BC-gauge}
\end{align}
for each gauge field, where $(y_0, y_1) = (0, L)$.  In terms of  
\begin{align}
P_{\bf 3}^{SU(3)}=I_3,\ 
P_{\bf 4}^{SO(5)}=\mbox{diag}\left(I_{2},-I_{2}\right),\
P_{\bf 5}^{SO(5)}=\mbox{diag}\left(I_{4},-I_{1}\right), 
\label{Eq:SO5-BCs}
\end{align}
$P_0=P_1= P_{\bf 3}^{SU(3)}$ for  $A_M^{SU(3)_C}$  and 
$P_0=P_1= 1$ for $A_M^{U(1)_X}$.  
$P_0=P_1 = P_{\bf 5}^{SO(5)}$ for $A_M^{SO(5)}$ in the vector
representation and $P_{\bf 4}^{SO(5)}$ in the spinor representation,
respectively.
The orbifold BCs $P_{\bf 4}^{SO(5)}$ and $P_{\bf 5}^{SO(5)}$ break
$SO(5)$ to $SO(4) \simeq SU(2)_L \times SU(2)_R$.
$W$, $Z$ bosons and $\gamma$ (photon) are zero modes  in the $SO(4)$ part
of $A_\mu^{SO(5)}$, whereas the 4D Higgs boson is a zero mode in the 
$SO(5)/SO(4)$ part of $A_y^{SO(5)}$.
In the GHU model, extra neutral gauge bosons $Z'$ correspond to
KK photons $\gamma^{(n)}$,  KK $Z$ bosons $Z^{(n)}$,
and   KK $Z_R$ bosons $Z_R^{(n)}$ ($n \ge 1$),
where the $\gamma$, and $Z$, $Z_R$ bosons are the mass eigen states of
the electro-magnetic $U(1)_{\rm EM}$ neutral gauge bosons of
$SU(2)_L$, $SU(2)_R$, and $U(1)_X$.

Matter fields are introduced both in the 5D bulk and on the UV brane.
They are listed in Table~\ref{Tab:matter}.
The SM quark and lepton multiples are identified with 
the zero modes of the quark and lepton multiplets
$\Psi_{({\bf 3,4})}^{\alpha}$ $(\alpha=1,2,3)$,
$\Psi_{({\bf 3,1})}^{\pm \alpha}$, and
$\Psi_{({\bf 1,4})}^{\alpha}$
in Table~\ref{Tab:parity}.
These fields obey the following BCs:
\begin{align}
&\Psi_{({\bf 3,4})}^{\alpha} (x, y_j - y) = 
 - P_{\bf 4}^{SO(5)} \gamma^5 \Psi_{({\bf 3,4})}^{\alpha} (x, y_j + y),
 \nonumber\\
&\Psi_{({\bf 3,1})}^{\pm \alpha}  (x, y_j - y) =
\mp \gamma^5 \Psi_{({\bf 3,1})}^{\pm \alpha}  (x, y_j + y),\nonumber\\
&\Psi_{({\bf 1,4})}^{\alpha} (x, y_j - y) = 
 - P_{\bf 4}^{SO(5)} \gamma^5 \Psi_{({\bf 1,4})}^{\alpha} (x, y_j + y).
\label{quarkBC1}
\end{align}
With BCs~(\ref{quarkBC1}), the parity assignment of quarks and leptons
are summarized in Table~\ref{Tab:parity}.
(See Refs.~\cite{Funatsu:2019xwr,Funatsu:2019fry,Funatsu:2020znj} in
detail.)

\begin{table}[tbh]
\begin{center}
\begin{tabular}{|c|c|c|}
\hline
 \rowcolor[gray]{0.9}
&{B-model} &{A-model}\\
\hline
Quark
 &$({\bf 3}, {\bf 4})_{\frac{1}{6}}\ \ 
   ({\bf 3}, {\bf 1})_{-\frac{1}{3}}^+ \ \ 
   ({\bf 3}, {\bf 1})_{-\frac{1}{3}}^-$
 &$({\bf 3}, {\bf 5})_{\frac{2}{3}}\ \
   ({\bf 3}, {\bf 5})_{-\frac{1}{3}}$ \\
Lepton
 &$\strut ({\bf 1}, {\bf 4})_{-\frac{1}{2}}$ 
 &$({\bf 1}, {\bf 5})_{0} ~ ({\bf 1}, {\bf 5})_{-1}$  \\
\hline
Dark fermion
 & $({\bf 3}, {\bf 4})_{\frac{1}{6}}\ \
    ({\bf 1}, {\bf 5})_{0}^+ \ \
    ({\bf 1}, {\bf 5})_{0}^-$ 
 &$({\bf 1}, {\bf 4})_{\frac{1}{2}}$ \\
\hline
Brane fermion
 &$({\bf 1}, {\bf 1})_{0} $ 
 &$\begin{matrix} ({\bf 3}, [{\bf 2,1}])_{\frac{7}{6}, \frac{1}{6}, -\frac{5}{6}} \cr
({\bf 1}, [{\bf 2,1}])_{\frac{1}{2}, -\frac{1}{2}, -\frac{3}{2}} \end{matrix}$\\
\hline
Brane scalar &$({\bf 1}, {\bf 4})_{\frac{1}{2}} $ 
&$({\bf 1}, [{\bf 1,2}])_{\frac{1}{2}}$ \\
\hline
\end{tabular}
\caption{\small
The $SU(3)_C\times SO(5) \times U(1)_X$ content of matter fields  is shown
in the  GUT inspired model (B-model) and the previous model (A-model).
The B-model is analyzed in the present paper.
}
\label{Tab:matter}
\end{center}
\end{table}

\begin{table}[tbh]
\begin{center}
\begin{tabular}{|c|c|c|c|c|c|}
\hline
 \rowcolor[gray]{0.9}
 Field & $(SU(3)_C\times SO(5))_{X}$&$G_{22}$ &Left-handed &Right-handed &Name\\
\hline
$\Psi_{({\bf 3,4})}^{\alpha}$ &$({\bf 3,4})_{\frac{1}{6}}$&$[{\bf 2,1}]$
&$(+,+)$ &$(-,-)$ &$\begin{matrix} u & c & t \cr d & s & b\end{matrix}$\\
\cline{3-6}
&&$[{\bf 1,2}]$
&$(-,-)$ &$(+,+)$ &$\begin{matrix} u'  & c' & t' \cr d' & s' & b' \end{matrix}$\\
\hline
$\Psi_{({\bf 3,1})}^{\pm \alpha}$ &$({\bf 3,1})_{-\frac{1}{3}}$&$[{\bf 1,1}]$
&$(\pm ,\pm )$ &$(\mp , \mp )$ &$D^{\pm}_d ~ D^{\pm}_s ~ D^{\pm}_b$\\
\hline
$\Psi_{({\bf 1,4})}^{\alpha}$ &$({\bf 1,4})_{-\frac{1}{2}}$&$[{\bf 2,1}]$ 
&$(+,+)$ &$(-,-)$ &$\begin{matrix} \nu_e  & \nu_\mu & \nu_\tau \cr e & \mu & \tau \end{matrix}$\\
\cline{3-6}
&&$[{\bf 1,2}]$
&$(-,-)$ &$(+,+)$ &$\begin{matrix} \nu_e'  & \nu_\mu' & \nu_\tau' \cr e' & \mu' & \tau' \end{matrix}$\\
\hline
\end{tabular}
\caption{\small
 Parity assignment $(P_0, P_1)$ of quark and lepton multiplets in the
 bulk is shown. $G_{22}$ stands for $SU(2)_L\times SU(2)_R(\subset SO(5))$.
}
\label{Tab:parity}
\end{center}
\end{table}

The brane scalar field $\Phi_{({\bf 1}, {\bf 4})}(x)$ in
Table~\ref{Tab:matter} is responsible for breaking $SO(5)\times U(1)_X$
to $SU(2)_L\times U(1)_Y$. 
A spinor {\bf 4} of $SO(5)$ is decomposed into
$[{\bf 2}, {\bf 1}] \oplus [{\bf 1}, {\bf 2}]$ of
$SO(4) \simeq SU(2)_L \times SU(2)_R$.
The $\Phi_{({\bf 1}, {\bf 4})}$ develops a nonvanishing vacuum expectation
value (VEV):
\begin{align}
\Phi_{({\bf 1,4})} =
\begin{pmatrix} \Phi_{[{\bf 2,1}]} \cr \Phi_{[{\bf 1,2}]} \end{pmatrix},
 \ \ 
\la  \Phi_{[{\bf 1,2}]} \ra = \begin{pmatrix} 0 \cr w \end{pmatrix},
\label{scalarVEV}
\end{align}
which reduces the symmetry
$SU(3)_C\times SO(4) \times U(1)_X$ to 
the SM gauge group  $G_{\rm SM}\equiv SU(3)_C\times SU(2)_L\times U(1)_Y$.
It is assumed that $w \gg m_\KK$,  which ensures that orbifold BCs
for the 4D components of gauge fields corresponding to broken generators
in the breaking $SU(2)_R \times U(1)_X \go  U(1)_Y$ obey effectively
Dirichlet conditions at the UV brane for low-lying KK
modes\cite{Furui:2016owe}. 
Accordingly the mass of the neutral physical mode of
$\Phi_{({\bf 1,4})}$ is much larger  than $m_\KK$.

The $U(1)_Y$ gauge boson is a mixed state of $U(1)_R(\subset SU(2)_R)$
and $U(1)_X$ gauge bosons. The $U(1)_Y$ gauge field $B_M^Y$ is
given in terms of the $SU(2)_R$ gauge fields $A_M^{a_R}$
$(a_R=1_R,2_R,3_R)$ and the $U(1)_X$ gauge field $B_M$  by 
\begin{align}
&B_M^Y = s_\phi A_M^{3_R} + c_\phi  B_M ~.
\end{align}
Here the mixing angle $\phi$ between $U(1)_R$ and $U(1)_X$ is given by 
$c_\phi = \cos \phi \equiv {g_A}/{\sqrt{g_A^2+g_B^2}}$ and
$s_\phi = \sin \phi \equiv {g_B}/{\sqrt{g_A^2+g_B^2}}$ where 
$g_A$ and $g_B$ are gauge couplings in $SO(5)$ and $U(1)_X$, respectively. 
The 4D $SU(2)_L$ gauge coupling is given by $g_w = g_A/\sqrt{L}$.  
The 5D gauge coupling $g_Y^{\rm 5D}$ of $U(1)_{Y}$ and the 4D bare
Weinberg angle at the tree level, $\theta_W^0$, are given by
\begin{align}
&g_Y^{\rm 5D} =\frac{g_Ag_B}{\sqrt{g_A^2+g_B^2}}, \ \ \
\sin \theta_W^0 = \frac{s_\phi}{\sqrt{\smash[b]{1 + s_\phi^2}}}.
\label{Eq:gY-sW}
\end{align}

The 4D Higgs boson doublet $\phi_H(x)$ is the zero mode contained in the 
$A_z = (kz)^{-1} A_y$ component:
\begin{align}
A_z^{(j5)} (x, z) &= \frac{1}{\sqrt{k}} \, \phi_j (x) u_H (z) + \cdots,\
u_H (z) = \sqrt{ \frac{2}{z_L^2 -1} } \, z ~, \cr
\noalign{\kern 5pt}
\phi_H(x) &= \frac{1}{\sqrt{2}} \begin{pmatrix} \phi_2 + i \phi_1 \cr \phi_4 - i\phi_3 \end{pmatrix} .
\label{4dHiggs}
\end{align}
Without loss of generality, we assume
$\la \phi_1 \ra , \la \phi_2 \ra , \la \phi_3 \ra  =0$ and  
$\la \phi_4 \ra \not= 0$, 
which is related to the Aharonov-Bohm (AB) phase $\theta_H$ in the fifth dimension by
$\la \phi_4 \ra  = \theta_H f_H$, where
\begin{align}
&f_H  = \frac{2}{g_w} \sqrt{ \frac{k}{L(z_L^2 -1)}} ~.
\label{fH1}
\end{align}

The gauge symmetry breaking pattern of
$SU(3)_C\times SO(5)\times U(1)_X$ is given as 
\begin{align}
&SU(3)_C\times SO(5)\times U(1)_X \cr
\noalign{\kern 5pt}
&\hskip 0.5cm
\underset{BC}{\to} ~
SU(3)_C\times SU(2)_L\times SU(2)_R\times U(1)_X
 \ \ \mbox{at $y=0,L$}  \cr
\noalign{\kern 5pt}
&\hskip 0.5cm
\underset{\langle\Phi\rangle}{\to} ~
SU(3)_C\times SU(2)_L\times U(1)_Y
 \ \ \mbox{by\ the VEV $\langle\Phi_{({\bf 1},{\bf 4})}\rangle\not=0$  at\ $y=0$} \cr
\noalign{\kern 5pt}
&\hskip 0.5cm
\underset{\theta_H}{\to} ~
SU(3)_C\times U(1)_{EM}\ \ \ \  \mbox{by\ the\ Hosotani\ mechanism,}
\end{align}
where BC stands for orbifold boundary conditions.

\section{Observables}
\label{Sec:Observables}

Here we summarize formulas of several observables in the $s$-channel
scattering processes of $e^-e^+\to f\bar{f}$ mediated by only neutral
vector bosons $V_i$ such as $\gamma$ and $Z$ where $f\bar{f}\not=e^-e^+$.
For $e^-e^+\to e^-e^+$, there are  contributions not only  from 
the $s$-channel scattering process but also from the $t$-channel scattering process.
The  formulas given in this section must be modified   when 
the intermediate state of the $s$-channel scattering process contains scalar fields.
In GHU $Z'$ bosons, $\gamma^{(n)}$, $Z^{(n)}$ and $Z_R^{(n)}$ ($n \ge 1$), give additional
contributions to the $e^-e^+\to f\bar{f}$ processes, which can be observed in future 
$e^-e^+$ collider experiments.

\subsection{Cross section}

The differential cross section for the $e^-e^+\to f\bar{f}$ process
is given by 
\begin{align}
&\frac{d\sigma^{f\bar{f}}}{d\cos\theta}(P_{e^-},P_{e^+},\cos\theta) \cr
\noalign{\kern 5pt}
&\hskip 1.cm
=\left(1 - P_{e^-} P_{e^+}\right)
 \frac{1}{4}\biggl\{
 (1 - P_{\rm eff})\frac{d\sigma_{LR}^{f\bar{f}}}{d\cos\theta}(\cos\theta)
+(1 + P_{\rm eff})\frac{d\sigma_{RL}^{f\bar{f}}}{d\cos\theta}(\cos\theta)
\biggr\}
\label{Eq:dsigma_P}
\end{align}
where $P_{e^\pm}$ denotes  longitudinal polarization of $e^\pm$.
$P_{e^\pm} = +1$ corresponds to purely right-handed $e^\pm$.
$P_{\rm eff}$ is defined as
\begin{align}
 P_{\rm eff}\equiv \frac{P_{e^-}-P_{e^+}}{1-P_{e^-}P_{e^+}} ~.
\label{Eq:P_eff}
\end{align}
${d\sigma_{LR}}/{d\cos\theta}$ and
${d\sigma_{RL}}/{d\cos\theta}$ are
differential cross sections for
$e_L^-e_R^+\to f\bar{f}$ and $e_R^-e_L^+\to f\bar{f}$:
\begin{align}
\frac{d\sigma_{LR}^{f\bar{f}}}{d\cos\theta}(\cos\theta)
&= \frac{\beta s}{32\pi}\biggl\{
 [1 + \beta^{2} \cos^{2}\theta] 
 \left\{ |Q_{e_{L} f_{L}}|^{2}  
 + |Q_{e_{L} f_{R}}|^{2} \right\}
\nonumber\\& \qquad\qquad
+ 2\beta\cos\theta \left\{ 
 |Q_{e_{L} f_{L}}|^{2} 
 -  |Q_{e_{L} f_{R}}|^{2}   \right\}
+ 8 \frac{m_{f}^{2}}{s} \left[
 \Re (  Q_{e_{L} f_{L}} Q_{e_{L} f_{R}}^{*})
 \right]
\biggr\},
\allowdisplaybreaks[1]\nonumber\\
\frac{d\sigma_{RL}^{f\bar{f}}}{d\cos\theta}(\cos\theta)
&= \frac{\beta s}{32\pi}\biggl\{
 [1 + \beta^{2} \cos^{2}\theta] 
 \left\{|Q_{e_{R} f_{R}}|^{2} 
  + |Q_{e_{R} f_{L}}|^{2} \right\}
\nonumber\\& \qquad\qquad
+ 2\beta\cos\theta \left\{ 
|Q_{e_{R} f_{R}}|^{2} 
 -  |Q_{e_{R} f_{L}}|^{2} \right\}
+ 8 \frac{m_{f}^{2}}{s} \left[
 \Re( Q_{e_{R} f_{L}} Q_{e_{R} f_{R}}^{*})
\right]
\biggr\},
\label{Eq:dsigma_LR-RL}
\end{align}
where $s$ is the square of the center-of-mass energy, $m_f$ is the mass of
the final state fermion, and $\beta \equiv  \sqrt{1 - (4m_f^2/s)}$.
$Q_{e_{L} f_{R}}$ etc.\ are given by
\begin{align}
Q_{e_{L} f_{L}}
&\equiv 
\sum_{i} \frac{g_{V_{i}e}^{L} g_{V_{i}f}^{L} }{(s-m_{V_{i}}^{2}) + i m_{V_{i}}\Gamma_{V_{i}}},
&
Q_{e_{L} f_{R}}
&\equiv 
\sum_{i} \frac{g_{V_{i}e}^{L} g_{V_{i}f}^{R} }{(s-m_{V_{i}}^{2}) + i m_{V_{i}}\Gamma_{V_{i}}},
\nonumber\\
Q_{e_{R} f_{L}}
&\equiv 
\sum_{i} \frac{g_{V_{i}e}^{R} g_{V_{i}f}^{L} }{(s-m_{V_{i}}^{2}) + i m_{V_{i}}\Gamma_{V_{i}}},
&
Q_{e_{R} f_{R}}
&\equiv 
\sum_{i} \frac{g_{V_{i}e}^{R} g_{V_{i}f}^{R} }{(s-m_{V_{i}}^{2}) + i m_{V_{i}}\Gamma_{V_{i}}},
\label{Eq:Qs}
\end{align}
where $g_{V_{i}f}^{L/R}$ are  couplings of the left- and right-handed 
fermion  $f$ to the vector boson $V_{i}$, and  $m_{V_{i}}$ and $\Gamma_{V_{i}}$ are
the mass and total decay width of $V_{i}$.
For $\sqrt{s}\gg m_f$ $(\beta\simeq 1)$, the differential
cross sections in Eq.~(\ref{Eq:dsigma_LR-RL}) are approximated  by
\begin{align}
\frac{d\sigma_{LR}^{f\bar{f}}}{d\cos\theta}(\cos\theta)
&\simeq \frac{s}{32\pi}\biggl\{
 \left(1+\cos\theta\right)^2|Q_{e_{L} f_{L}}|^{2}  
+\left(1-\cos\theta\right)^2|Q_{e_{L} f_{R}}|^{2}  
\biggr\},
\nonumber\\
\frac{d\sigma_{RL}^{f\bar{f}}}{d\cos\theta}(\cos\theta)
&\simeq \frac{s}{32\pi}\biggl\{
 \left(1+\cos\theta\right)^2|Q_{e_{R} f_{R}}|^{2}  
+\left(1-\cos\theta\right)^2|Q_{e_{R} f_{L}}|^{2}  
\biggr\}.
\label{Eq:dsigma_LR-RL-mf=0}
\end{align}

We define $\sigma^{f\bar{f}}(P_{e^-},P_{e^+},[\cos\theta_1,\cos\theta_2])$
as the differential cross section integrated over the angle
$\theta=[\theta_1,\theta_2]$: 
\begin{align}
 \sigma^{f\bar{f}}(P_{e^-},P_{e^+},[\cos\theta_1,\cos\theta_2])\equiv 
 \int_{\cos\theta_1}^{\cos\theta_2}
 \frac{d\sigma^{f\bar{f}}}{d\cos\theta}(P_{e^-},P_{e^+},\cos\theta)
 d\cos\theta,
\label{Eq:sigma-integral}
\end{align}
where
$\frac{d\sigma^{f\bar{f}}}{d\cos\theta}(P_{e^-},P_{e^+},\cos\theta)$ is 
given in Eq.~(\ref{Eq:dsigma_P}).
The observed total cross section $\sigma^{f\bar{f}}_{\rm tot}(P_{e^-},P_{e^+})$
is given by
\begin{align}
 \sigma^{f\bar{f}}_{\rm tot}(P_{e^-},P_{e^+})=
 \sigma^{f\bar{f}}
 (P_{e^-},P_{e^+},[-\cos\theta_{\rm max},+\cos\theta_{\rm max}]),
\end{align}
where the available value of $\theta_{\rm max}$ depends on each experiment. 
By using the cross sections for
$e_L^-e_R^+\to f\bar{f}$ and $e_R^-e_L^+\to f\bar{f}$,
the cross section $\sigma^{f\bar{f}}_{\rm tot}(P_{e^-},P_{e^+})$
can be written by
\begin{align}
\sigma_{\rm tot}^{f\bar{f}}(P_{e^-}, P_{e^+})
 &=(1 - P_{e^-} P_{e^+}) \cdot
 \frac{1}{4}   
\biggl\{
  (1 - P_{\rm eff}) \sigma_{LR}^{f\bar{f}}
 +(1 + P_{\rm eff}) \sigma_{RL}^{f\bar{f}} \biggr\} ~.
\label{Eq:sigma_P}
\end{align}
$\sigma_{LR}^{f\bar{f}}$ and $\sigma_{RL}^{f\bar{f}}$ are given by 
\begin{align}
 \sigma_{LR}^{f\bar{f}}&=
 \int_{-\cos\theta_{\rm max}}^{+\cos\theta_{\rm max}}
\frac{d\sigma_{LR}^{f\bar{f}}}{d\cos\theta}(\cos\theta)\
 d\cos\theta,
 \nonumber\\
 \sigma_{RL}^{f\bar{f}}&=
 \int_{-\cos\theta_{\rm max}}^{+\cos\theta_{\rm max}}
\frac{d\sigma_{RL}^{f\bar{f}}}{d\cos\theta}(\cos\theta)\
 d\cos\theta.
\end{align}
For $\cos\theta_{\rm max}=1$
\begin{align}
\sigma_{LR}^{f\bar{f}}
&= \frac{\beta s}{32\pi}\biggl\{
\left[2 + \frac{2}{3}\beta^{2} \right] \{ |Q_{e_{L} f_{L}}|^{2} + |Q_{e_{L} f_{R}}|^{2} \}
+ 16 \frac{m_{f}^{2}}{s}\Re[ Q_{e_{L}f_{L}} Q_{e_{L} f_{R}}^{*}]
\biggr\},
\allowdisplaybreaks[1]\nonumber\\
\sigma_{RL}^{f\bar{f}}
&= \frac{\beta s}{32\pi}\biggl\{
\left[2 + \frac{2}{3}\beta^{2} \right] \{ |Q_{e_{R} f_{R}}|^{2} + |Q_{e_{R} f_{L}}|^{2} \}
+ 16 \frac{m_{f}^{2}}{s}\Re[ Q_{e_{R}f_{L}} Q_{e_{R} f_{R}}^{*}]
\biggr\}.
\label{Eq:sigma_LR-RL}
\end{align}
Further, for $\sqrt{s}\gg m_f$ 
\begin{align}
\sigma_{LR}^{f\bar{f}}
 &\simeq \frac{s}{12\pi}
 \left(|Q_{e_{L} f_{L}}|^{2} + |Q_{e_{L} f_{R}}|^{2}\right),
\allowdisplaybreaks[1]\nonumber\\
\sigma_{RL}^{f\bar{f}}
&\simeq \frac{s}{12\pi}
\left(|Q_{e_{R} f_{R}}|^{2} + |Q_{e_{R} f_{L}}|^{2}\right).
\label{Eq:sigma_LR-RL_mf=0}
\end{align}

The statistical error of the cross section $\Delta \sigma^{f\bar{f}}$
is given by 
\begin{align}
&\Delta \sigma^{f\bar{f}}(P_{e^-},P_{e^+},[\cos\theta_1,\cos\theta_2])=
 \frac{\sigma^{f\bar{f}}(P_{e^-},P_{e^+},[\cos\theta_1,\cos\theta_2])}{\sqrt{N^{f\bar{f}} }} ~, \cr
\noalign{\kern 5pt}
& N^{f\bar{f}} =  L_{\rm int} \cdot
 \sigma^{f\bar{f}}(P_{e^-},P_{e^+},[\cos\theta_1,\cos\theta_2]) ~,
\label{Eq:stat-error-sigma}
\end{align}
where $L_{\rm int}$ is integrated luminosity.
The amount of the deviation from the SM in  the differential cross section for $e^-e^+\to f\bar{f}$
is characterized by
\begin{align}
 \Delta_{d\sigma}^{f\bar{f}}(P_{e^-},P_{e^+},\cos\theta)\equiv 
 \frac{\myfrac{d\sigma_{\rm GHU}^{f\bar{f}}}
 {d\cos\theta}(P_{e^-},P_{e^+},\cos\theta)}
 {\myfrac{d\sigma_{\rm SM}^{f\bar{f}}}{d\cos\theta}(P_{e^-},P_{e^+},\cos\theta)}-1 ~. 
\label{Eq:Delta_dsigma}
\end{align}
Similarly, for the total cross section we introduce
\begin{align}
 \Delta_{\sigma}^{f\bar{f}}(P_{e^-},P_{e^+})\equiv 
 \frac{\sigma_{\rm GHU}^{f\bar{f}}(P_{e^-},P_{e^+})}
 {\sigma_{\rm SM}^{f\bar{f}}(P_{e^-},P_{e^+})}-1 ~.
\label{Eq:Delta_sigma}
\end{align}

\subsection{Forward-backward asymmetry}

The forward-backward asymmetry $A_{FB}^{{f\bar{f}}}(P_{e^-},P_{e^+})$
\cite{Schrempp:1987zy,Kennedy:1988rt} is given by
\begin{align}
A_{FB}^{{f\bar{f}}}(P_{e^-},P_{e^+})
&=\frac{\sigma_F^{{f\bar{f}}}(P_{e^-},P_{e^+})
  -\sigma_B^{{f\bar{f}}}(P_{e^-},P_{e^+})}
 {\sigma_F^{{f\bar{f}}}(P_{e^-},P_{e^+})
 +\sigma_B^{{f\bar{f}}}(P_{e^-},P_{e^+})} ~, \cr
\noalign{\kern 5pt}
\sigma_F^{f\bar{f}}(P_{e^-},P_{e^+})&=
 \sigma^{f\bar{f}}(P_{e^-},P_{e^+},[0,+\cos\theta_{\rm max}]) ~, \cr
\noalign{\kern 5pt}
\sigma_B^{{f\bar{f}}}(P_{e^-},P_{e^+})&=
 \sigma^{f\bar{f}}(P_{e^-},P_{e^+},[-\cos\theta_{\rm max},0]) ~,
\label{Eq:A_FB}
\end{align}
where the available value of $\theta_{\rm max}$ depends on each experiment.
For $\sqrt{s}\gg m_f$ and $\cos \theta_{\rm max} =1$
\begin{align}
&A_{FB}^{f\bar{f}}(P_{e^-},P_{e^+}) \simeq \frac{3}{4} \frac{B_1 - B_2}{B_1 + B_2} ~, \cr
\noalign{\kern 5pt}
&\quad 
B_1 = (1+P_{\rm eff}) |Q_{e_{R} f_{R}}|^{2} + (1-P_{\rm eff})|Q_{e_{L} f_{L}}|^{2} ~, \cr
\noalign{\kern 5pt}
&\quad 
B_2 = (1+P_{\rm eff}) |Q_{e_{R} f_{L}}|^{2} + (1-P_{\rm eff})|Q_{e_{L} f_{R}}|^{2} ~, 
\label{Eq:A_FB-mf=0}
\end{align}
where $P_{\rm eff}$ is given in Eq.~(\ref{Eq:P_eff}).

The statistical error of the forward-backward asymmetry $\Delta
A_{FB}^{f\bar{f}}$ is given by
\begin{align}
 \Delta A_{FB}^{f\bar{f}} &=
 2 \frac{\sqrt{ n_1 n_2}
 \left(\sqrt{n_1}+\sqrt{n_2}\right)}
 {(n_1 + n_2)^2}  \cr
\noalign{\kern 5pt}
& =\frac{2\sqrt{n_1 n_2}}
 {(n_1 + n_2)
 \left(\sqrt{n_1}-\sqrt{n_2}\right)} \, 
A_{FB}^{f\bar{f}} ~, \cr 
\noalign{\kern 5pt}
&(n_1, n_2) = (N_F^{f\bar{f}} , N_B^{f\bar{f}} ) ~, 
\label{Eq:Error-A_FB}
\end{align}
where
$N_{F/B}^{f\bar{f}}=
L_{\rm int} \cdot  \sigma_{F/B}^{f\bar{f}}(P_{e^-},P_{e^+})$ is the number of   events.
The amount of the deviation from the SM is characterized by
\begin{align}
 \Delta_{A_{FB}}^{f\bar{f}}\equiv 
 \frac{A_{FB,{\rm GHU}}^{f\bar{f}}}{A_{FB,{\rm SM}}^{f\bar{f}}}-1 ~.
\label{Eq:Delta_A_FB}
\end{align}

\subsection{Left-right asymmetry}

The left-right asymmetry
\cite{Schrempp:1987zy,Kennedy:1988rt,MoortgatPick:2005cw}
is given by
\begin{align}
A_{LR}^{{f\bar{f}}}(\cos\theta)
 =  \frac{
 \sigma_{LR}^{f\bar{f}}(\cos\theta)
 -\sigma_{RL}^{f\bar{f}}(\cos\theta)
 }
 {
 \sigma_{LR}^{f\bar{f}}(\cos\theta)
 +\sigma_{RL}^{f\bar{f}}(\cos\theta)
 },
\label{Eq:A_LR-cos}
\end{align}
where
$\sigma_{LR}^{f\bar{f}}(\cos\theta)$ and
$\sigma_{RL}^{f\bar{f}}(\cos\theta)$ stand for
$\frac{d\sigma_{LR}^{f\bar{f}}}{d\cos\theta}(\cos\theta)$ and 
$\frac{d\sigma_{RL}^{f\bar{f}}}{d\cos\theta}(\cos\theta)$
in Eq.~(\ref{Eq:dsigma_LR-RL}), respectively.
For $\sqrt{s}\gg m_f$, 
\begin{align}
A_{LR}^{f\bar{f}}(\cos\theta)
 &\simeq \frac{(1+\cos\theta)^2
 \left(|Q_{e_{L} f_{L}}|^{2}-|Q_{e_{R} f_{R}}|^{2}\right)
 +(1-\cos\theta)^2
 \left(|Q_{e_{L} f_{R}}|^{2}-|Q_{e_{R} f_{L}}|^{2}\right)
}
{(1+\cos\theta)^2
 \left(|Q_{e_{L} f_{L}}|^{2}+|Q_{e_{R} f_{R}}|^{2}\right)
 +(1-\cos\theta)^2
 \left(|Q_{e_{L} f_{R}}|^{2}+|Q_{e_{R} f_{L}}|^{2}\right)
 }.
\label{Eq:A_LR-cos-mf=0} 
\end{align}
The observable left-right asymmetry is given by
\begin{align}
A_{LR}^{{f\bar{f}}}(P_{e^-},P_{e^+},\cos\theta)
& =  \frac{
 \sigma^{f\bar{f}}(P_{e^-},P_{e^+},\cos\theta)
 -\sigma^{f\bar{f}}(-P_{e^-},-P_{e^+},\cos\theta)
 }
 {
 \sigma^{f\bar{f}}(P_{e^-},P_{e^+},\cos\theta)
 +\sigma^{f\bar{f}}(-P_{e^-},-P_{e^+},\cos\theta)
 }
\label{Eq:A_LR-cos-obs}
\end{align}
for $P_{e^-}<0$ and $|P_{e^-}|>|P_{e^+}|$,
where
$\sigma^{f\bar{f}}(P_{e^-},P_{e^+},\cos\theta)$ and 
$\sigma^{f\bar{f}}(-P_{e^-},-P_{e^+},\cos\theta)$ stand for
$\frac{d\sigma^{f\bar{f}}}{d\cos\theta}(P_{e^-},P_{e^+},\cos\theta)$ and 
$\frac{d\sigma^{f\bar{f}}}{d\cos\theta}(-P_{e^-},-P_{e^+},\cos\theta)$
in Eq.~(\ref{Eq:dsigma_P}), respectively.
(\ref{Eq:A_LR-cos-obs}) is related to (\ref{Eq:A_LR-cos}) by
\begin{align}
A_{LR}^{{f\bar{f}}}(\cos\theta)=\frac{1}{P_{\rm eff}}
A_{LR}^{{f\bar{f}}}(P_{e^-},P_{e^+},\cos\theta) ~.
\label{A_LR_cos}
\end{align}

The {integrated left-right asymmetry
$A_{LR}^{f\bar{f}}$
\cite{Schrempp:1987zy,Kennedy:1988rt}
is given by
\begin{align}
 A_{LR}^{{f\bar{f}}}
 &= \frac{\sigma_{LR}^{{f\bar{f}}}-\sigma_{RL}^{{f\bar{f}}}}
 {\sigma_{LR}^{{f\bar{f}}}+ \sigma_{RL}^{{f\bar{f}}}}.
\label{Eq:A_LR}
\end{align}
In terms of  $Q_{e_Xf_Y} (X,Y=L,R)$ in Eq.~(\ref{Eq:Qs}),
$A_{LR}^{{f\bar{f}}}$ is expressed as
\begin{align}
 A_{LR}^{{f\bar{f}}} &= \frac{C_-}{C_+} ~, \cr
 \noalign{\kern 5pt}
 C_\pm &= \Big( 1 + \frac{1}{3}\beta^{2} \Big) 
 \Big\{  \big[ |Q_{e_{L} f_{L}}|^{2} + |Q_{e_{L} f_{R}}|^{2} \big]
 \pm  \big[ |Q_{e_{R} f_{R}}|^{2} + | Q_{e_{R} f_{L}}|^{2} \big] \Big\} \cr
\noalign{\kern 5pt}
&\hskip 1.cm
+ 8 \frac{m_{f}^{2}}{s} \,  \Big\{  \Re(Q_{e_{L}f_{L}} Q_{e_{L} f_{R}}^{*})
\pm \Re( Q_{e_{R} f_{R}} Q_{e_{R} f_{L}}^{*}) \Big\} ~.
\label{Eq:A_LR-Qs}
\end{align}
For $\sqrt{s}\gg m_{f}$,
\begin{align}
 A_{LR}^{{f\bar{f}}}
 &\simeq 
\frac{
[|Q_{e_{L} f_{L}}|^{2} + |Q_{e_{L} f_{R}}|^{2}]
 - [|Q_{e_{R} f_{R}}|^{2} + | Q_{e_{R} f_{L}}|^{2}] 
}{
[|Q_{e_{L} f_{L}}|^{2} + |Q_{e_{L} f_{R}}|^{2}]
 + [|Q_{e_{R} f_{R}}|^{2} + | Q_{e_{R} f_{L}}|^{2}] 
}.
\label{Eq:A_LR-mf=0}
\end{align}
The observable left-right asymmetry is given by 
\begin{align}
A_{LR}^{{f\bar{f}}}(P_{e^-},P_{e^+})
 =  \frac{
 \sigma^{f\bar{f}}(P_{e^-},P_{e^+})
 -\sigma^{f\bar{f}}(-P_{e^-},-P_{e^+})
 }
 {
 \sigma^{f\bar{f}}(P_{e^-},P_{e^+})
 +\sigma^{f\bar{f}}(-P_{e^-},-P_{e^+})
 }
\label{Eq:A_LR-obs}
\end{align}
for $P_{e^-}<0$ and $|P_{e^-}|>|P_{e^+}|$.
It is related to (\ref{Eq:A_LR}) by
\begin{align}
A_{LR}^{{f\bar{f}}}=\frac{1}{P_{\rm eff}}
A_{LR}^{{f\bar{f}}}(P_{e^-},P_{e^+}) ~.
\end{align}

The statistical error of the left-right asymmetry $\Delta A_{LR}^{f\bar{f}}$ is given by
\begin{align}
 \Delta A_{LR}^{f\bar{f}} &=
 2\frac{\sqrt{N_{LR}^{f\bar{f}}N_{RL}^{f\bar{f}}}
 \left(\sqrt{N_{LR}^{f\bar{f}}}+\sqrt{N_{RL}^{f\bar{f}}}\right)}
 {(N_{LR}^{f\bar{f}}+N_{RL}^{f\bar{f}})^2} \cr
 \noalign{\kern 5pt}
 &
 =\frac{2\sqrt{N_{LR}^{f\bar{f}}N_{RL}^{f\bar{f}}}}
 {(N_{LR}^{f\bar{f}}+N_{RL}^{f\bar{f}})
 \left(\sqrt{N_{LR}^{f\bar{f}}}-\sqrt{N_{RL}^{f\bar{f}}}\right)}
 A_{LR}^{f\bar{f}} ~,
\label{Eq:Error-A_LR}
\end{align}
where $N_{LR}^{f\bar{f}}=L_{\rm int} \, \sigma_{LR}^{f\bar{f}}$
and $N_{RL}^{f\bar{f}}=L_{\rm int} \, \sigma_{RL}^{f\bar{f}}$
are the numbers of the events.
The amount of the deviation from the SM in (\ref{A_LR_cos}) and (\ref{Eq:A_LR}) is 
characterized by
\begin{align}
\Delta_{A_{LR}}^{f\bar{f}}(\cos\theta) &\equiv 
\frac{A_{LR,{\rm GHU}}^{{f\bar{f}}}(\cos\theta)}{A_{LR,{\rm SM}}^{{f\bar{f}}}(\cos\theta)}  -1 ~, \cr
\noalign{\kern 5pt}
\Delta_{A_{LR}}^{f\bar{f}} &\equiv 
 \frac{A_{LR,{\rm GHU}}^{f\bar{f}}}{A_{LR,{\rm SM}}^{f\bar{f}}}-1 ~.
\label{Eq:Delta_A_LR}
\end{align}

\subsection{Left-right forward-backward asymmetry}

The left-right forward-backward asymmetry
\cite{Blondel:1987gp,Kennedy:1988rt,Abe:1994bj,Abe:1994bm,Abe:1995yh}
is given by
\begin{align}
A_{LR,FB}^{f\bar{f}}(\cos\theta)&=
\frac{\left[\sigma_{LR}^{f\bar{f}}(\cos\theta)
- \sigma_{RL}^{f\bar{f}}(\cos\theta)\right]
- \left[\sigma_{LR}^{f\bar{f}}(-\cos\theta)
- \sigma_{RL}^{f\bar{f}}(-\cos\theta)\right]
}{
\left[\sigma_{LR}^{f\bar{f}}(\cos\theta)
+ \sigma_{RL}^{f\bar{f}}(\cos\theta)\right]
+ \left[\sigma_{LR}^{f\bar{f}}(-\cos\theta)
+ \sigma_{RL}^{f\bar{f}}(-\cos\theta)\right]} ~.
\label{Eq:A_LRFB}
\end{align}
In terms of  $Q_{e_Xf_Y} (X,Y=L,R)$ in Eq.~(\ref{Eq:Qs}),
$A_{LR,FB}^{{f\bar{f}}}$ is expressed as
\begin{align}
&A_{LR,FB}^{f\bar{f}}(\cos\theta) \cr
\noalign{\kern 5pt}
&= \frac{ 2\beta \cos\theta \, D_-} 
{ (1+\beta^{2}\cos^{2}\theta) D_+ + 8(m_{f}^{2}/s)  \big[ \Re(Q_{e_{L} f_{L}} Q_{e_{L} f_{R}}^{*})
  +\Re(Q_{e_{R} f_{R}} Q_{e_{R} f_{L}}^{*}) \big]}  ~, \cr
\noalign{\kern 5pt}
&D_\pm =  \big( |Q_{e_{L} f_{L}}|^{2} + |Q_{e_{R} f_{L}}|^{2} \big) 
          \pm \big(  |Q_{e_{L} f_{R}}|^{2} + |Q_{e_{R} f_{R}}|^{2} \big)  ~.  
\end{align}
For $\sqrt{s}\gg m_f$,
\begin{align}
A_{LR,FB}^{f\bar{f}}(\cos\theta)&\simeq
 \frac{2 \cos\theta}{1+\cos^2\theta} \frac{D_-}{D_+} 
\label{Eq:A_LRFB-mf=0}
\end{align}
The observable left-right forward-backward asymmetry is given by
\begin{align}
&A_{LR,FB}^{{f\bar{f}}}(P_{e^-},P_{e^+},\cos\theta) = \frac{E_-}{E_+}~, \cr
\noalign{\kern 5pt}
&E_\pm =  \big[  \sigma^{f\bar{f}}(P_{e^-},P_{e^+},\cos\theta) + \sigma^{f\bar{f}}(-P_{e^-},-P_{e^+},-\cos\theta) \big] \cr
\noalign{\kern 5pt}
&\hskip 1.2cm
\pm  \big[ \sigma^{f\bar{f}}(-P_{e^-},-P_{e^+},\cos\theta) +\sigma^{f\bar{f}}(P_{e^-},P_{e^+},-\cos\theta) \big] 
\label{Eq:A_LRFB-cos-obs}
\end{align}
for $P_{e^-}<0$ and $|P_{e^-}|>|P_{e^+}|$.
The relation between $A_{LR,FB}^{{f\bar{f}}}(\cos\theta)$ in
Eq.~(\ref{Eq:A_LRFB}) and
$A_{LR,FB}^{{f\bar{f}}}(P_{e^-},P_{e^+},\cos\theta)$ in 
Eq.~(\ref{Eq:A_LRFB-cos-obs}) is given by
\begin{align}
A_{LR,FB}^{{f\bar{f}}}(\cos\theta)=\frac{1}{P_{\rm eff}}
A_{LR,FB}^{{f\bar{f}}}(P_{e^-},P_{e^+},\cos\theta) ~. 
\end{align}

The statistical error of the left-right forward-backward asymmetry
$\Delta A_{LR,FB}$ is given by
\begin{align}
\Delta A_{LR,FB}
& =2\frac{(n_3+ n_2) \left(\sqrt{n_1} + \sqrt{n_4} \right)
 +(n_1 + n_4)  \left(\sqrt{n_3} + \sqrt{n_2}\right) }{(n_1 + n_3 + n_2 + n_4)^2}
 \cr
\noalign{\kern 5pt} 
 &=2\frac{(n_3+ n_2) \left(\sqrt{n_1} + \sqrt{n_4} \right)
 +(n_1 + n_4)  \left(\sqrt{n_3} + \sqrt{n_2}\right) }{(n_1 + n_4)^2 - (n_3 + n_2)^2} \,  A_{LR,FB}   ~,  \cr
\noalign{\kern 5pt}
&\hskip -1.5cm 
(n_1, n_2, n_3, n_4) = (N_{LRF}^{f\bar{f}},  N_{RLF}^{f\bar{f}}, N_{LRB}^{f\bar{f}},  N_{RLB}^{f\bar{f}} ) ~, 
\label{Eq:Error-A_LRFB}
\end{align}
where
$N_{XF}^{f\bar{f}}
=L_{\rm int} \cdot \sigma_{X}^{f\bar{f}}([\cos\theta_1,\cos\theta_2])$
and
$N_{XB}^{f\bar{f}}
=L_{\rm int} \cdot \sigma_{X}^{f\bar{f}}([-\cos\theta_2,-\cos\theta_1])$
$(X=LR,RL; \, 0< \cos \theta_1< \cos \theta_2 )$
are the numbers of the events.
The amount of the deviation in $A_{LR,FB}$ from the SM is characterized by
\begin{align}
 \Delta_{A_{LR,FB}}^{f\bar{f}}(\cos\theta)\equiv 
 \frac{A_{LR,FB,{\rm GHU}}^{f\bar{f}}(\cos\theta)}
{A_{LR,FB,{\rm SM}}^{f\bar{f}}(\cos\theta)} -1 ~.
\label{Eq:Delta-A_LRFB}
\end{align}

\section{Fermion pair production via $Z'$ mediation}
\label{Sec:Reults}

In this section we calculate various observables of the $s$-channel scattering process
of $e^-e^+\to f\bar{f}$ mediated by  neutral vector bosons $V$ in
the GHU, where $V =\gamma,Z,Z^{(n)},Z_R^{(n)},\gamma^{(n)}$ ($n \ge 1$), and
$f\bar{f}=\mu^-\mu^+$, $c\bar{c}$, $b\bar{b}$, $t\bar{t}$.

\subsection{Parameter sets}

Parameters of the model are determined in the steps described in 
Refs.~\cite{Funatsu:2019xwr, Funatsu:2019fry,Funatsu:2020znj}.

\noindent
(i) We pick the values of $\theta_H$ and $m_{\KK}=  \pi k (z_L -1)^{-1}$. 

\noindent
(ii) $k$ is determined in order for the $Z$ boson mass $m_Z$ to be reproduced, 
which fixes the warped factor $z_L$ as well.

\noindent
(iii) The bare Weinberg angle $\theta_W^0$ in Eq.~(\ref{Eq:gY-sW}) with given $\theta_H$ 
is not known beforehand.  It is determined self-consistently to  fit the observed 
forward-backward asymmetry $A_{FB}(e^-e^+\to\mu^-\mu^+)=0.0169\pm0.0013$
at $\sqrt{s}=m_Z$\cite{ALEPH:2005ema,Tanabashi:2018oca}, after evaluating the lepton gauge couplings 
with the procedure described below.   We have checked that self-consistent value of $\theta_W^0$ is 
found after a couple of iterations of this process.
For instance, for $\theta_H=0.10$ and $m_\KK=13 \,$TeV,  $\sin \theta_W^0=0.2305$ yields
$A_{FB}(e^-e^+\to\mu^-\mu^+)=0.01693$ at $\sqrt{s}=m_Z$.   
If one chooses $\sin\theta_W^0=0.2313 ~ (0.2298)$ instead, then one finds 
$A_{FB}(e^-e^+\to\mu^-\mu^+)=0.01562 ~(0.01821)$.
It has been shown in \cite{Funatsu:2014fda,Funatsu:2016uvi} that $\sin\theta_W^0=0.2305$
yields $W$ and $Z$ coupling constants of quarks and leptons which are nearly the same as
those in the SM with $\sin^2 \theta_W = 0.2312$.
In our analysis, we will use the values of $\sin\theta_W^0$
for each set of $\theta_H$ and $m_\KK$ that reproduce the central value
of $A_{FB}(e^-e^+\to\mu^-\mu^+)$.

\noindent
(iv) With given $\sin \theta_W^0$, wave functions of gauge bosons  are fixed.

\noindent
(v)  The bulk mass parameters of $\Psi_{({\bf 3,4})}^\alpha$ and
$\Psi_{({\bf 1,4})}^\alpha$ are fixed from the masses of up-type quarks and charged leptons.

\noindent
(vi) The bulk mass parameters of $\Psi_{({\bf 3,1})}^{\pm \alpha}$ and brane interaction coefficients
in the down-quark sector are determined so as to reproduce the masses of down-type quarks. 
Similarly the Majorana mass terms  and brane interactions in the neutrino sector are determined so as
to reproduce neutrino masses.
We use the masses of quarks and leptons given by
$m_u=20\,$MeV, $m_c=619\,$MeV, $m_t=172.9\,$GeV, 
$m_d=2.9\,$MeV, $m_s=55\,$MeV, $m_b=2.89\,$GeV, 
$m_e=0.486\,$MeV, $m_\mu=102.7\,$MeV, $m_\tau=1.746\,$GeV,
$m_{\nu_e}=m_{\nu_\mu}=m_{\nu_\tau}=10^{-12}\,$GeV.
{As discussed in Ref.~\cite{Funatsu:2019xwr},
left-handed and right-handed up- and down-type
quarks $(u, d, u', d')$, $(c, s, c', s')$, $(t, b, t', b')$
belong to the same multiplet $\Psi_{({\bf 3,4})}^{\alpha}$ shown in Table~\ref{Tab:parity} 
in each generation so that the up- and down-type quarks have a degenerate
mass in each generation in the absence of mixing among 
$(d,  d'),  (s,  s'),  (b,  b')$ and  $D_{d}^{\pm} ,  D_{s}^{\pm} ,  D_{b}^{\pm}$, respectively.
The mixing resolves the degeneracy between up- and down-type quarks
in each generation, but always makes the down-type quark lighter than the up-type quark.
For this reason we adopt the value $m_u > m_d$ at the moment.
It is left as a future task to explain the observed $m_u$ in the GUT inspired GHU.}

With these parameters fixed, wave functions of quarks and leptons are determined.
In the present paper we mostly ignore the flavor mixing in the quark and lepton gauge 
couplings
\cite{Cabibbo:1963yz,Kobayashi:1973fv,Maki:1962mu, Cacciapaglia:2007fw,Adachi:2010cc,Adachi:2011tn,Adachi:2011cb,Funatsu:2019fry}.
{It has been shown that the Cabibbo-Kobayashi-Maskawa (CKM) mixing matrix can be incorporated in GHU with
naturally suppressed FCNCs (flavor changing neutral currents)\cite{Funatsu:2019fry}.
FCNC couplings are suppressed  by a factor of $O(10^{-6})$.
There arise  flavor changing couplings of $Z'$ bosons in the down-type quark sector.
For $\theta_H = 0.1$ and $m_\KK = 13\,$TeV, the $Z^{(1)}$ couplings
in the down-type quark sector, for instance,  are given by
\begin{align}
g^L_{Z^{(1)} d} &= \begin{pmatrix} 
-2.6792&	-0.0215  &	-0.0001 \cr
-0.0215 &	-2.5907 &	-0.0018 \cr
-0.0001 &	-0.0018 &	-2.1284 \end{pmatrix}  g_w,  \cr
\noalign{\kern 5pt}
g^R_{Z^{(1)} d} &= \begin{pmatrix} 
0.1907 &	-0.0420 &	0.0144 \cr
-0.0420 &	0.0301 &	-0.0436 \cr
0.0144 &	-0.0436 &	0.2786 \end{pmatrix} g_w
\label{FCZprime}
\end{align}
with typical brane interactions yielding the CKM matrix approximately.
Flavor changing $Z'$ couplings in the left-handed components are very small
compared to diagonal ones.  Flavor changing $Z'$ couplings in the right-handed components
are slightly bigger, but their magnitude is small.
In the processes $e^-e^+ \to f \bar f$, the effect of flavor changing $Z'$ couplings remains
very small  for $\sqrt{s}< 3\,$TeV.  
In the following analysis we shall safely ignore these flavor changing $Z'$ couplings in the down-type quark sector.
}
%We remark that FCNCs through $Z'$ bosons can be observed in the process 
% $e^-e^+ \to b \bar s,  s \bar b$ for $\sqrt{s} \sim m_{Z'}$.

With the parameter set given,  the $Z'$ coupling constants to the SM fermions, etc. are determined.   
To evaluate the cross section and other quantities in the processes  $e^-e^+ \to f \bar f$, 
we need to know the four-dimensional $Z'$ couplings of quarks and leptons. They are 
obtained from the five-dimensional gauge interaction terms by inserting
wave functions of gauge bosons and quarks or leptons and integrating
over the fifth-dimensional coordinate\cite{Funatsu:2014fda,Funatsu:2015xba,Funatsu:2016uvi}.
Decay widths of $Z'$ bosons are calculated by using the
formulas in Appendix~\ref{Sec:Decay-width} with masses and various couplings of  $Z'$ bosons.
(For the total decay widths of $Z'$s, we take into account   the two body
decays at tree level approximation.) 
The masses and widths of $\gamma$, $Z$ boson, and the first neutral KK
vector bosons $Z^{(1)}$, $Z_R^{(1)}$, $\gamma^{(1)}$ are listed in 
Table~\ref{Table:Mass-Width-Vector-Bosons}.
The coupling constants of $Z$ boson and the first neutral KK vector
bosons $Z^{(1)}$, $Z_R^{(1)}$, $\gamma^{(1)}$ to quarks and leptons 
are listed in Tables~\ref{Table:Couplings-Zprime_thetaH=010-mKK=13},
\ref{Table:Couplings-Zprime_thetaH=010-mKK=11},
\ref{Table:Couplings-Zprime_thetaH=010-mKK=15},
\ref{Table:Couplings-Zprime_thetaH=011-mKK=13},
\ref{Table:Couplings-Zprime_thetaH=009-mKK=13}.
In Table~\ref{Table:Mass_Couplings-Zprime_thetaH=010-mKK=13-Higher-KK},
masses of neutral higher KK vector bosons $Z^{(2k-1)}$, $Z^{(2k)}$,
$Z_R^{(k)}$, $\gamma^{(k)}$ $(k=1,2,\cdots,10)$ and their couplings
constants to left- and right-handed electrons are summarized.
We note that possible values of $z_L$ is restricted with given  $\theta_H$.
It has been shown in Ref.~\cite{Funatsu:2020znj} that for $\theta_H=0.10$
the top quark mass can be reproduced only if $z_L\geq 10^{8.1}$ and
dynamical electroweak symmetry breaking is achieved only if
$z_L\leq 10^{15.5}$, the values of which correspond to $m_{\rm KK}\simeq[11,15] \,$TeV.

\begin{table}[thb]
{\footnotesize
\begin{center}
\begin{tabular}{c|cc|cccccccc|c}
\hline
 \rowcolor[gray]{0.9}
 &&&&&&&&&&&\\[-0.75em]
\rowcolor[gray]{0.9}
 Name&$\theta_H$&$m_{\KK}$&$z_L$&$k$
 &$m_{\gamma^{(1)}}$&$\Gamma_{\gamma^{(1)}}$
 &$m_{Z^{(1)}}$&$\Gamma_{Z^{(1)}}$
 &$m_{Z_R^{(1)}}$&$\Gamma_{Z_R^{(1)}}$
 &Table\\
\rowcolor[gray]{0.9}
 &\mbox{[rad.]}&[TeV]&&[GeV]&[TeV]&[TeV]&[TeV]&[TeV]&[TeV]&[TeV]
 &\\ 
\hline 
 &&&&&&&&&&\\[-0.75em]
 \hspace{0.5em}B$^{\rm L}$&0.10&11.00&1.980$\times10^{8\ }$&6.933$\times10^{11}$&8.715&2.080&8.713&4.773&8.420&0.603
 &\ref{Table:Couplings-Zprime_thetaH=010-mKK=11}\\
 B&0.10&13.00&3.865$\times10^{11}$ &1.599$\times10^{15}$&10.20&3.252&10.20&7.840&9.951&0.816
 &\ref{Table:Couplings-Zprime_thetaH=010-mKK=13}\\
 \hspace{0.5em}B$^{\rm H}$&0.10&15.00&2.667$\times10^{15}$&1.273$\times10^{19}$&11.69&4.885&11.69&11.82&11.48&1.253
 &\ref{Table:Couplings-Zprime_thetaH=010-mKK=15}\\
\hline
\end{tabular}\\[0.5em]
\begin{tabular}{c|cc|cccccccc|c}
\hline
 \rowcolor[gray]{0.9}
 &&&&&&&&&&&\\[-0.75em]
\rowcolor[gray]{0.9}
 Name&$\theta_H$&$m_{\KK}$&$z_L$&$k$
 &$m_{\gamma^{(1)}}$&$\Gamma_{\gamma^{(1)}}$
 &$m_{Z^{(1)}}$&$\Gamma_{Z^{(1)}}$
 &$m_{Z_R^{(1)}}$&$\Gamma_{Z_R^{(1)}}$
 &Table\\
\rowcolor[gray]{0.9}
 &\mbox{[rad.]}&[TeV]&&[GeV]&[TeV]&[TeV]&[TeV]&[TeV]&[TeV]&[TeV]
 &\\ 
\hline 
 &&&&&&&&&&\\[-0.75em]
\hspace{0.5em}B$^+$&0.11&13.00&
 1.021$\times10^{14}$&4.223$\times10^{17}$&10.15&3.836&10.15&9.374&9.951&0.924
 &\ref{Table:Couplings-Zprime_thetaH=011-mKK=13}\\
 B&0.10&13.00&
 3.865$\times10^{11}$&1.599$\times10^{15}$&10.20&3.252&10.20&7.840&9.951&0.816
 &\ref{Table:Couplings-Zprime_thetaH=010-mKK=13}\\
\hspace{0.5em}B$^-$&0.09&13.00&
 2.470$\times10^{9\ }$ &1.022$\times10^{13}$&10.26&2.723&10.26&6.413&9.951&0.732
 &\ref{Table:Couplings-Zprime_thetaH=009-mKK=13}\\
\hline
\end{tabular}
 \caption{\small
 Masses and widths of $Z'$ bosons ($Z^{(1)}$, $\gamma^{(1)}$, and
 $Z_R^{(1)}$) are listed for
 $\theta_H=0.10$ and three $m_{\rm KK}=11,13,15\,$TeV values  in the upper table,  and
 $m_{\rm KK}=13\,$TeV and three $\theta_H=0.11,0.10,0.09$ values in the  lower table.
 $m_{Z}=91.1876\,$GeV and $\Gamma_{Z}=2.4952 \,$GeV\cite{Tanabashi:2018oca}.
The column ``Name''   denotes each parameter set 
 and the column  ``Table'' indicate the table summarizing  coupling constants in each set.
 }
\label{Table:Mass-Width-Vector-Bosons}
\end{center}
}
\end{table}

\begin{table}[htb]
\begin{center}
\begin{tabular}{ccccccccc}
\hline
 \rowcolor[gray]{0.9}
 &&&&&&&&\\[-0.75em]
\rowcolor[gray]{0.9}
 $f$
 &$g_{Zf}^L$&$g_{Zf}^R$
 &$g_{Z^{(1)}f}^L$&$g_{Z^{(1)}f}^R$
 &$g_{Z_R^{(1)}f}^L$&$g_{Z_R^{(1)}f}^R$
 &$g_{\gamma^{(1)}f}^L$&$g_{\gamma^{(1)}f}^R$\\
 \hline
 $\nu_e$
 &\ \ 0.5687&0
 &\ \ 3.2774&0
 &$-$1.0322&0
 &0&0
 \\
 $\nu_\mu$
 &\ \ 0.5687&0
 &\ \ 3.1207&0
 &$-$0.9852&0
 &0&0
 \\
 $\nu_\tau$
 &\ \ 0.5687&0
 &\ \ 3.0165&0
 &$-$0.9539&0
 &0&0
 \\[0.5em]
 $e$
 &$-$0.3058&\ \ 0.2629
 &$-$1.7621&$-$0.0584
 &$-$1.0444&0
 &$-$2.7587&\ \ 0.1071
 \\
 $\mu$
 &$-$0.3058&\ \ 0.2629
 &$-$1.6778&$-$0.0584
 &$-$0.9969&0
 &$-$2.6268&\ \ 0.1071
 \\
 $\tau$
 &$-$0.3058&\ \ 0.2629
 &$-$1.6218&$-$0.0584
 &$-$0.9652&\ \ 0.0001
 &$-$2.5391&\ \ 0.1070
 \\[0.5em]
 $u$
 &\ \ 0.3934&$-$0.1753
 &\ \ 2.1951&\ \ 0.0390
 &\ \ 0.3415&0
 &\ \ 1.7807&$-$0.0714
 \\
 $c$
 &\ \ 0.3934&$-$0.1753
 &\ \ 2.1147&\ \ 0.0389
 &\ \ 0.3296&0
 &\ \ 1.7154&$-$0.0714
 \\
 $t$
 &\ \ 0.3938&$-$0.1749
 &\ \ 1.7406&$-$0.3269
 &\ \ 0.2740&$-$0.7395
 &\ \ 1.4121&\ \ 0.6017
 \\[0.5em]
 $d$
 &$-$0.4811&\ \ 0.0876
 &$-$2.6842&\ \ 0.1162
 &\ \ 0.3297&$-$0.1801
 &$-$0.8904&$-$0.2113
 \\
 $s$
 &$-$0.4811&\ \ 0.0876
 &$-$2.5858&\ \ 0.1460
 &\ \ 0.3182&$-$0.2197
 &$-$0.8577&$-$0.2657
 \\
 $b$
 &$-$0.4811&\ \ 0.0876
 &$-$2.1284&\ \ 0.2900
 &\ \ 0.2646&$-$0.4096
 &$-$0.7059&$-$0.5279
 \\
\hline
\end{tabular}
 \caption{\small
 Coupling constants of neutral vector bosons, $Z'$ bosons, to
 fermions in units of $g_w=e/\sin\theta_W^0$
 are listed  for $\theta_H=0.10$ and $m_{\KK}=13.00 \,$TeV 
 (B) in Table~\ref{Table:Mass-Width-Vector-Bosons},
 where $\sin^2\theta_W^0=0.2306$. 
 Their corresponding $Z$ boson coupling constants in the SM are
 $(g_{Z_{\nu}}^L,g_{Z_{\nu}}^R)=(0.5703,0)$, 
 $(g_{Z_{e}}^L,g_{Z_{e}}^R)=(-0.3065,0.2638)$, 
 $(g_{Z_{u}}^L,g_{Z_{u}}^R)=(0.3944,-0.1748)$, 
 $(g_{Z_{d}}^L,g_{Z_{d}}^R)=(-0.4823,0.0879)$.
 Their corresponding $\gamma$ boson coupling constants are the same as
 those in the SM. When the value is less than $10^{-4}$, we write $0$.
 }
\label{Table:Couplings-Zprime_thetaH=010-mKK=13}
\end{center}
\end{table}

\begin{table}[htb]
\begin{center}
\begin{tabular}{ccccccccc}
\hline
 \rowcolor[gray]{0.9}
 &&&&&&&&\\[-0.75em]
\rowcolor[gray]{0.9}
 $f$
 &$g_{Zf}^L$&$g_{Zf}^R$
 &$g_{Z^{(1)}f}^L$&$g_{Z^{(1)}f}^R$
 &$g_{Z_R^{(1)}f}^L$&$g_{Z_R^{(1)}f}^R$
 &$g_{\gamma^{(1)}f}^L$&$g_{\gamma^{(1)}f}^R$\\
 \hline
 $\nu_e$
 &\ \ 0.5688&0
 &\ \ 2.8639&0
 &$-$0.9037&0
 &0&0
 \\
 $\nu_\mu$
 &\ \ 0.5687&0
 &\ \ 2.7053&0
 &$-$0.8569&0
 &0&0
 \\
 $\nu_\tau$
 &\ \ 0.5687&0
 &\ \ 2.5929&0
 &$-$0.8237&0
 &0&0
 \\[0.5em]
 $e$
 &$-$0.3058&\ \ 0.2629
 &$-$1.5398&$-$0.0695
 &$-$0.9143&0
 &$-$2.4107&\ \ 0.1274
 \\
 $\mu$
 &$-$0.3058&\ \ 0.2629
 &$-$1.4545&$-$0.0695
 &$-$0.8670&0
 &$-$2.2772&\ \ 0.1274
 \\
 $\tau$
 &$-$0.3058&\ \ 0.2629
 &$-$1.3940&$-$0.0694
 &$-$0.8334&0
 &$-$2.1824&\ \ 0.1272
 \\[0.5em]
 $u$
 &\ \ 0.3934&$-$0.1753
 &\ \ 1.9092&\ \ 0.0463
 &\ \ 0.2979&0
 &\ \ 1.5487&$-$0.0849
 \\
 $c$
 &\ \ 0.3934&$-$0.1753
 &\ \ 1.8243&\ \ 0.0463
 &\ \ 0.2855&0
 &\ \ 1.4799&$-$0.0849
 \\
 $t$
 &\ \ 0.3940&$-$0.1747
 &\ \ 1.2374&$-$0.4429
 &\ \ 0.1993&$-$0.9777
 &\ \ 1.0041&\ \ 0.8145
 \\[0.5em]
 $d$
 &$-$0.4811&\ \ 0.0876
 &$-$2.3345&\ \ 0.1280
 &\ \ 0.2876&$-$0.1989
 &$-$0.7744&$-$0.2328
 \\
 $s$
 &$-$0.4811&\ \ 0.0876
 &$-$2.2308&\ \ 0.1280
 &\ \ 0.2756&$-$0.2394
 &$-$0.7399&$-$0.2892
 \\
 $b$
 &$-$0.4811&\ \ 0.0877
 &$-$1.5138&\ \ 0.3256
 &\ \ 0.1927&$-$0.4562
 &$-$0.5020&$-$0.5928
 \\
\hline
\end{tabular}
 \caption{\small
 Coupling constants of neutral vector bosons, $Z'$ bosons, to
 fermions in units of $g_w=e/\sin\theta_W^0$
 are listed  for
 $\theta_H=0.10$ and $m_{\KK}=11.00 \,$TeV
 (B$^{\rm L}$) in  Table~\ref{Table:Mass-Width-Vector-Bosons},
 where $\sin^2\theta_W^0=0.2306$.
Other information is the same as in
 Table~\ref{Table:Couplings-Zprime_thetaH=010-mKK=13}.
 }
\label{Table:Couplings-Zprime_thetaH=010-mKK=11}
\end{center}
\end{table}

\begin{table}[htb]
\begin{center}
\begin{tabular}{ccccccccc}
\hline
 \rowcolor[gray]{0.9}
 &&&&&&&&\\[-0.75em]
\rowcolor[gray]{0.9}
 $f$
 &$g_{Zf}^L$&$g_{Zf}^R$
 &$g_{Z^{(1)}f}^L$&$g_{Z^{(1)}f}^R$
 &$g_{Z_R^{(1)}f}^L$&$g_{Z_R^{(1)}f}^R$
 &$g_{\gamma^{(1)}f}^L$&$g_{\gamma^{(1)}f}^R$\\
 \hline
 $\nu_e$
 &\ \ 0.5687&0
 &\ \ 3.6903&0
 &$-$1.1603&0
 &0&0
 \\
 $\nu_\mu$
 &\ \ 0.5687&0
 &\ \ 3.5400&0
 &$-$1.1147&0
 &0&0
 \\
 $\nu_\tau$
 &\ \ 0.5687&0
 &\ \ 3.4442&0
 &$-$1.0857&0
 &0&0
 \\[0.5em]
 $e$
 &$-$0.3057&\ \ 0.2629
 &$-$1.9841&$-$0.0504
 &$-$1.1740&0
 &$-$3.1063&\ \ 0.0924
 \\
 $\mu$
 &$-$0.3057&\ \ 0.2629
 &$-$1.9033&$-$0.0504
 &$-$1.1279&0
 &$-$2.9780&\ \ 0.0924
 \\
 $\tau$
 &$-$0.3057&\ \ 0.2629
 &$-$1.8518&$-$0.0504
 &$-$1.0985&0
 &$-$2.8991&\ \ 0.0923
 \\[0.5em]
 $u$
 &\ \ 0.3934&$-$0.1753
 &\ \ 2.4831&\ \ 0.0336
 &\ \ 0.3855&0
 &\ \ 2.0143&$-$0.0616
 \\
 $c$
 &\ \ 0.3934&$-$0.1753
 &\ \ 2.4080&\ \ 0.0336
 &\ \ 0.3742&0
 &\ \ 1.9534&$-$0.0616
 \\
 $t$
 &\ \ 0.3937&$-$0.1750
 &\ \ 2.1069&$-$0.2768
 &\ \ 0.3291&$-$0.6311
 &\ \ 1.7092&\ \ 0.5096
 \\[0.5em]
 $d$
 &$-$0.4810&\ \ 0.0876
 &$-$3.0363&\ \ 0.1055
 &\ \ 0.3721&$-$0.1632
 &$-$1.0072&$-$0.1919
 \\
 $s$
 &$-$0.4810&\ \ 0.0876
 &$-$2.9446&\ \ 0.1337
 &\ \ 0.3613&$-$0.2009
 &$-$0.9767&$-$0.2433
 \\
 $b$
 &$-$0.4810&\ \ 0.0876
 &$-$2.5762&\ \ 0.1887
 &\ \ 0.3178 &$-$0.1691
 &$-$0.8545&$-$0.3440
 \\
\hline
\end{tabular}
 \caption{\small
 Coupling constants of neutral vector bosons, $Z'$ bosons, to
 fermions in units of $g_w=e/\sin\theta_W^0$
 are listed for  $\theta_H=0.10$ and $m_{\KK}=15.00\,$TeV
 (B$^{\rm H}$) in Table~\ref{Table:Mass-Width-Vector-Bosons},
 where $\sin^2\theta_W^0=0.2306$. 
Other information is the same as in
 Table~\ref{Table:Couplings-Zprime_thetaH=010-mKK=13}.
 }
\label{Table:Couplings-Zprime_thetaH=010-mKK=15}
\end{center}
\end{table}

\begin{table}[htb]
\begin{center}
\begin{tabular}{ccccccccc}
\hline
 \rowcolor[gray]{0.9}
 &&&&&&&&\\[-0.75em]
\rowcolor[gray]{0.9}
 $f$
 &$g_{Zf}^L$&$g_{Zf}^R$
 &$g_{Z^{(1)}f}^L$&$g_{Z^{(1)}f}^R$
 &$g_{Z_R^{(1)}f}^L$&$g_{Z_R^{(1)}f}^R$
 &$g_{\gamma^{(1)}f}^L$&$g_{\gamma^{(1)}f}^R$\\
 \hline
 $\nu_e$
 &\ \ 0.5684&0
 &\ \ 3.5449&0
 &$-$1.1125&0
 &0&0
 \\
 $\nu_\mu$
 &\ \ 0.5684&0
 &\ \ 3.3920&0
 &$-$1.0664&0
 &0&0
 \\
 $\nu_\tau$
 &\ \ 0.5684&0
 &\ \ 3.2933&0
 &$-$1.0367&0
 &0&0
 \\[0.5em]
 $e$
 &$-$0.3056&\ \ 0.2628
 &$-$1.9057&$-$0.0529
 &$-$1.1284&0
 &$-$2.9829&\ \ 0.0971
 \\
 $\mu$
 &$-$0.3056&\ \ 0.2628
 &$-$1.8235&$-$0.0529
 &$-$1.0817&0
 &$-$2.8543&\ \ 0.0971
 \\
 $\tau$
 &$-$0.3056&\ \ 0.2628
 &$-$1.7705&$-$0.0529
 &$-$1.0515&0
 &$-$2.7712&\ \ 0.0970
 \\[0.5em]
 $u$
 &\ \ 0.3932&$-$0.1752
 &\ \ 2.3814&\ \ 0.0353
 &\ \ 0.3709&0
 &\ \ 1.9313&$-$0.0647
 \\
 $c$
 &\ \ 0.3932&$-$0.1752
 &\ \ 2.3044&\ \ 0.0353
 &\ \ 0.3594&0
 &\ \ 1.8688&$-$0.0647
 \\
 $t$
 &\ \ 0.3935&$-$0.1748
 &\ \ 1.9823&$-$0.2910
 &\ \ 0.3111&$-$0.6640
 &\ \ 1.6078&\ \ 0.5369
 \\[0.5em]
 $d$
 &$-$0.4808&\ \ 0.0876
 &$-$2.9120&\ \ 0.1091
 &\ \ 0.3554&$-$0.1688
 &$-$0.9656&$-$0.1984
 \\
 $s$
 &$-$0.4808&\ \ 0.0876
 &$-$2.8179&\ \ 0.1380
 &\ \ 0.3444&$-$0.2071
 &$-$0.9344&$-$0.2509
 \\
 $b$
 &$-$0.4807&\ \ 0.0876
 &$-$2.4235&\ \ 0.2760
 &\ \ 0.2981&$-$0.3898
 &$-$0.8036&$-$0.5021
 \\
\hline
\end{tabular}
 \caption{\small
 Coupling constants of neutral vector bosons, $Z'$ bosons, to
 fermions in units of $g_w=e/\sin\theta_W^0$
 are listed  for $\theta_H=0.11$ and $m_{\KK}=13.00\,$TeV 
 (B$^+$) in Table~\ref{Table:Mass-Width-Vector-Bosons},
 where $\sin^2\theta_W^0=0.2305$. 
Other information is the same as in
 Table~\ref{Table:Couplings-Zprime_thetaH=010-mKK=13}.
 }
\label{Table:Couplings-Zprime_thetaH=011-mKK=13}
\end{center}
\end{table}

\begin{table}[htbh]
\begin{center}
\begin{tabular}{ccccccccc}
\hline
 \rowcolor[gray]{0.9}
 &&&&&&&&\\[-0.75em]
\rowcolor[gray]{0.9}
 $f$
 &$g_{Zf}^L$&$g_{Zf}^R$
 &$g_{Z^{(1)}f}^L$&$g_{Z^{(1)}f}^R$
 &$g_{Z_R^{(1)}f}^L$&$g_{Z_R^{(1)}f}^R$
 &$g_{\gamma^{(1)}f}^L$&$g_{\gamma^{(1)}f}^R$\\
 \hline
 $\nu_e$
 &\ \ 0.5690&0
 &\ \ 3.0096&0
 &$-$0.9511&0
 &0&0
 \\
 $\nu_\mu$
 &\ \ 0.5690&0
 &\ \ 2.8509&0
 &$-$0.9039&0
 &0&0
 \\
 $\nu_\tau$
 &\ \ 0.5690&0
 &\ \ 2.7412&0
 &$-$0.8712&0
 &0&0
 \\[0.5em]
 $e$
 &$-$0.3059&\ \ 0.2630
 &$-$1.6181&$-$0.0652
 &$-$0.9602&0
 &$-$2.5341&\ \ 0.1194
 \\
 $\mu$
 &$-$0.3059&\ \ 0.2630
 &$-$1.5328&$-$0.0652
 &$-$0.9125&0
 &$-$2.4004&\ \ 0.1194
 \\
 $\tau$
 &$-$0.3059&\ \ 0.2630
 &$-$1.4739&$-$0.0652
 &$-$0.8795&\ \ 0.0001
 &$-$2.3080&\ \ 0.1193
 \\[0.5em]
 $u$
 &\ \ 0.3936&$-$0.1754
 &\ \ 2.0096&\ \ 0.0435
 &\ \ 0.3125&0
 &\ \ 1.6306&$-$0.0796
 \\
 $c$
 &\ \ 0.3936&$-$0.1754
 &\ \ 1.9260&\ \ 0.0435
 &\ \ 0.3003&0
 &\ \ 1.5628&$-$0.0796
 \\
 $t$
 &\ \ 0.3940&$-$0.1750
 &\ \ 1.4605&$-$0.3819
 &\ \ 0.2318&$-$0.8528
 &\ \ 1.1853&\ \ 0.7016
 \\[0.5em]
 $d$
 &$-$0.4813&\ \ 0.0877
 &$-$2.4572&\ \ 0.1237
 &\ \ 0.3037&$-$0.1923
 &$-$0.8153&$-$0.2252
 \\
 $s$
 &$-$0.4813&\ \ 0.0877
 &$-$2.3551&\ \ 0.1544
 &\ \ 0.2919&$-$0.2327
 &$-$0.7814&$-$0.2810
 \\
 $b$
 &$-$0.4813&\ \ 0.0877
 &$-$1.7861&\ \ 0.3075
 &\ \ 0.2254&$-$0.4333
 &$-$0.5926&$-$0.5600
 \\
\hline
\end{tabular}
 \caption{\small
 Coupling constants of neutral vector bosons, $Z'$ bosons, to
 fermions in units of $g_w=e/\sin\theta_W^0$
 are listed  for $\theta_H=0.09$ and $m_{\KK}=13.00 \,$TeV 
 (B$^-$) in Table~\ref{Table:Mass-Width-Vector-Bosons},
 where $\sin^2\theta_W^0=0.2307$. 
Other information is the same as in
 Table~\ref{Table:Couplings-Zprime_thetaH=010-mKK=13}.
 }
\label{Table:Couplings-Zprime_thetaH=009-mKK=13}
\end{center}
\end{table}

\begin{table}[htb]
\begin{center}
\begin{tabular}{c|cccccc}
\hline
 \rowcolor[gray]{0.9}
 &&&&&&\\[-0.75em]
 \rowcolor[gray]{0.9}
 $k$&$m_{Z^{(2k-1)}}$&$g_{Z^{(2k-1)}e}^L$&$g_{Z^{(2k-1)}e}^R$
 &$m_{Z^{(2k)}}$&$g_{Z^{(2k)}e}^L$&$g_{Z^{(2k)}e}^R$\\
 \rowcolor[gray]{0.9}
 &[TeV]&&&[TeV]&&\\
 \hline
  1&10.20&$-$1.7621&$-$0.0584&15.86&$-$0.0064&$-$0.0040\\
  2&23.09&$-$0.6931&\ \ 0.0403&29.03&$-$0.0021&\ \ 0.0030\\
  3&36.07&$-$0.2514&$-$0.0329&42.10&$-$0.0010&$-$0.0025\\
  4&49.06&$-$0.1480&\ \ 0.0286&55.14&$-$0.0006&\ \ 0.0022\\
  5&62.05&$-$0.0882&$-$0.0257&68.16&$-$0.0004&$-$0.0020\\
  6&75.05&$-$0.0626&\ \ 0.0235&81.17&$-$0.0003&\ \ 0.0018\\
  7&88.05&$-$0.0443&$-$0.0219&94.18&$-$0.0002&$-$0.0017\\
  8&101.0&$-$0.0344&\ \ 0.0205&107.2&$-$0.0002&\ \ 0.0016\\
  9&114.0&$-$0.0265&$-$0.0194&120.2&$-$0.0001&$-$0.0015\\
 10&127.0&$-$0.0217&\ \ 0.0185&133.2&$-$0.0001&\ \ 0.0015\\
\hline
\end{tabular}\\[0.5em]
\begin{tabular}{c|ccccccc}
\hline
 \rowcolor[gray]{0.9}
 &&&&&&\\[-0.75em]
\rowcolor[gray]{0.9}
 $k$&\ \ $m_{Z_R^{(k)}}$\ \ &\ \ \  $g_{Z_R^{(k)}e}^L$\ \ \ &\ \ $g_{Z_R^{(k)}e}^R$\ \
 &$m_{\gamma}^{(k)}$&$g_{\gamma^{(k)}e}^L$&$g_{\gamma^{(k)}e}^R$\\
 \rowcolor[gray]{0.9}
 &[TeV]&&&[TeV]&&\\
 \hline
  1&9.951&$-$1.0444&0&10.20&$-$2.7587&\ \ 0.1071\\
  2&22.84&$-$0.4158&0&23.10&$-$1.0851&$-$0.0739\\
  3&35.81&$-$0.1494&0&36.07&$-$0.3936&\ \ 0.0603\\
  4&48.79&$-$0.0877&0&49.06&$-$0.2318&$-$0.0524\\
  5&61.78&$-$0.0521&0&62.05&$-$0.1380&\ \ 0.0470\\
  6&74.78&$-$0.0370&0&75.05&$-$0.0981&$-$0.0431\\
  7&87.77&$-$0.0261&0&88.05&$-$0.0693&\ \ 0.0401\\
  8&100.8&$-$0.0203&0&101.0&$-$0.0539&$-$0.0376\\
  9&113.8&$-$0.0156&0&114.0&$-$0.0415&\ \ 0.0356\\
 10&126.8&$-$0.0128&0&127.0&$-$0.0340&$-$0.0339\\
\hline
\end{tabular}
 \caption{\small
 Masses of neutral KK vector bosons $Z^{(2k-1)}$, $Z^{(2k)}$,
 $Z_R^{(k)}$, $\gamma^{(k)}$ $(k=1,2,\cdots,10)$ 
 and their couplings constants to left- and right-handed electrons
 in units of $g_w=e/\sin\theta_W^0$ are listed  for $\theta_H=0.10$ and
 $m_{\KK}=13.00 \,$TeV 
 (B) in Table~\ref{Table:Mass-Width-Vector-Bosons},
 where $\sin^2\theta_W^0=0.2306$.
Other information is the same as in
 Table~\ref{Table:Couplings-Zprime_thetaH=010-mKK=13}.
 }
\label{Table:Mass_Couplings-Zprime_thetaH=010-mKK=13-Higher-KK}
\end{center}
\end{table}

It is seen from Table~\ref{Table:Mass-Width-Vector-Bosons}  that
for the same KK mass scale $m_{\KK}$ and different $\theta_H$, the masses of
the first neutral KK vector bosons $Z^{(1)}$, $Z_R^{(1)}$,
$\gamma^{(1)}$ are almost the same, while 
the decay widths of $Z^{(1)}$, $Z_R^{(1)}$, and $\gamma^{(1)}$
become smaller for smaller $\theta_H$.
For the same $\theta_H$, the masses and decay widths of the first
neutral KK vector bosons $Z^{(1)}$, $Z_R^{(1)}$, and $\gamma^{(1)}$
become larger for larger $m_{\rm KK}$.
The total decay widths satisfy the relation 
$\Gamma_{Z^{(1)}}>\Gamma_{\gamma^{(1)}}\gg\Gamma_{Z_R^{(1)}}$.

From Tables~\ref{Table:Couplings-Zprime_thetaH=010-mKK=13},
\ref{Table:Couplings-Zprime_thetaH=010-mKK=11},
\ref{Table:Couplings-Zprime_thetaH=010-mKK=15},
\ref{Table:Couplings-Zprime_thetaH=011-mKK=13},
\ref{Table:Couplings-Zprime_thetaH=009-mKK=13},
we find that the coupling constants of the first neutral KK vector 
bosons $Z^{(1)}$, $Z_R^{(1)}$, $\gamma^{(1)}$ to quarks and leptons
are larger than those of the right-handed fermions 
except for $Z_R^{(1)}$ couplings to the top and bottom quarks.

In Table~\ref{Table:Mass_Couplings-Zprime_thetaH=010-mKK=13-Higher-KK},
the masses of neutral higher KK vector bosons $Z^{(2k-1)}$, $Z^{(2k)}$,
$Z_R^{(k)}$, and $\gamma^{(k)}$ $(k=1,2,\cdots,10)$ almost linearly increase as  $k$.
For instance, $m_{Z^{(n)}}/m_\KK = 0.784, 1.220$, 1.777, 2.233 2.775, 3.238, $\cdots$ for
$n=1, 2, 3, \cdots$.
The couplings constants of them to left- and right-handed electrons
is decreasing when $k$ is increasing.
In Figure~\ref{Figure:sigma-emu-2ndKK}, total cross section
$\sigma(e^-e^+\to{\mu^-\mu^+})$ with and without the contribution from
the second KK modes for $\theta_H = 0.10$ and $m_\KK = 13\,$TeV (B) is shown.
The coupling constants of the 1st KK bosons to the SM fermions are
listed in  Table~\ref{Table:Couplings-Zprime_thetaH=010-mKK=13}.
The masses and widths of the second KK bosons are given by
$(m_{Z^{(2)}},\Gamma_{Z^{(2)}})=(15.86,0.876)$,
$(m_{Z^{(3)}},\Gamma_{Z^{(3)}})=(23.10,1.498)$,
$(m_{Z_R^{(2)}},\Gamma_{Z_R^{(2)}})=(22.84,0.160)$,
$(m_{\gamma^{(2)}},\Gamma_{\gamma^{(2)}})=(23.10,0.645)$
in units of TeV,
where the decay widths include only the final states of the SM
fermions and bosons. 
The coupling constants of the second KK bosons to $e$ are found 
in Table~\ref{Table:Mass_Couplings-Zprime_thetaH=010-mKK=13-Higher-KK}.
The coupling constants of the second KK bosons to $\mu$ are
$(g_{Z^{(2)}\mu}^L,g_{Z^{(2)}\mu}^R)=(-0.0057,-0.0040)$, 
$(g_{Z^{(3)}\mu}^L,g_{Z^{(3)}\mu}^R)=(-0.5301,+0.0403)$, 
$(g_{Z_R^{(2)}\mu}^L,g_{Z_R^{(2)}\mu}^R)=(-0.3198,0)$, 
$(g_{\gamma^{(2)}\mu}^L,g_{\gamma^{(2)}\mu}^R)=(-0.8299,-0.0739)$.
The contribution for the low-energy observables from 
each higher KK vector boson $Z^{(k)}$, $Z_R^{(k)}$, $\gamma^{(k)}$
$(k\geq 2)$ is sub-dominant.
In the following, we consider  contributions for the low-energy
observables only from the first KK bosons $Z^{(1)}$, $Z_R^{(1)}$,
and $\gamma^{(1)}$.

\begin{figure}[thb]
\begin{center}
\includegraphics[bb=0 0 504 335,height=5cm]{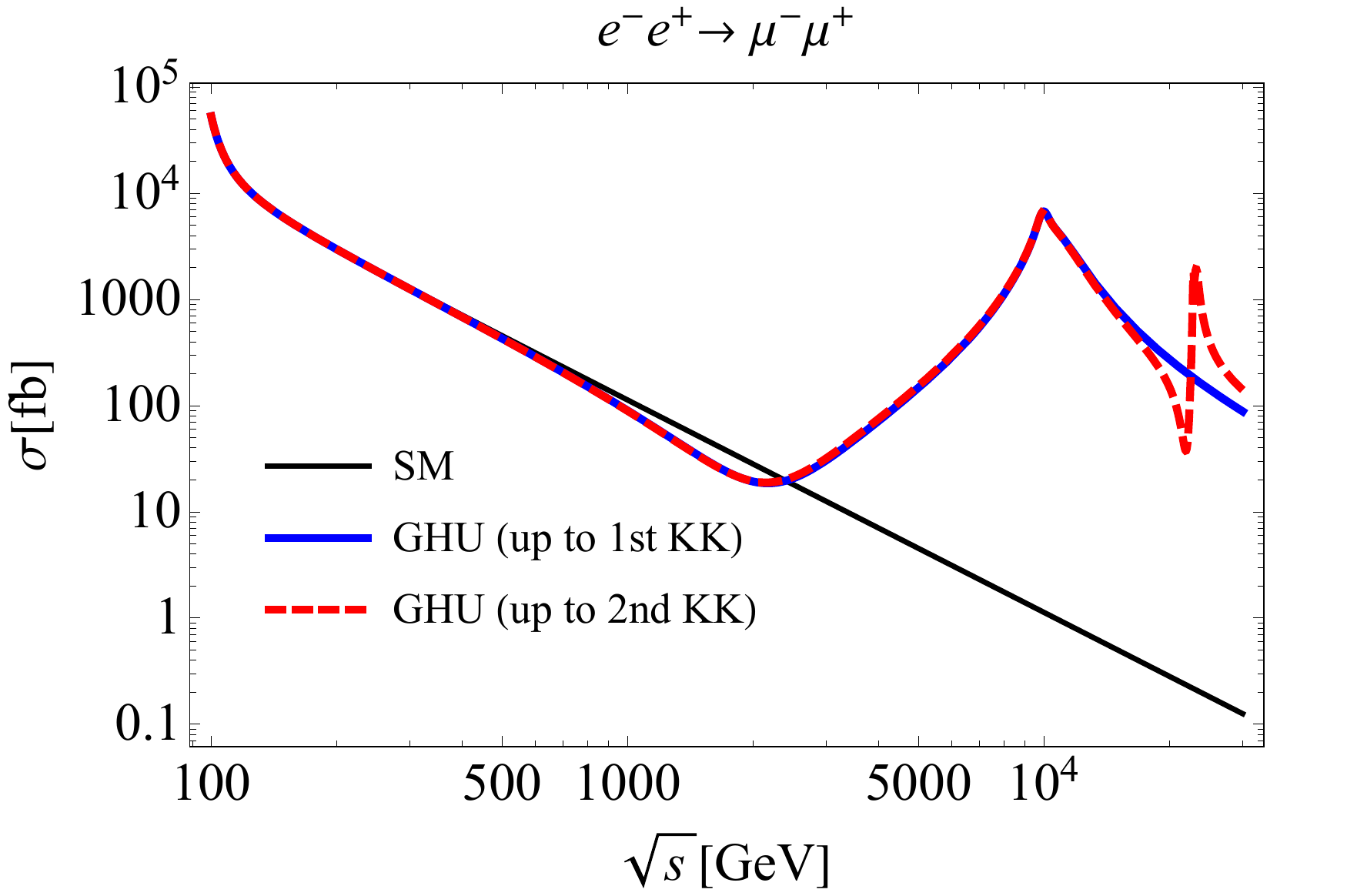}
\includegraphics[bb=0 0 504 330,height=5cm]{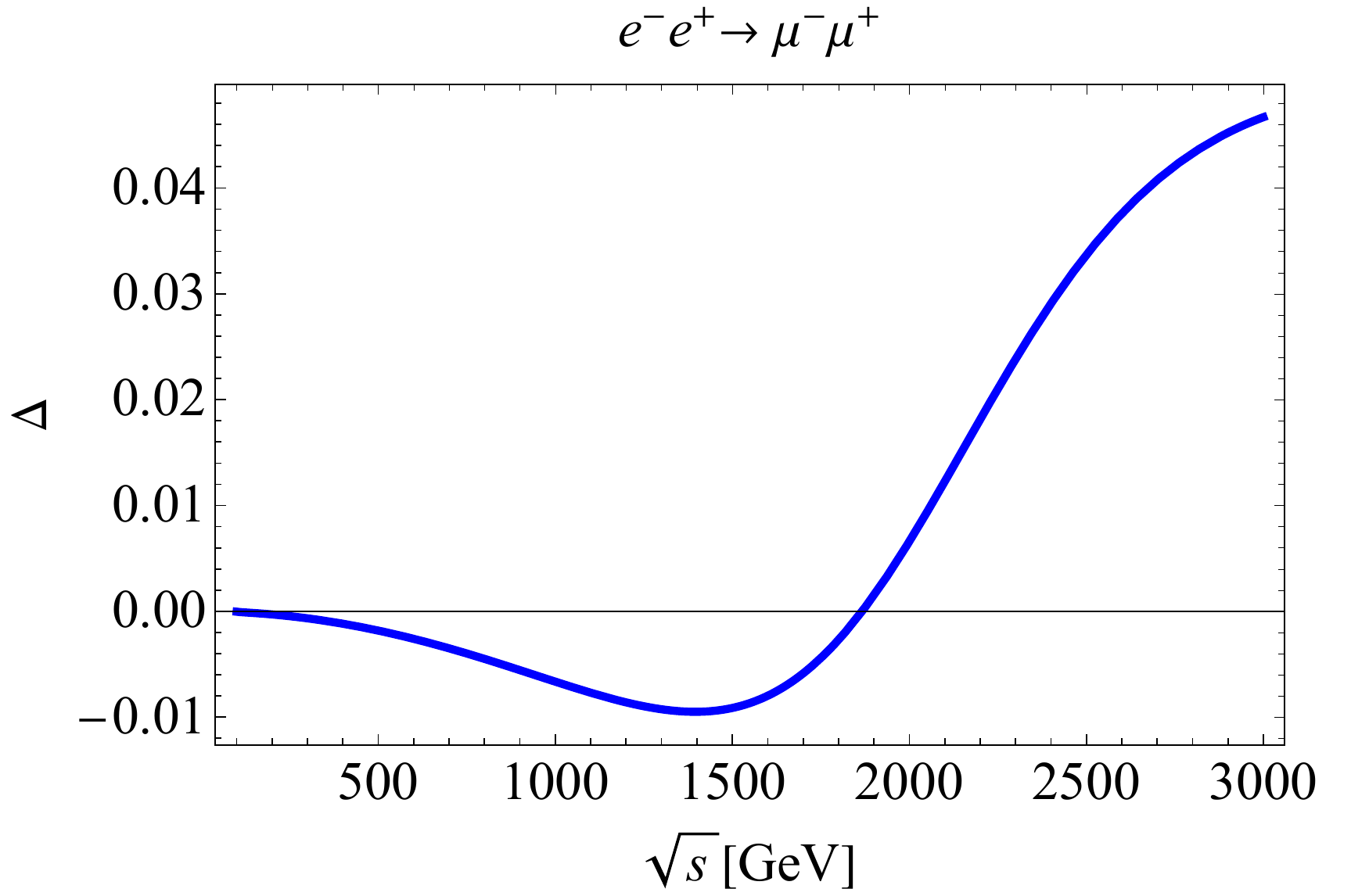}
 \caption{\small
 Total cross section $\sigma(e^-e^+\to{\mu^-\mu^+})$
 with and without the contribution from the ``second KK modes''
 $(\gamma^{(2)},Z^{(2)},Z^{(3)},Z_R^{(2)})$ is shown.
 The left figure shows the total cross section $\sigma (e^-e^+ \to \mu^-\mu^+)$
 with unpolarized electron and positron beams in the SM
 and the GHU (B) model in Table~\ref{Table:Mass-Width-Vector-Bosons}
up to  $\sqrt{s} =  30\,$TeV.
The right figure shows the proportion of the contribution from the second KK modes,
$\Delta = \sigma (\hbox{up to 2nd KK})/\sigma (\hbox{up to 1st KK}) -1$.
The  contribution from the second KK modes remains small  for $\sqrt{s} <  3\,$TeV.}
 \label{Figure:sigma-emu-2ndKK}
\end{center}
\end{figure}

\subsection{Cross section}

Total cross sections $\sigma^{f\bar{f}}$ for $e^- e^+ \go f \bar f$  $(f\bar{f}=\mu^-\mu^+,c\bar{c},b\bar{b},t\bar{t})$
are plotted with various polarization $(P_{e^-},P_{e^+}) =(0,0)$, $(-0.8,+0.3)$, $(+0.8,-0.3)$ 
in Figures~\ref{Figure:sigma-ef-Peff-theta} and \ref{Figure:sigma-ef-mKK-theta}. 
On the left side in Figure~\ref{Figure:sigma-ef-Peff-theta} the $\sqrt{s}$ dependence is shown.
On the right side the amount of the deviation from the SM, $\Delta_\sigma^{f\bar{f}}$ defined in
 Eq.~(\ref{Eq:Delta_sigma}), is shown. 
 One can see large deviation for $(P_{e^-},P_{e^+})=(-0.8,+0.3)$ in the B-model.
 It is due to the fact that the coupling constants of the
left-handed electron and $\mu$ to $Z'$ bosons are much larger than those
of the right-handed ones as seen in
Table~\ref{Table:Couplings-Zprime_thetaH=010-mKK=13}. 
Distinct signals of GHU can be clearly observed in the $e^- e^+$ collision experiments
at $\sqrt{s}=250 \,$GeV even with 250$\,$fb$^{-1}$ data by examining polarization dependence.
$\sigma^{f\bar{f}}(s)$ in wider range of $\sqrt{s}$ is displayed in Figure~\ref{Figure:sigma-ef-mKK-theta}.

\begin{figure}[thb]
\begin{center}
\includegraphics[bb=0 0 504 341,height=5cm]{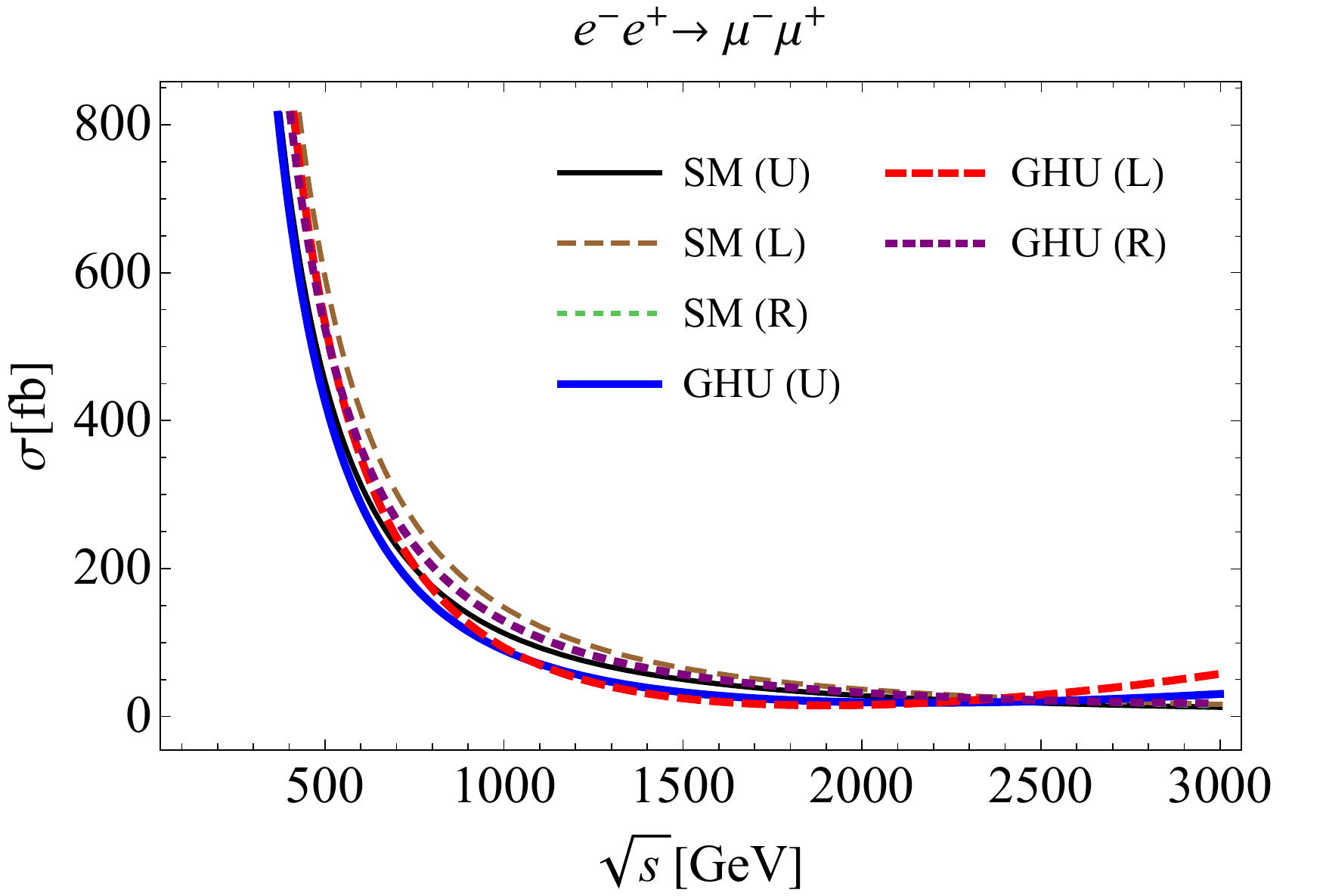}
\includegraphics[bb=0 0 504 327,height=5cm]{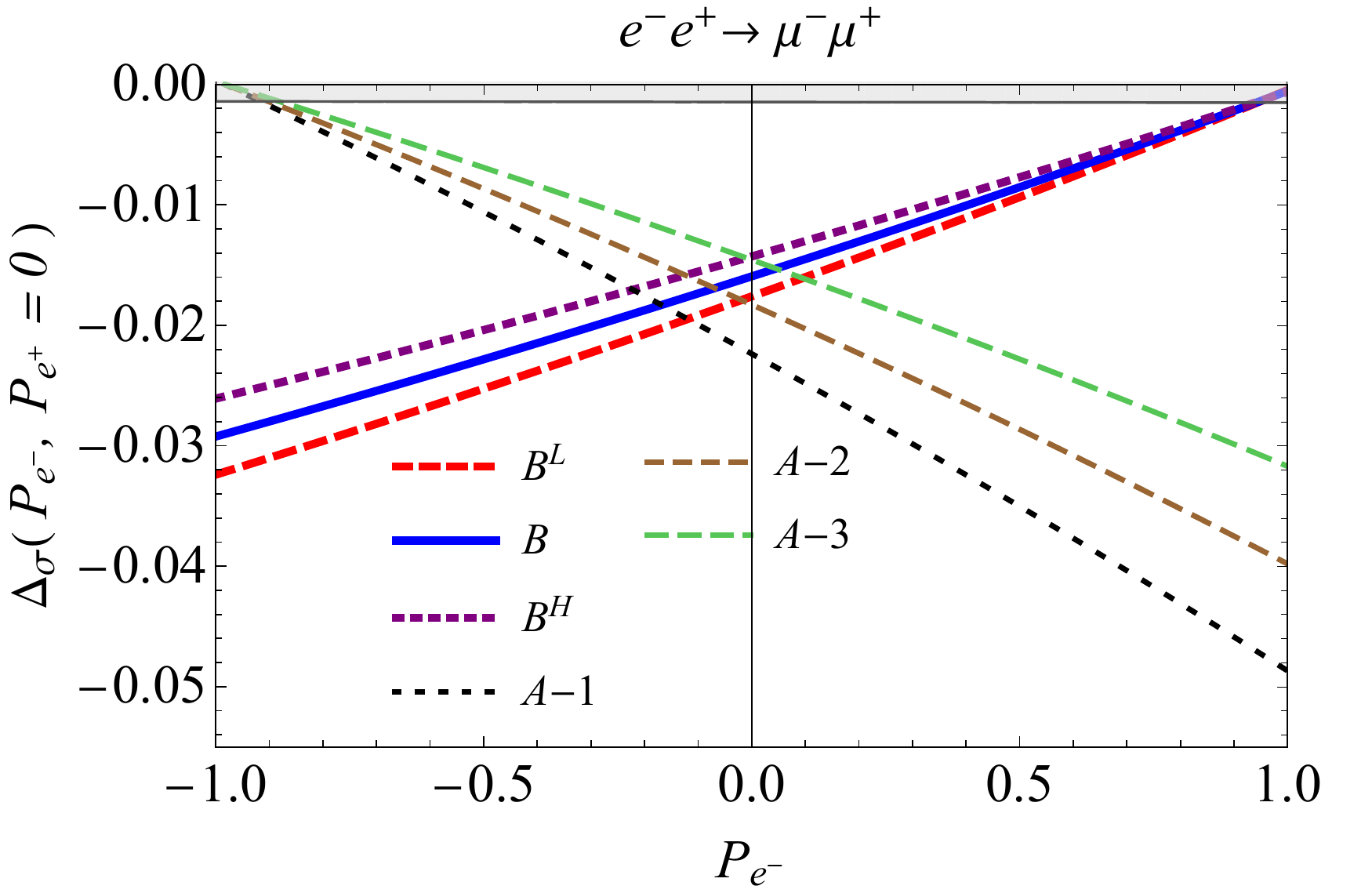}\\[0.5em]
\includegraphics[bb=0 0 504 341,height=5cm]{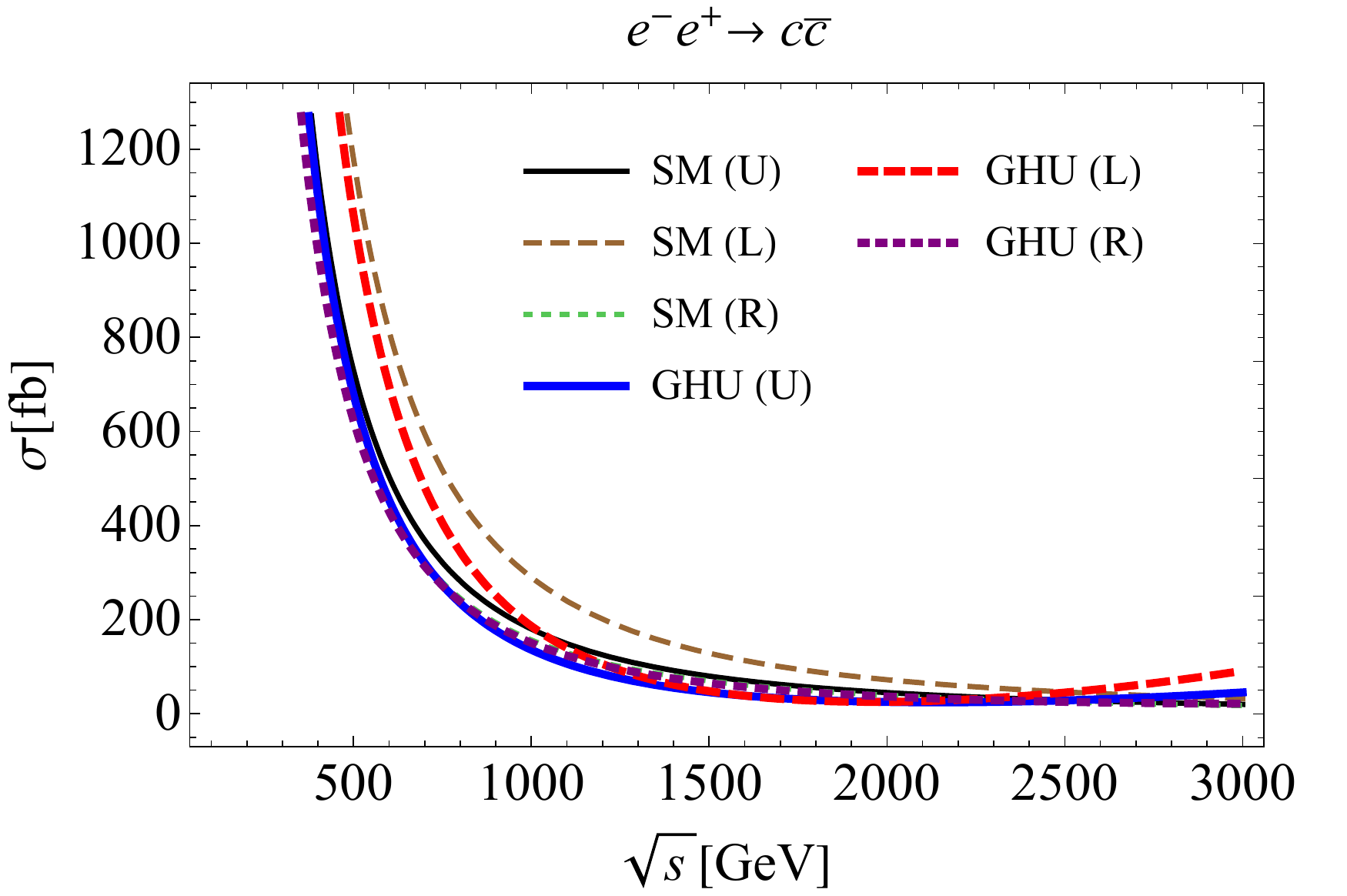}
\includegraphics[bb=0 0 504 327,height=5cm]{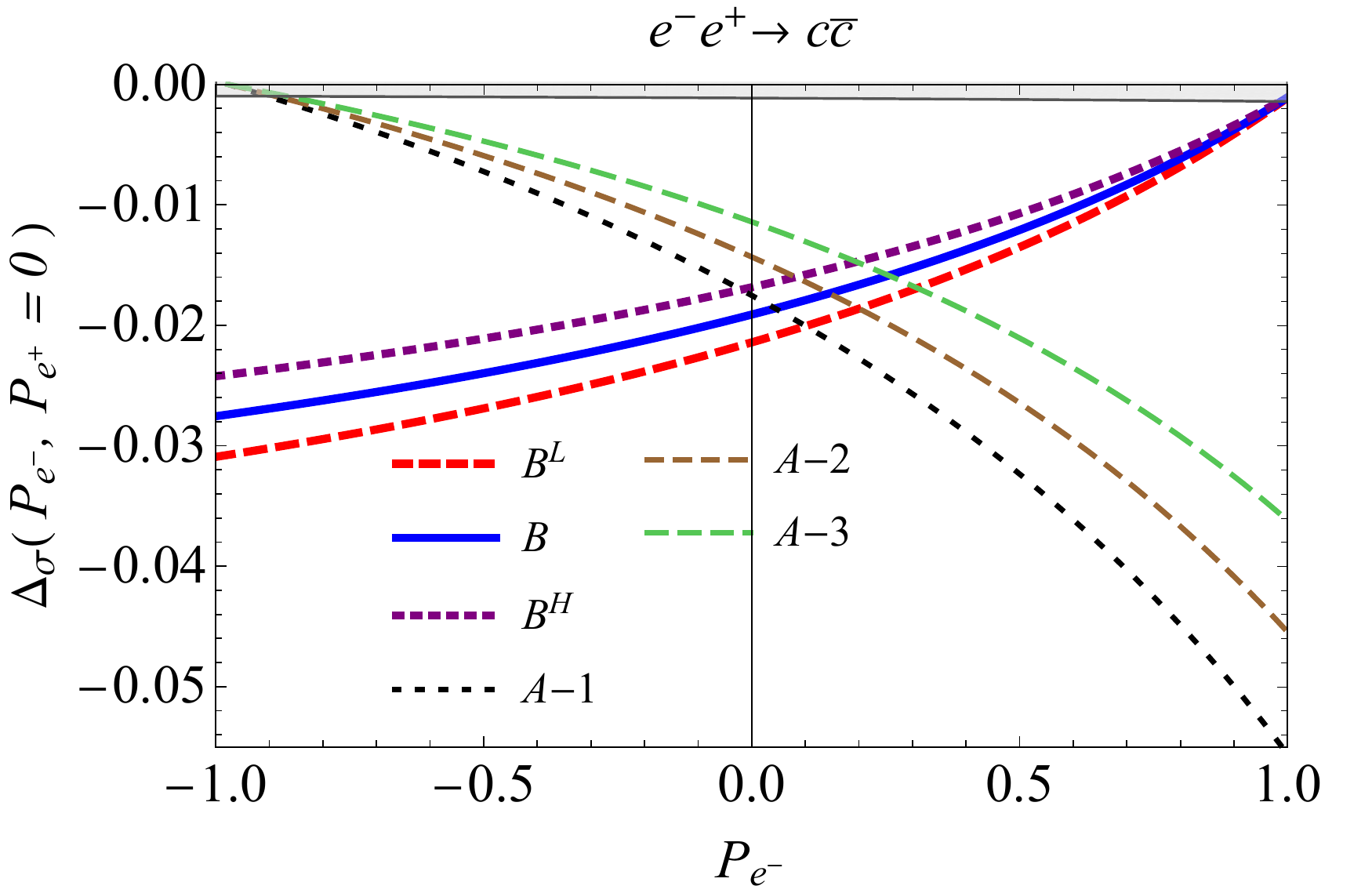}\\[0.5em]
\includegraphics[bb=0 0 504 341,height=5cm]{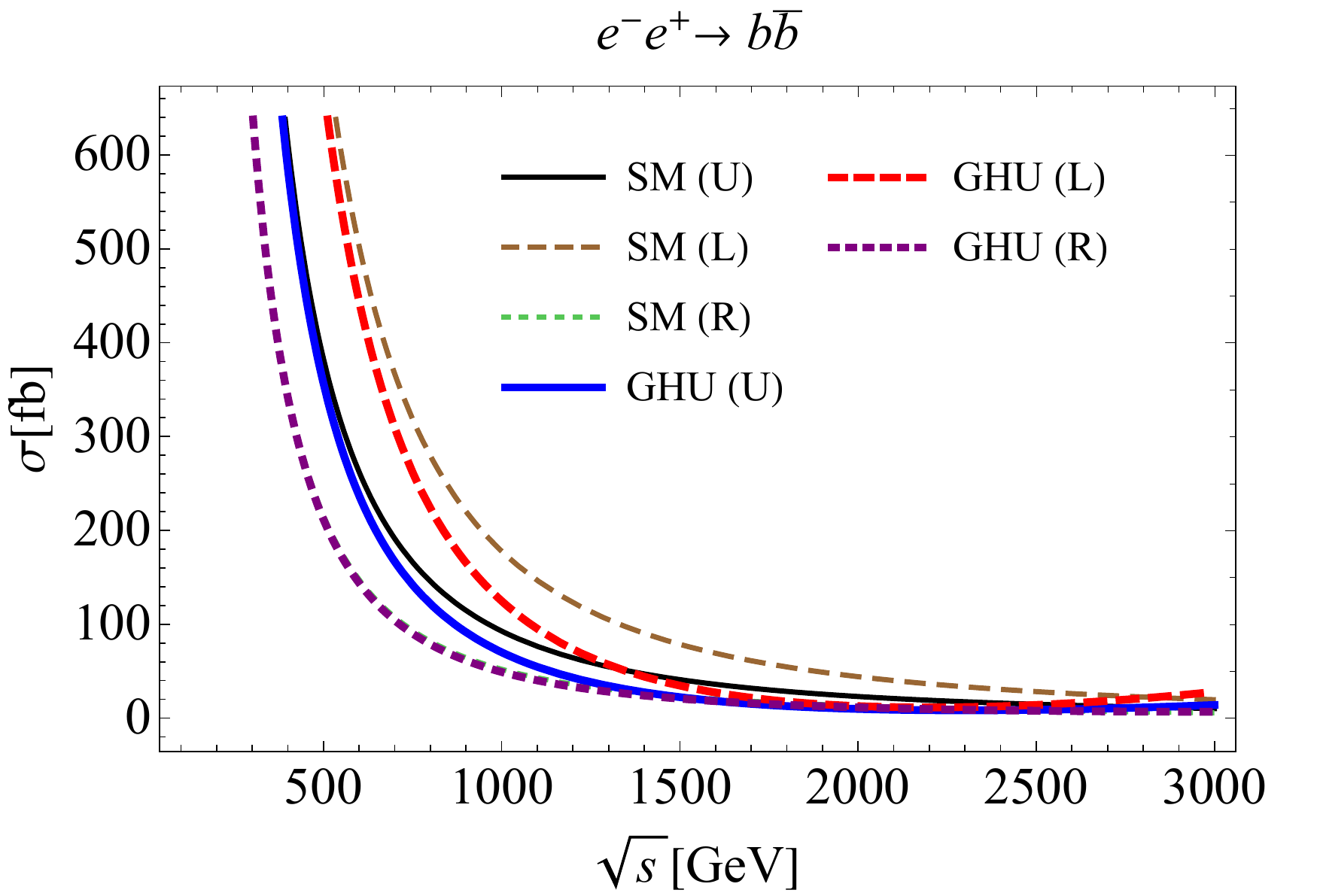}
\includegraphics[bb=0 0 504 327,height=5cm]{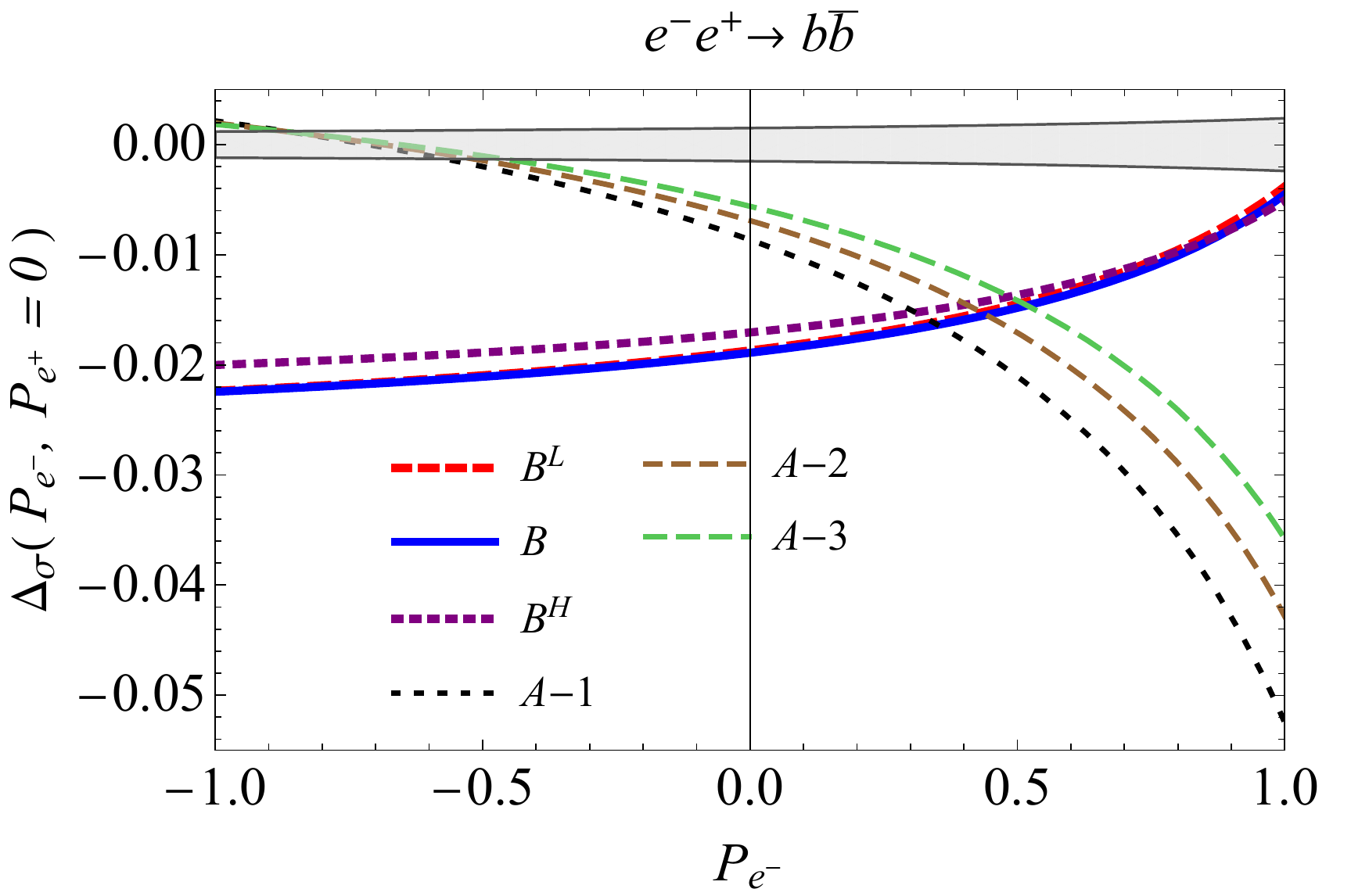}\\[0.5em]
\includegraphics[bb=0 0 504 341,height=5cm]{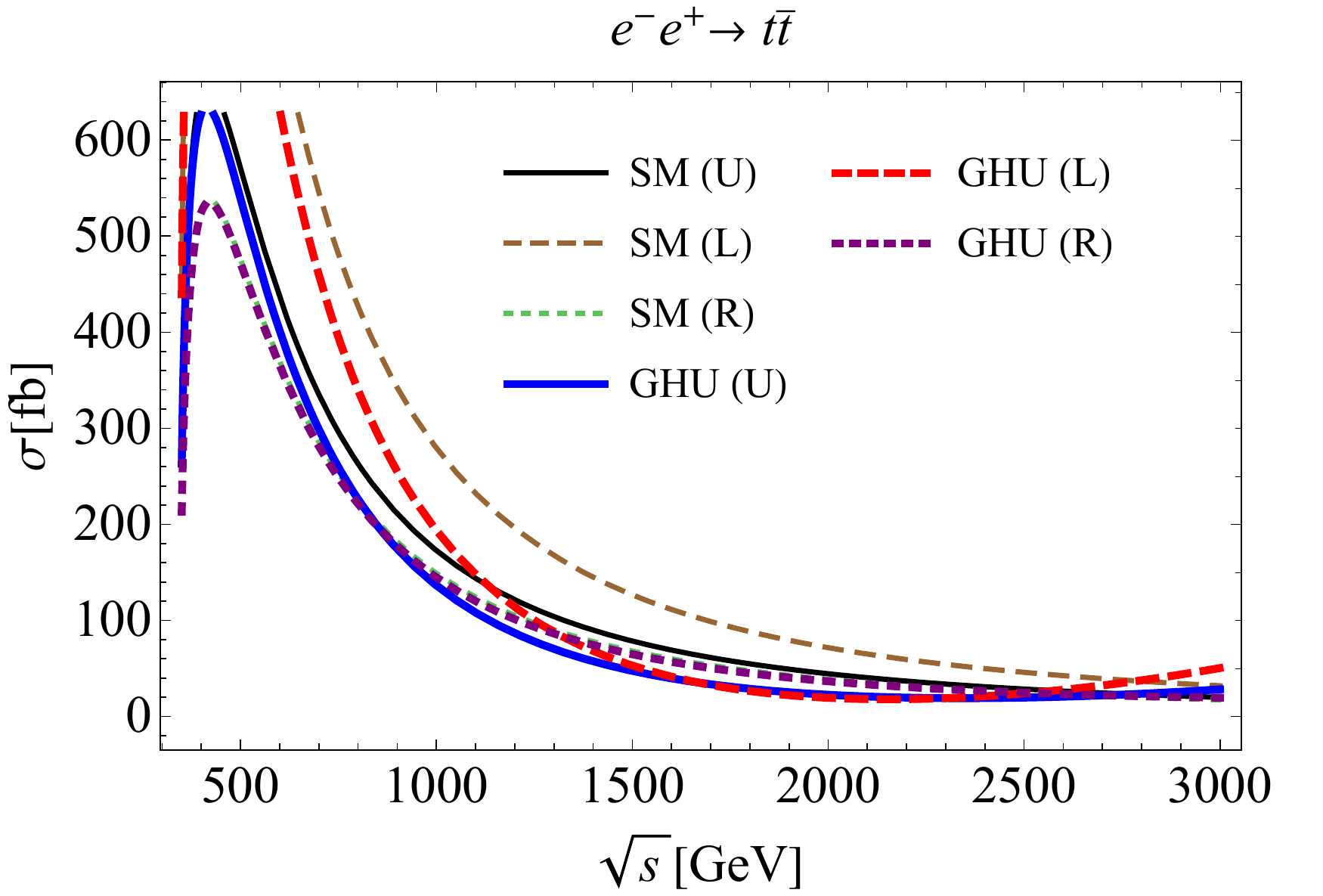}
\includegraphics[bb=0 0 504 327,height=5cm]{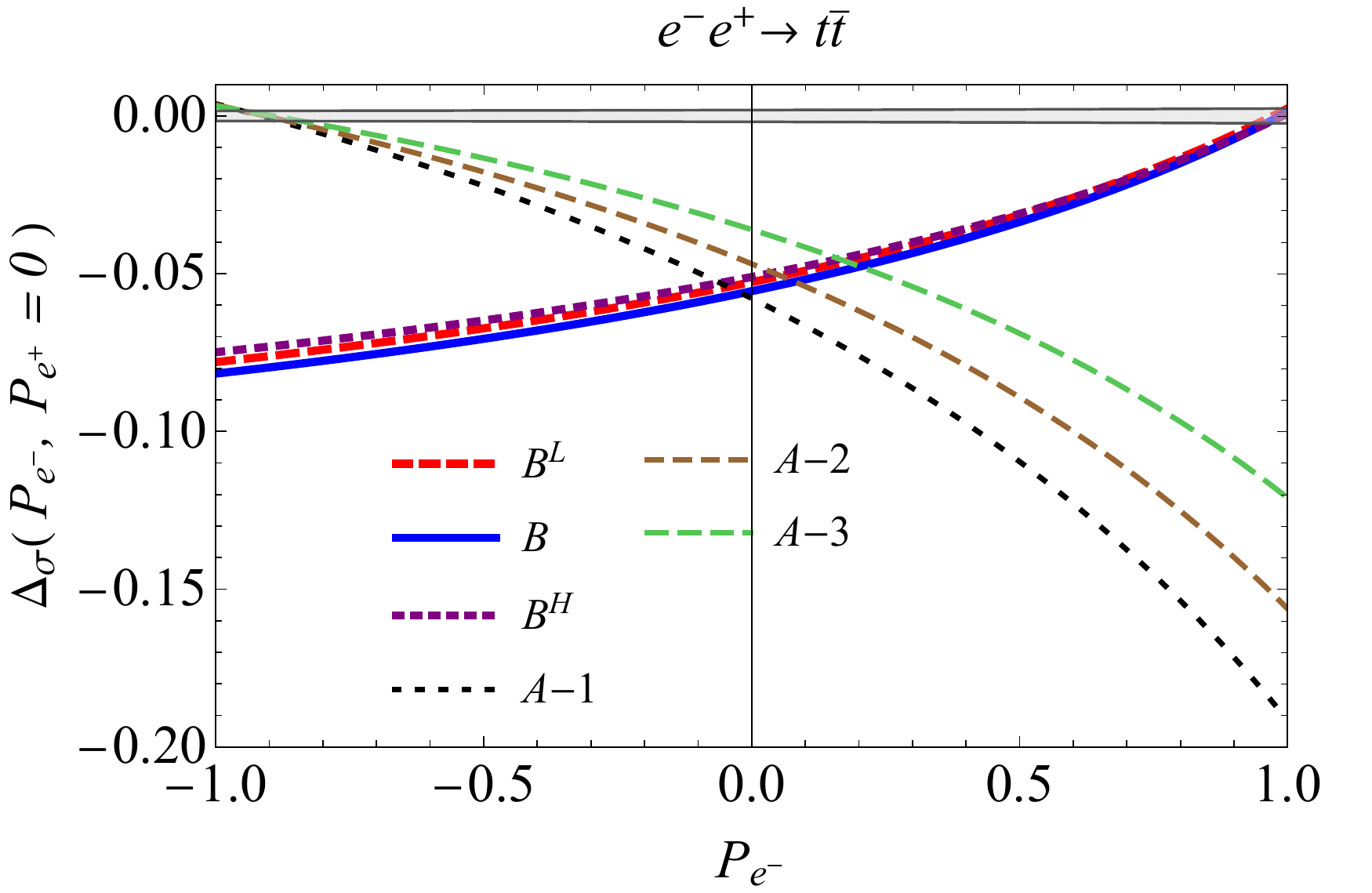}\\
 \caption{\small
Total cross sections $\sigma^{f\bar{f}}$ 
 $(f\bar{f}=\mu^-\mu^+,c\bar{c},b\bar{b},t\bar{t})$ are shown.
On the left side  the $\sqrt{s}$ dependence of
$\sigma^{f\bar{f}}$
in the SM and the GHU (B) in Table~\ref{Table:Mass-Width-Vector-Bosons}
 with 
 $(P_{e^-},P_{e^+})=(0,0), (-0.8,+0.3), (+0.8,-0.3)$, which are
 referred to as (U), (L), (R), respectively, is shown.
On the right side the electron polarization $P_{e^-}$ dependence of the amount of the deviation
from the SM, $\Delta_\sigma^{f\bar{f}}$  in Eq.~(\ref{Eq:Delta_sigma}), is shown
for both the GUT ({\it B-model}) (B$^{\rm L}$), (B), (B$^{\rm H}$) 
 and the previous GHU ({\it A-model})
 $(\theta_H=0.10, m_{\rm KK}=8.1\, \mbox{TeV})$,
 $(\theta_H=0.09, m_{\rm KK}=8.7\, \mbox{TeV})$,
 $(\theta_H=0.08, m_{\rm KK}=9.5\, \mbox{TeV})$
 which are referred to as $A$-1, $A$-2 and $A$-3, respectively.
 The gray band represents the  statistical error in the SM 
 at $\sqrt{s}=250 \,$GeV with 250$\,$fb$^{-1}$ data for $P_{e^-}=P_{e^+}=0$.
For the A-model, the masses and decay widths of the KK bosons and the
coupling constants are listed in Ref.~\cite{Funatsu:2017nfm}. 
 } 
 \label{Figure:sigma-ef-Peff-theta}
\end{center}
\end{figure}

\begin{figure}[thb]
\begin{center}
\includegraphics[bb=0 0 504 335,height=3.45cm]{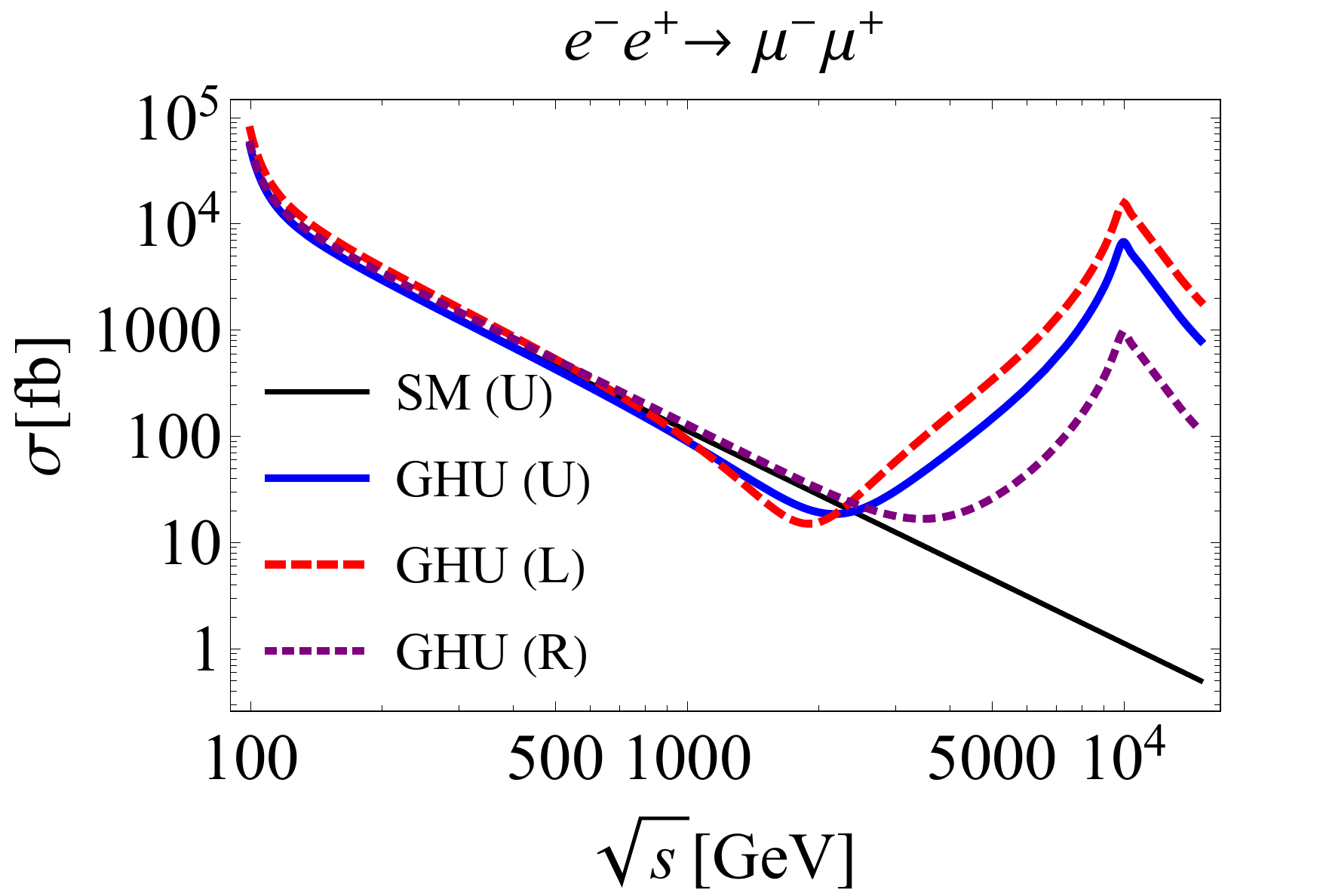}
\includegraphics[bb=0 0 504 335,height=3.45cm]{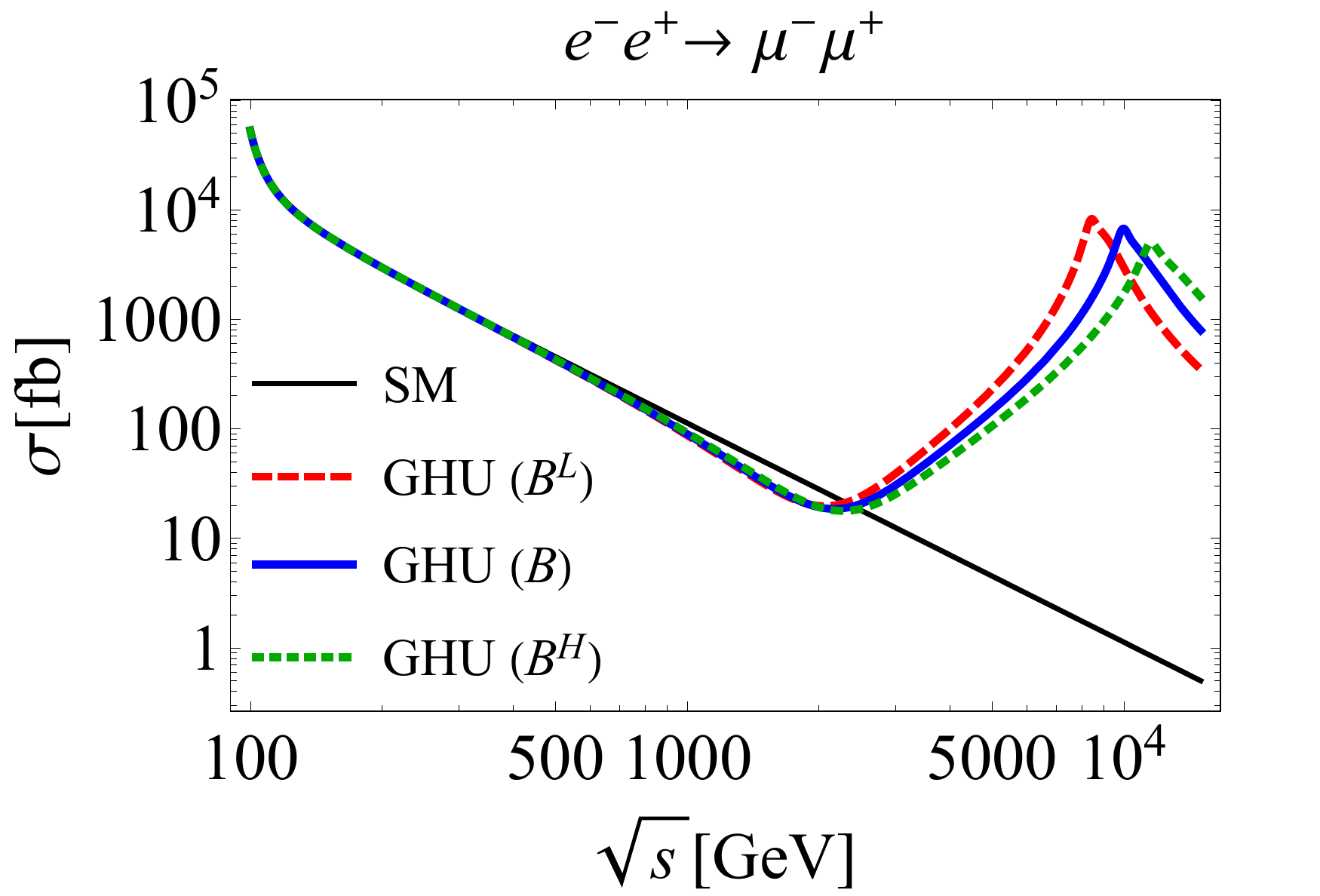}
\includegraphics[bb=0 0 504 335,height=3.45cm]{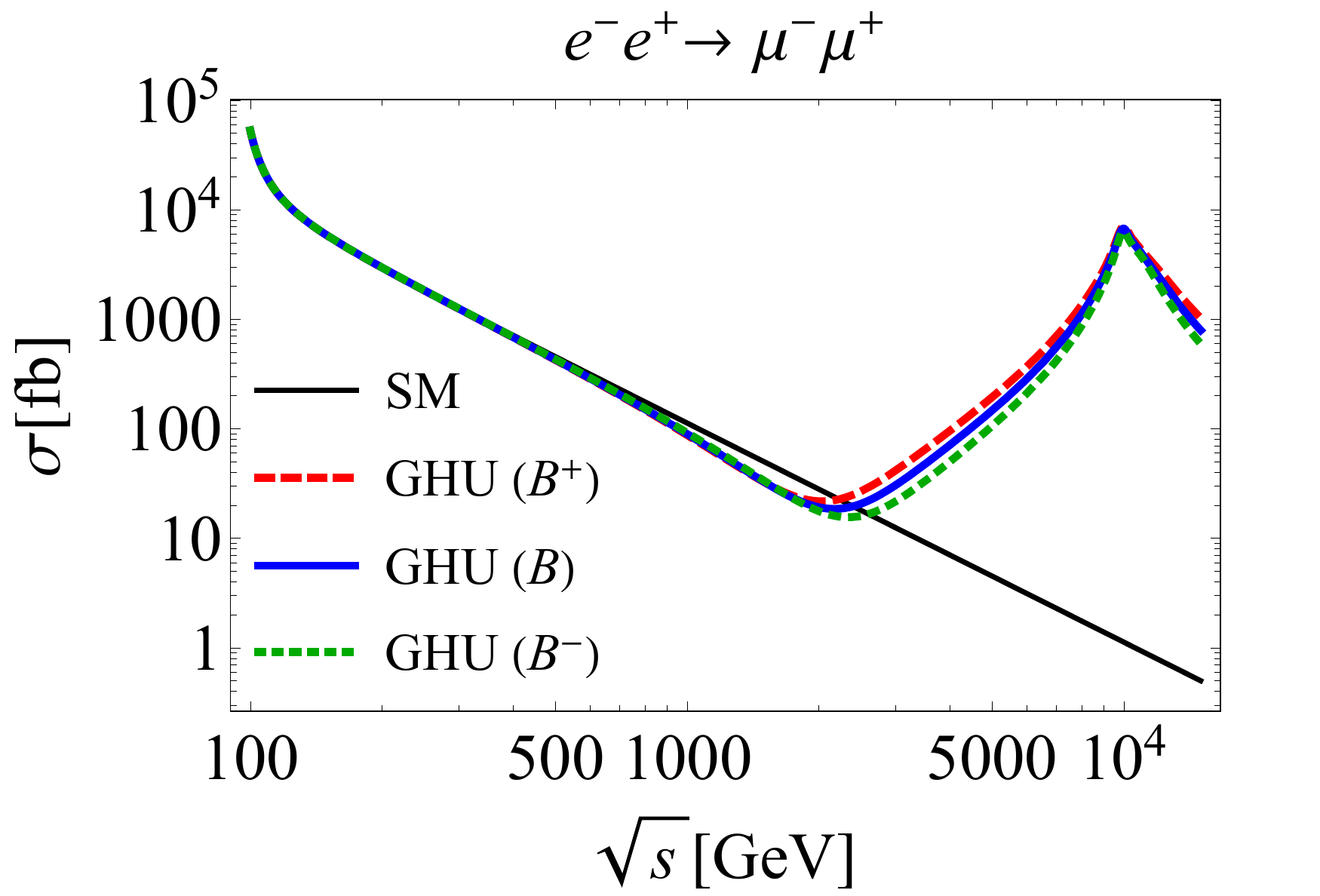}\\[1em]
\includegraphics[bb=0 0 504 335,height=3.45cm]{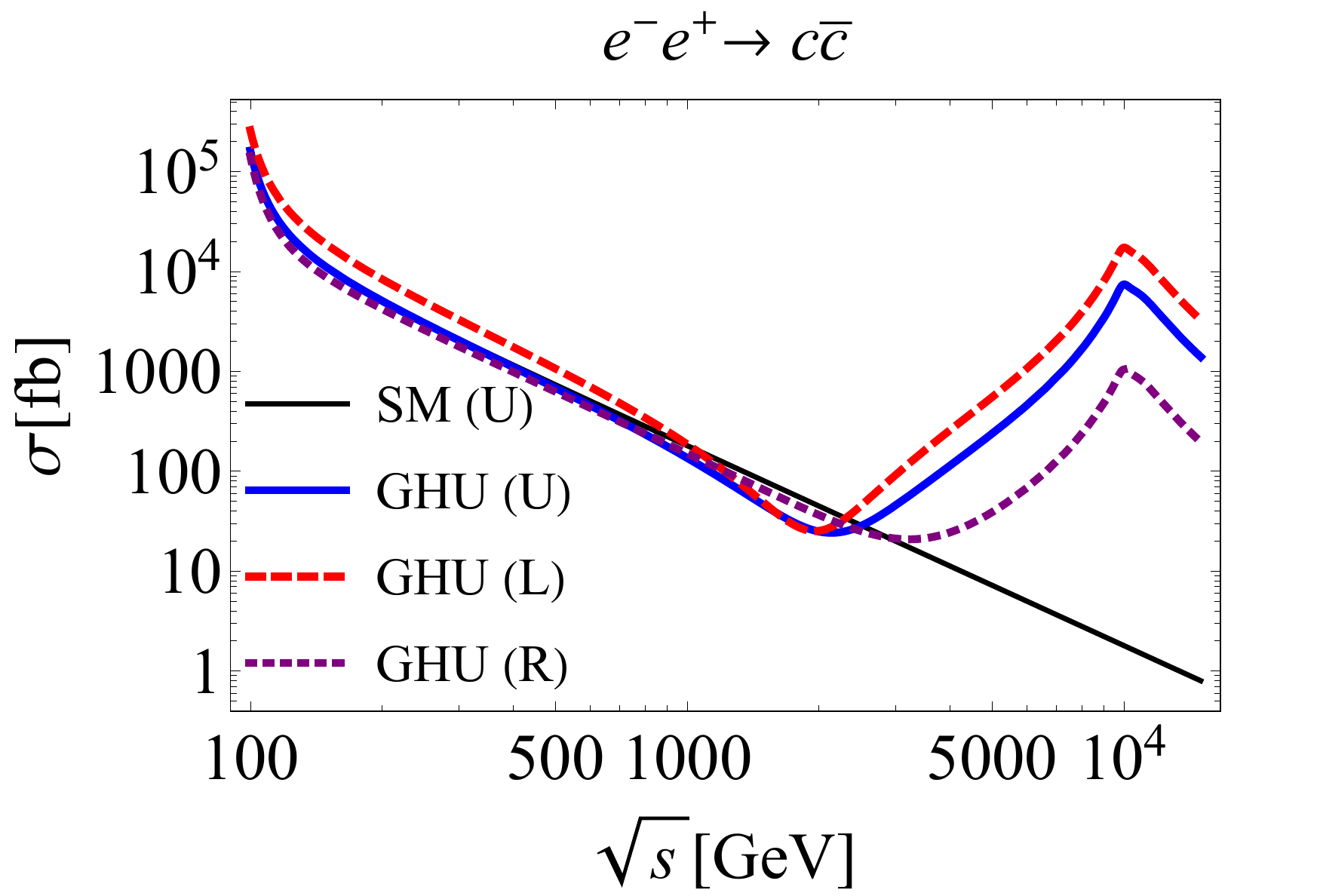}
\includegraphics[bb=0 0 504 335,height=3.45cm]{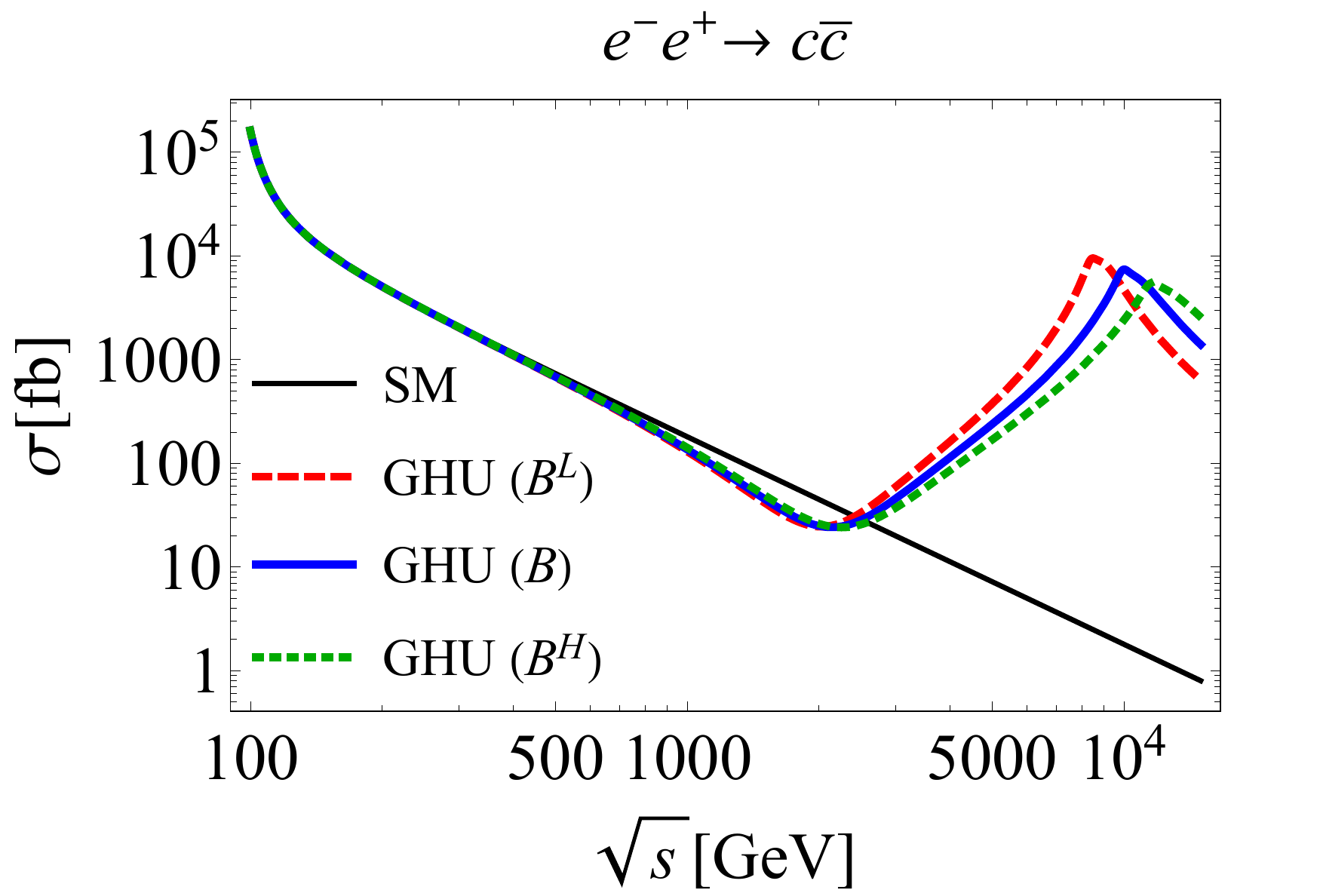}
\includegraphics[bb=0 0 504 335,height=3.45cm]{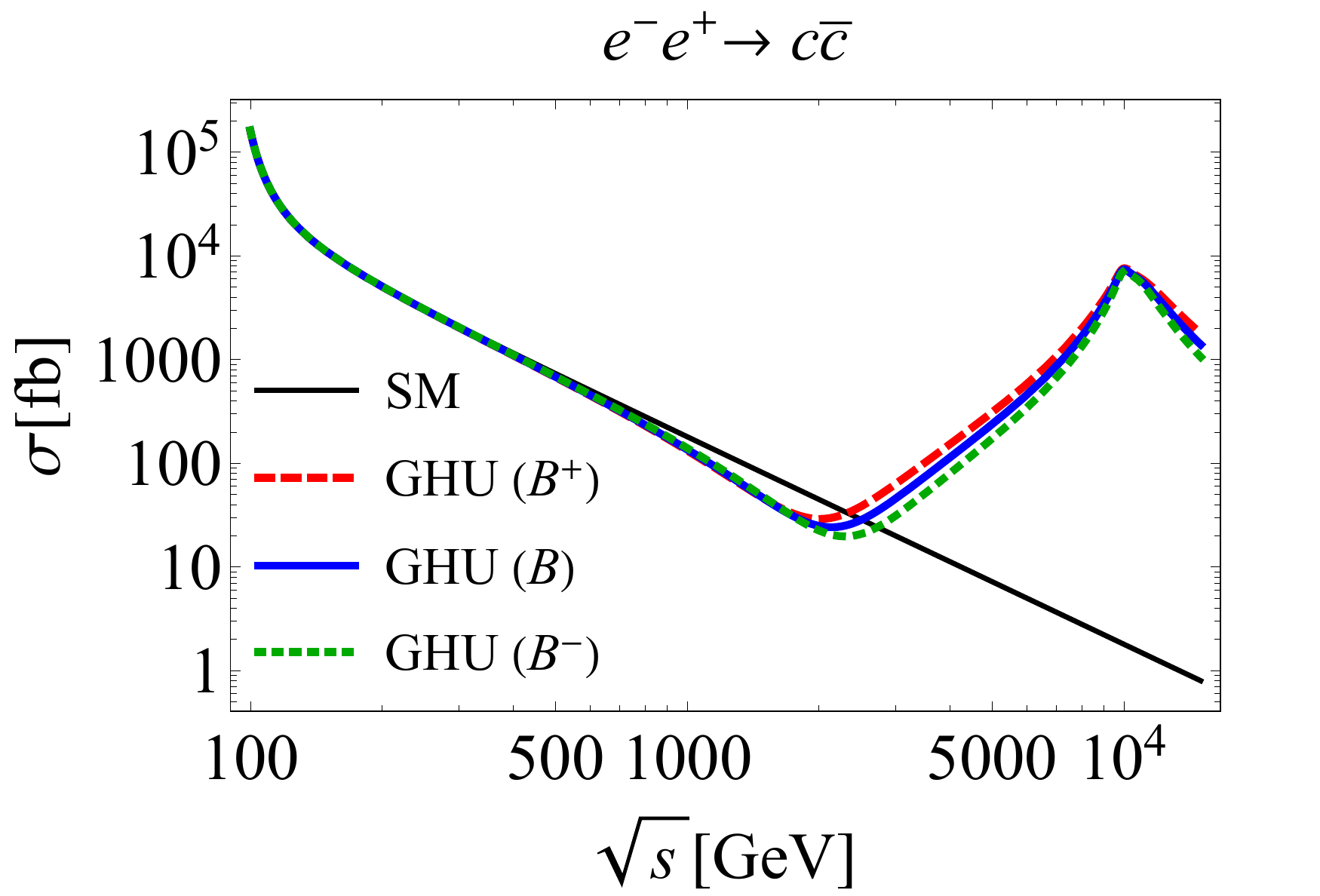}\\[1em]
\includegraphics[bb=0 0 504 335,height=3.45cm]{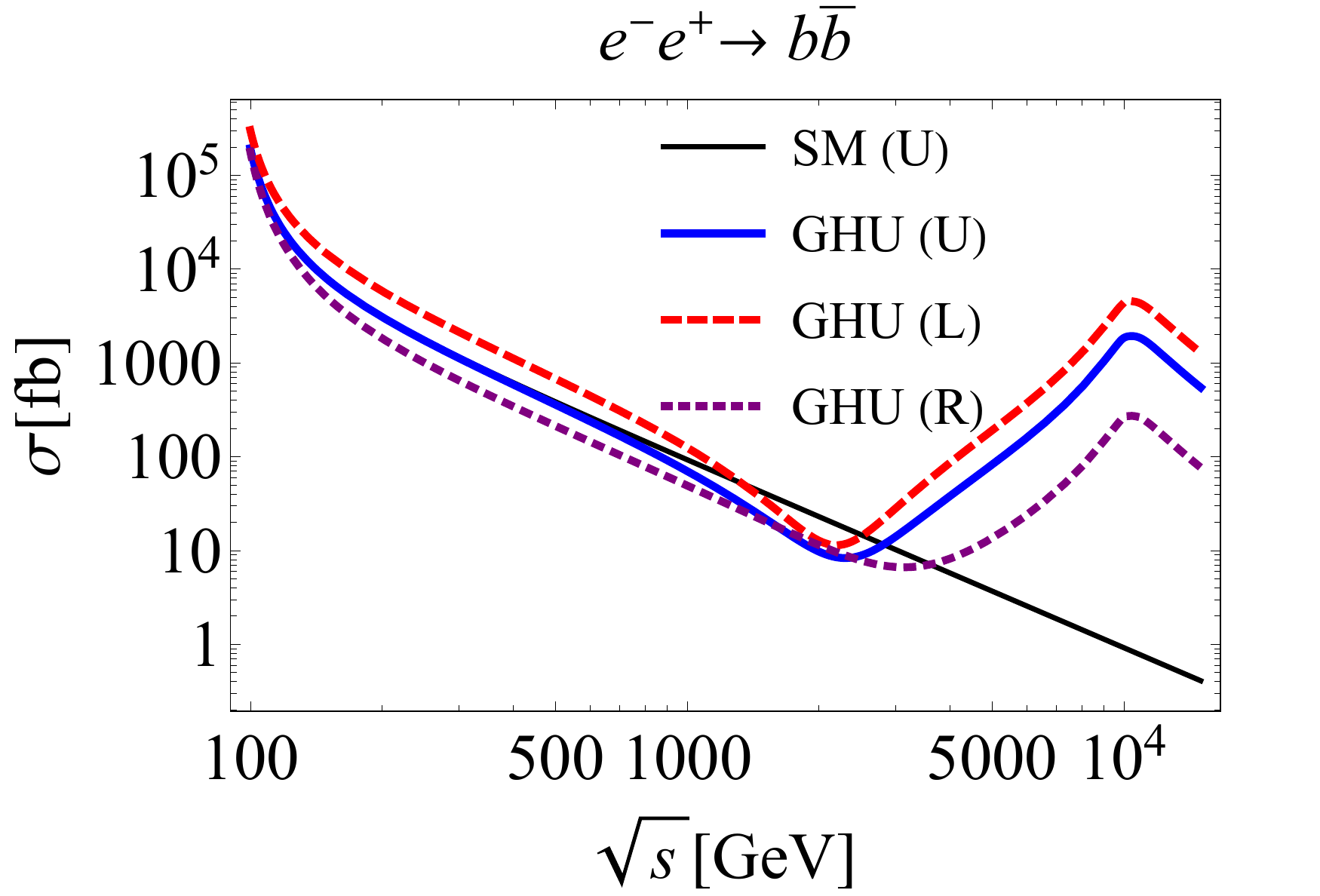}
\includegraphics[bb=0 0 504 335,height=3.45cm]{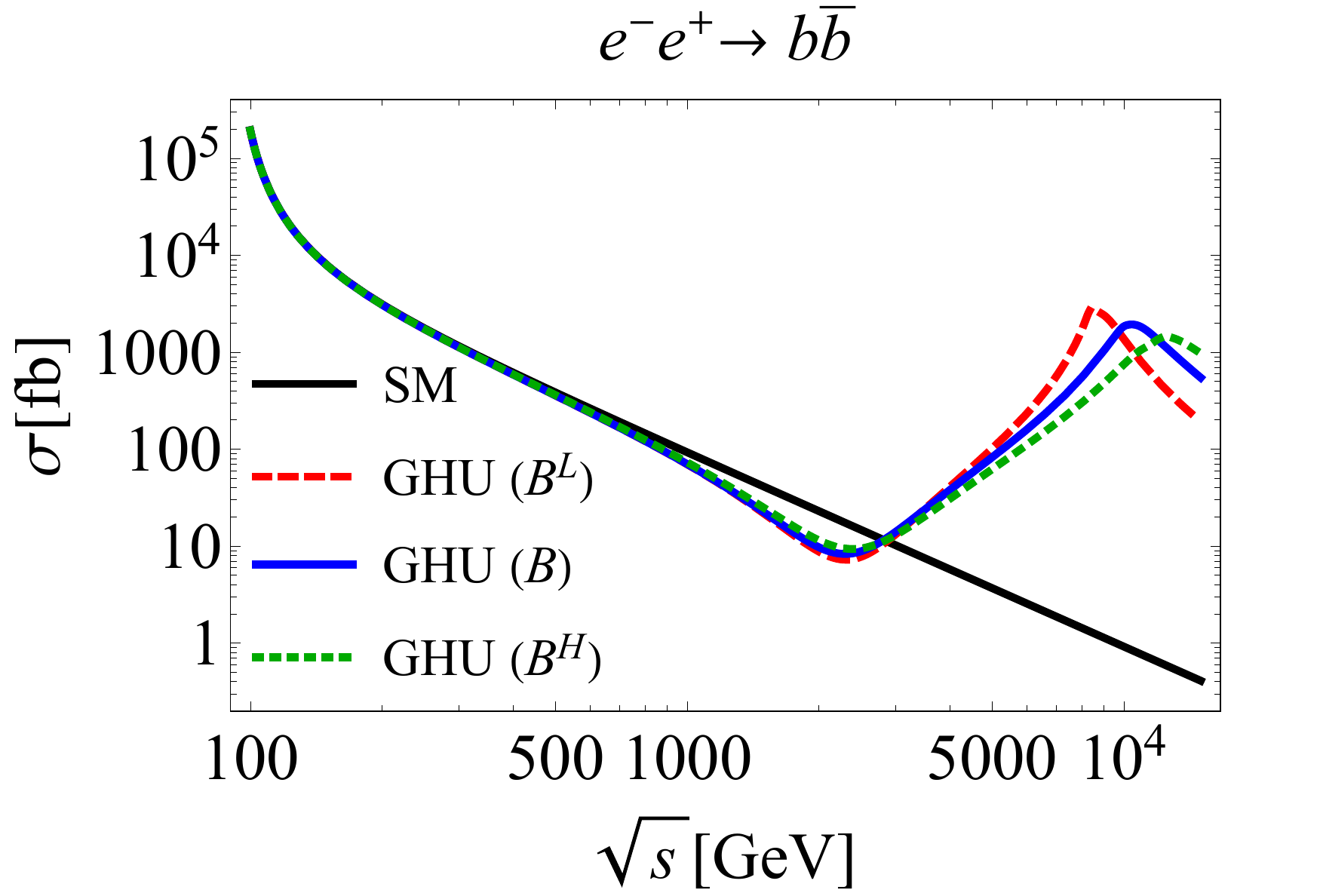}
\includegraphics[bb=0 0 504 335,height=3.45cm]{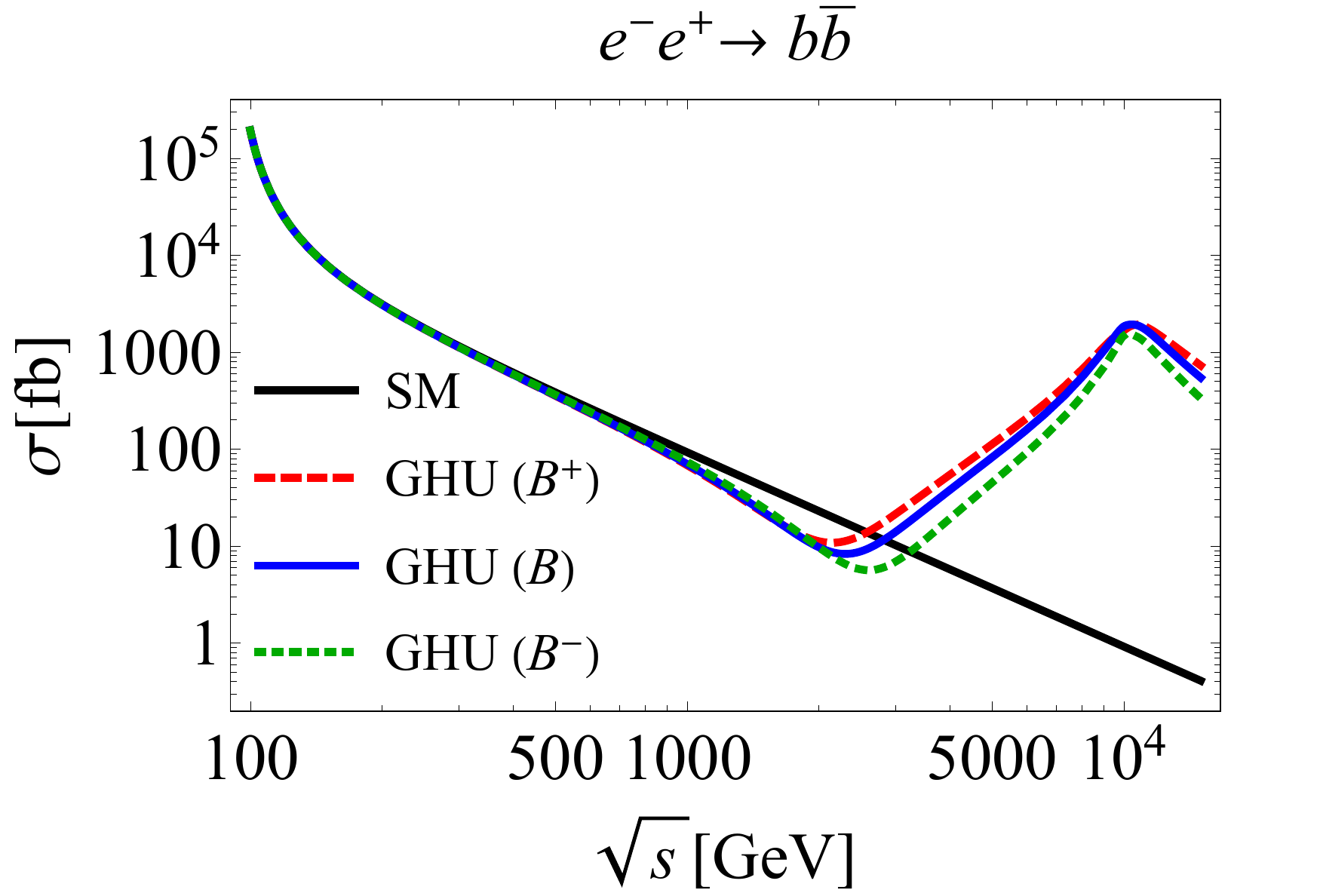}\\[1em]
\includegraphics[bb=0 0 504 335,height=3.45cm]{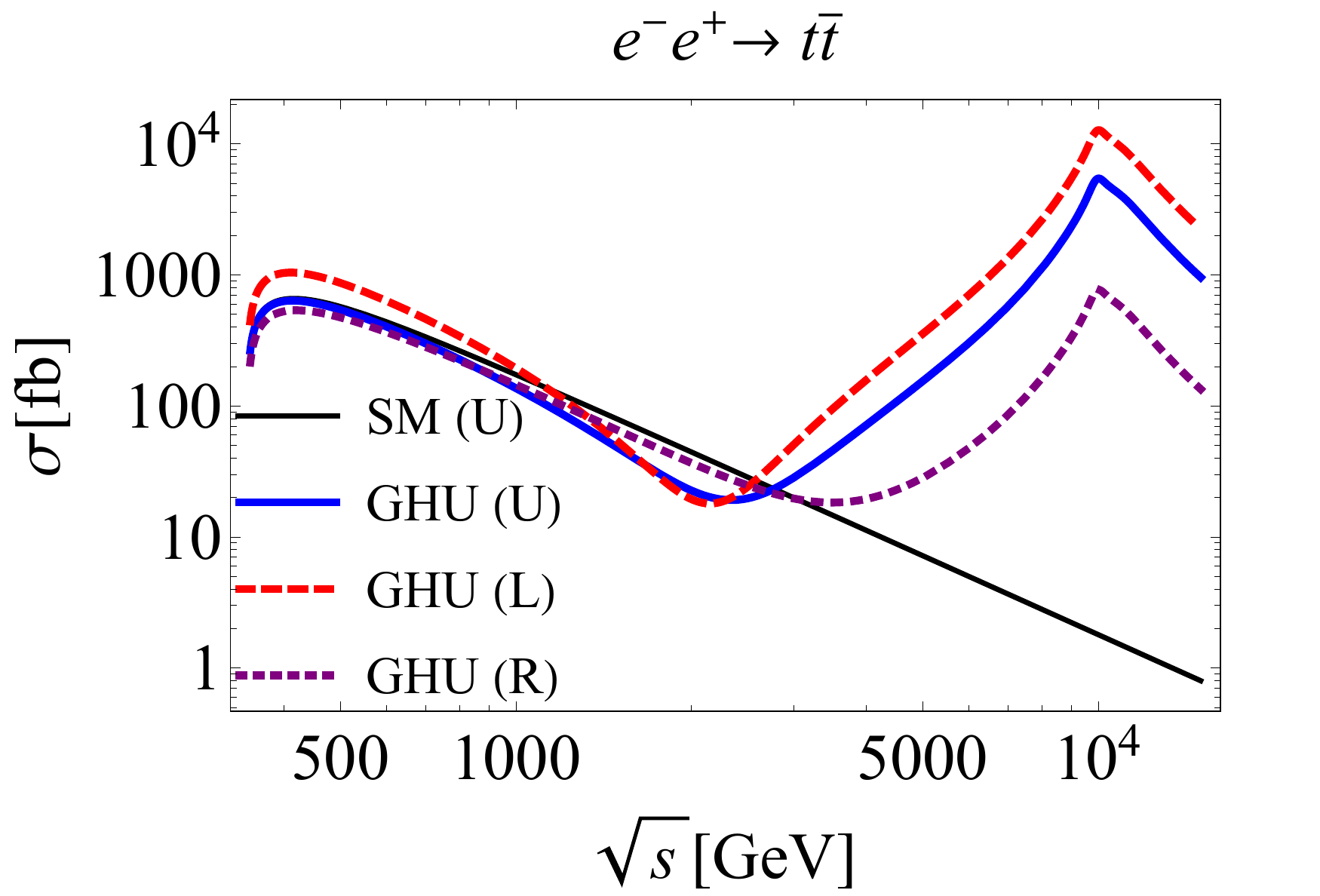}
\includegraphics[bb=0 0 504 335,height=3.45cm]{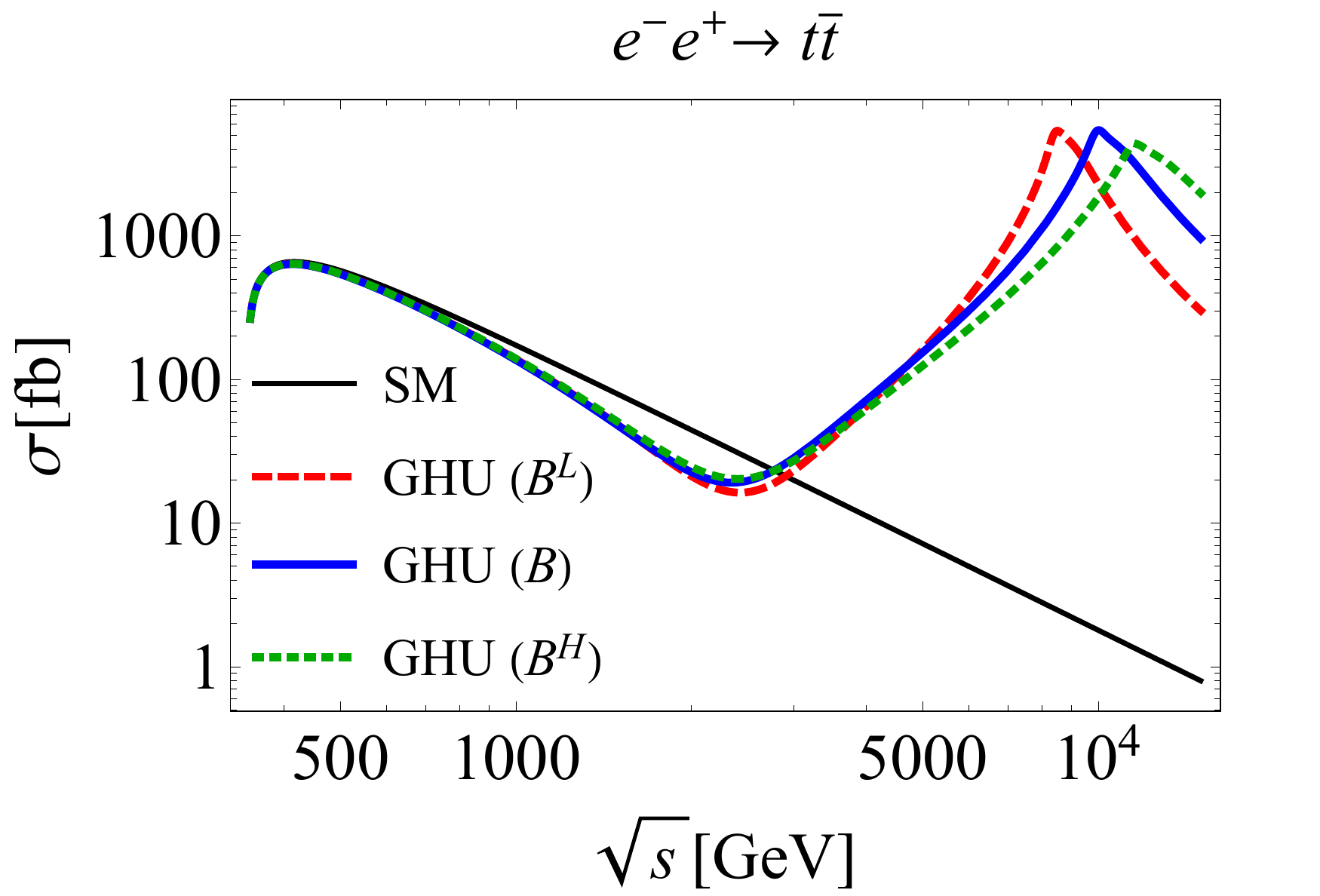}
\includegraphics[bb=0 0 504 335,height=3.45cm]{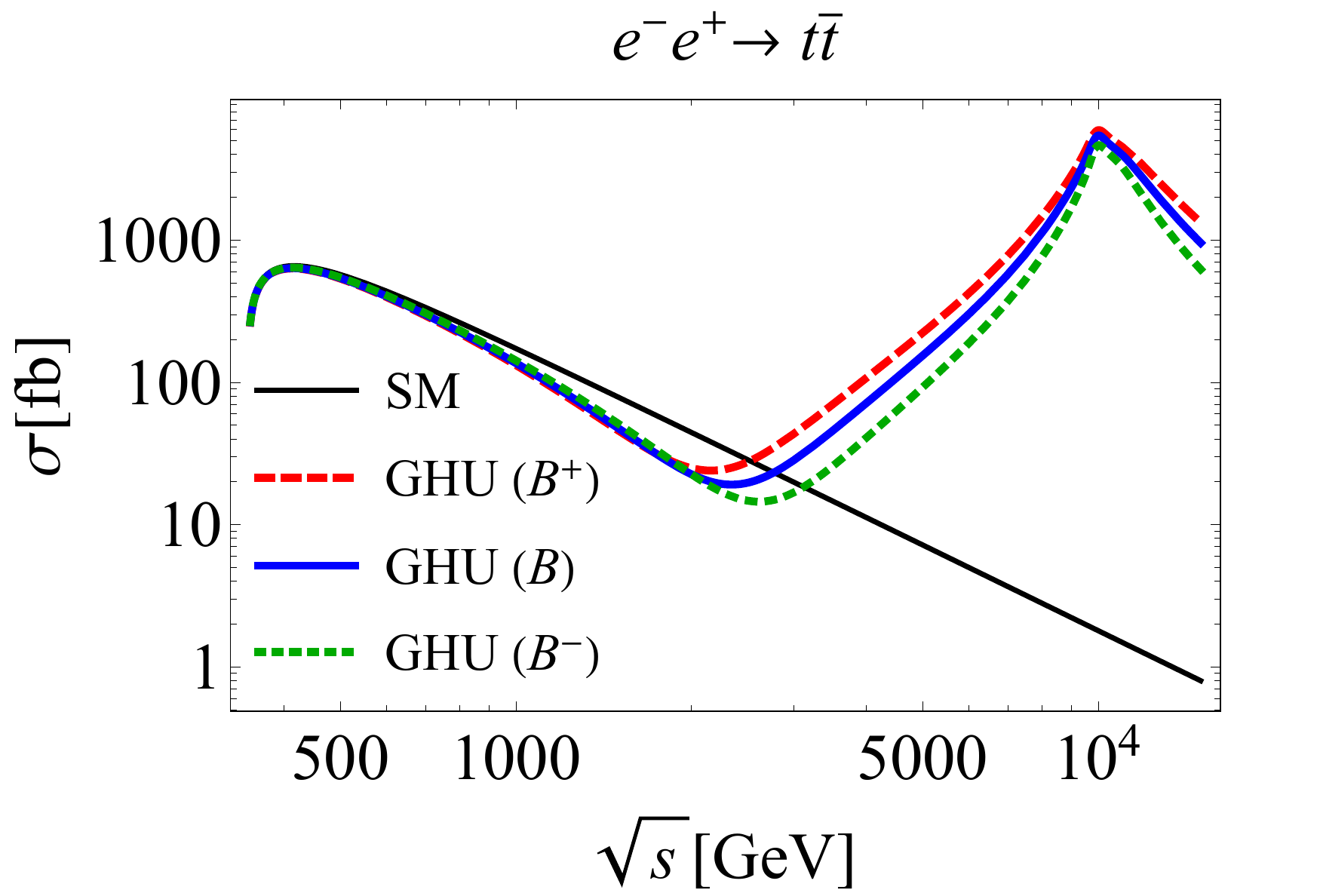}\\
 \caption{\small
Total cross sections $\sigma^{f\bar{f}}$ 
 $(f\bar{f}=\mu^-\mu^+,c\bar{c},b\bar{b},t\bar{t})$ are displayed in wider range of $\sqrt{s}$.
 In the left column $\sigma^{f\bar{f}}$ in GHU (B) is shown with  polarized and unpolarized 
 $e^\mp$  beams with $(P_{e^-},P_{e^+})=(0,0), (-0.8,+0.3), (+0.8,-0.3)$
which are  referred to as (U), (L), (R), respectively. 
In the middle and right  columns $\sigma^{f\bar{f}}$ with $(P_{e^-},P_{e^+})=(0,0)$
is shown in the GHU (B$^{\rm L}$), (B), (B$^{\rm H}$)  and 
in the GHU (B$^+$), (B), (B$^-$) in Table~\ref{Table:Mass-Width-Vector-Bosons}.
 } 
 \label{Figure:sigma-ef-mKK-theta}
\end{center}
\end{figure}

Cross sections are determined in terms of $Q_{e_{X} f_{Y}}$ ($X, Y = L, R$) in (\ref{Eq:Qs}).
In Figure~\ref{Figure:sQ_ef-s_dependence} $\sqrt{s}$-dependence of  $s |Q_{e_{X} f_{Y}}|$
is displayed.  In the SM, for $e^-e^+\to\mu^-\mu^+$ for instance,
\begin{align}
Q_{e_{X} \mu_{Y}}^\SM
&= \frac{e^2}{s}  +\frac{g_{Ze}^Xg_{Z\mu}^Yg_{w}^2}{(s-m_{Z}^{2}) + i m_{Z}\Gamma_{Z}} ~, \cr
\noalign{\kern 5pt}
\big| s Q_{e_X\mu_Y}^\SM \big|^2 &=
e^4   +\frac{(g_{Ze}^Xg_{Z\mu}^Yg_w^2)^2 ~ s^2} {(s-m_Z^2)^2+m_Z^2\Gamma_Z^2}
 + \frac{2  e^2 g_{Ze}^Xg_{Z\mu}^Yg_w^2 ~ s (s-m_Z^2)}{(s-m_Z^2)^2+m_Z^2\Gamma_Z^2}.
\end{align}
$s |Q_{e_{X} \mu_{Y}}|$ has peak at $\sqrt{s}=m_Z$ and $Q_{e_{L} \mu_{R}} = Q_{e_{R} \mu_{L}}$.
$Q_{e_{L} \mu_{L}} =Q_{e_{R} \mu_{R}}$ becomes smaller below $\sqrt{s}=m_Z$ and 
$Q_{e_{L} \mu_{R}}$ and $Q_{e_{R} \mu_{L}}$ become smaller above $\sqrt{s}=m_Z$
as a result  of the interference of the $\gamma$ and $Z$ amplitudes.
We also note that $sQ_{e_Xf_Y} \simeq e^2+g_{Ze}^Xg_{Z\mu}^Yg_w^2$ for $\sqrt{s}\gg m_{Z}$.

\ignore{
The first and second terms are always positive values, while
the last term can be positive or negative values depending on
the sign of the coupling constants $g_{Ze}^{X}$ and $g_{Z\mu}^Y$ and $(s-m_Z^2)$.
The sign of $g_{Ze}^{L}g_{Z\mu}^{L}$ and $g_{Ze}^{R}g_{Z\mu}^{R}$ is positive, while
the sign of $g_{Ze}^{L}g_{Z\mu}^{R}$ and $g_{Ze}^{R}g_{Z\mu}^{L}$ is negative.
Therefore, the last term of $Q_{e_L\mu_L}$ and $Q_{e_R\mu_R}$ is negative below $\sqrt{s}=m_Z$;
that of $Q_{e_L\mu_R}$ and $Q_{e_R\mu_L}$ is negative above $\sqrt{s}=m_Z$.}

\begin{figure}[thb]
\begin{center}
\includegraphics[bb=0 0 504 344,height=5cm]{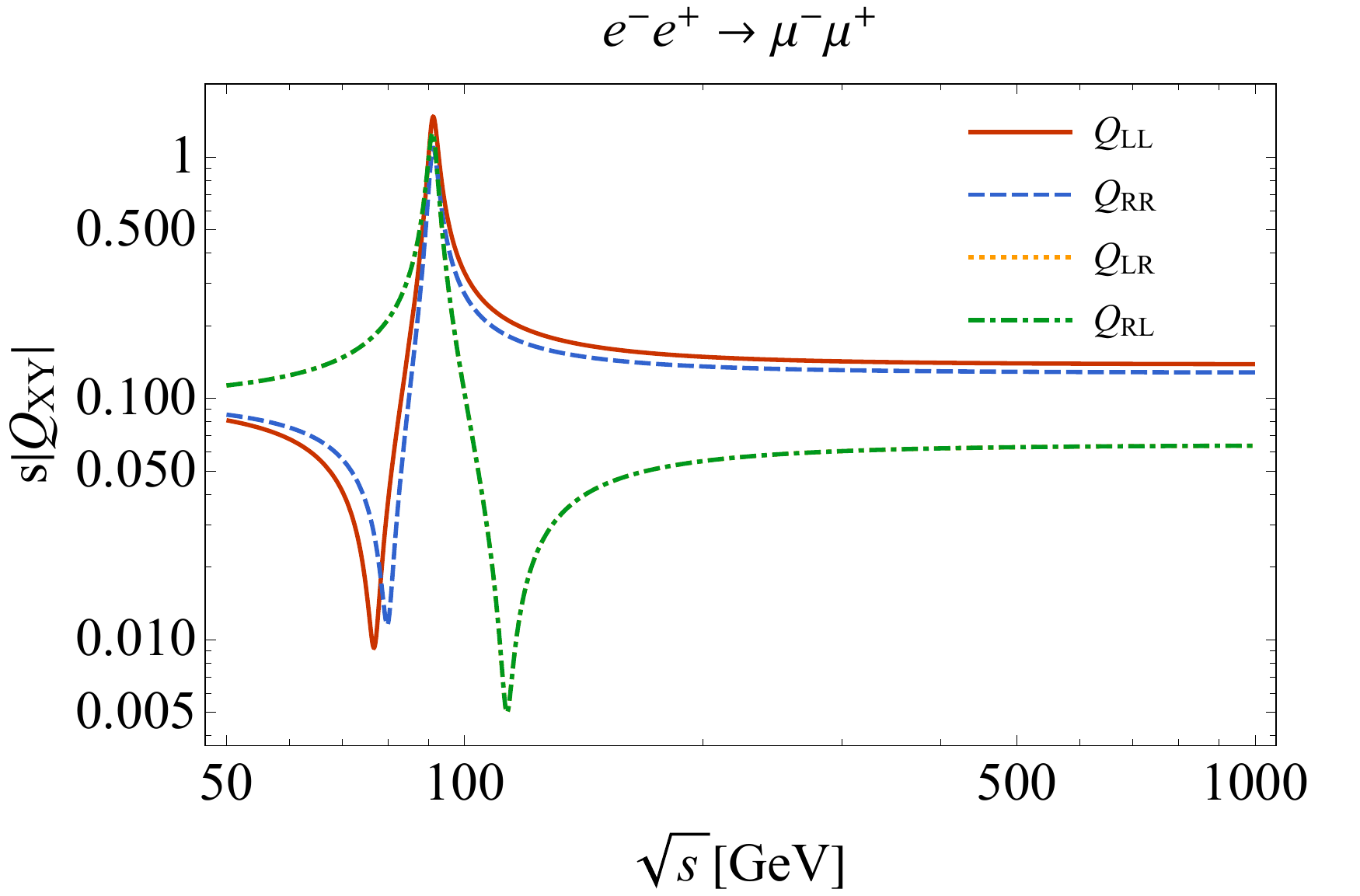}
\includegraphics[bb=0 0 504 338,height=5cm]{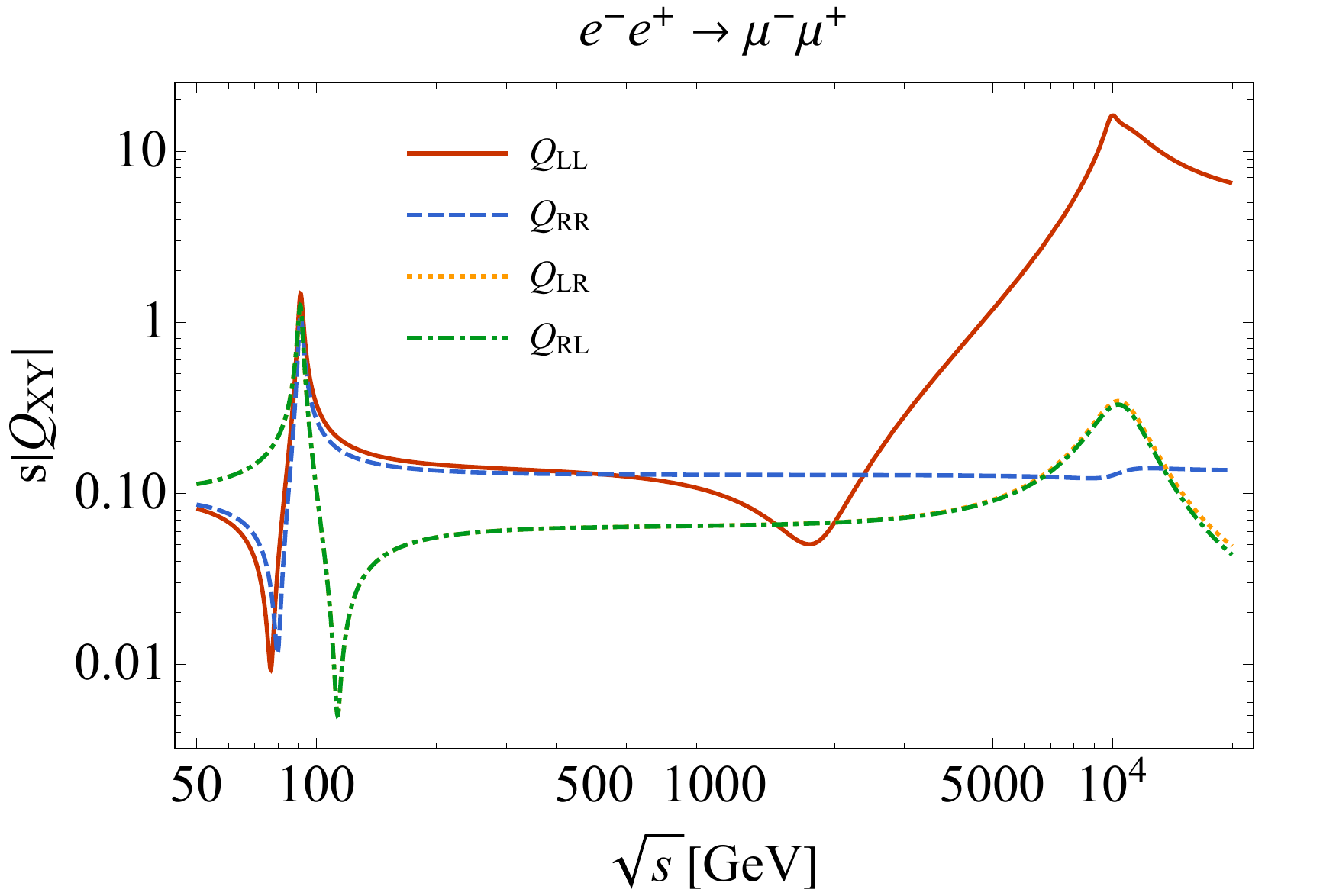}\\[1em]
\includegraphics[bb=0 0 504 344,height=5cm]{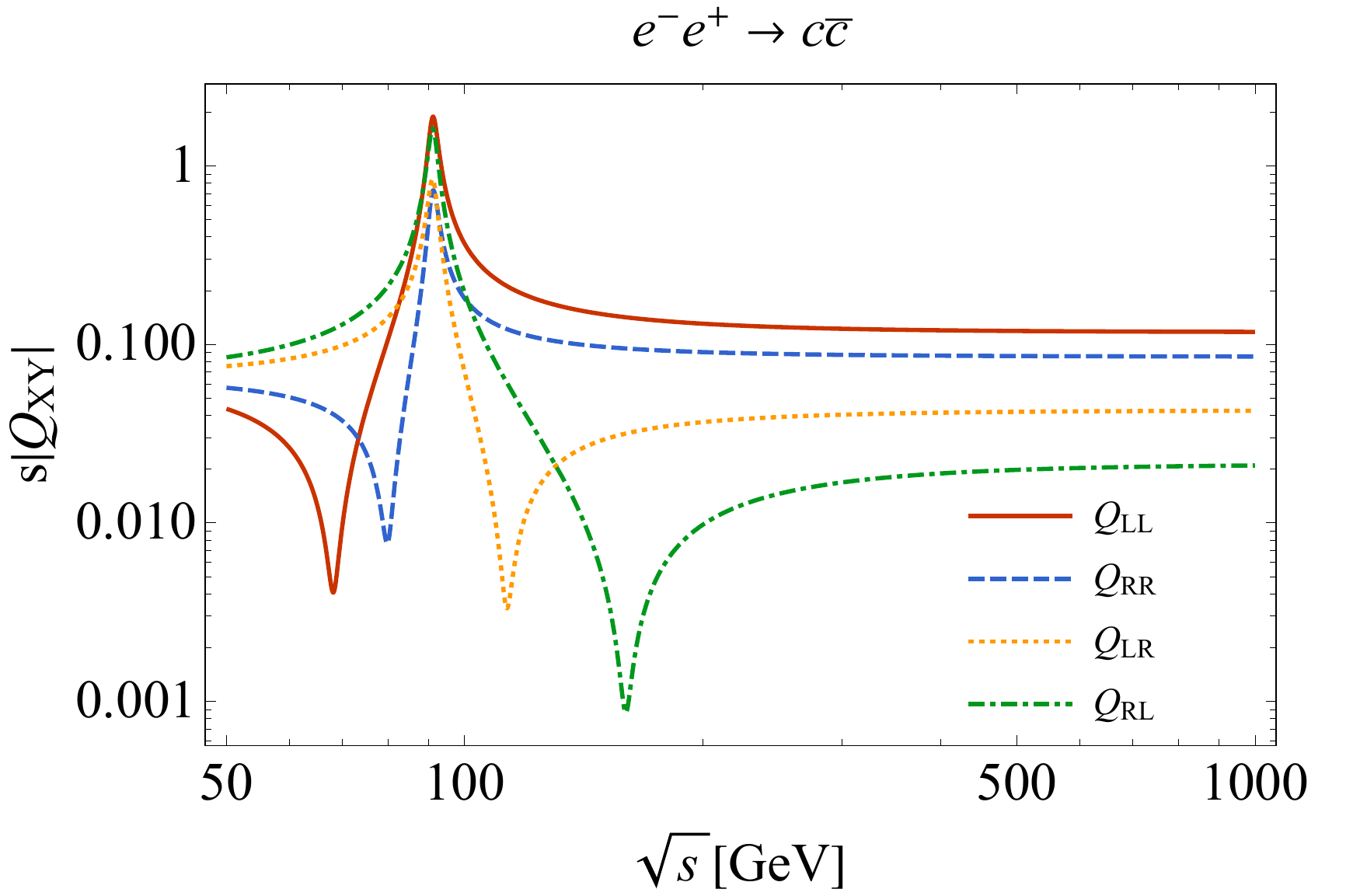}
\includegraphics[bb=0 0 504 332,height=5cm]{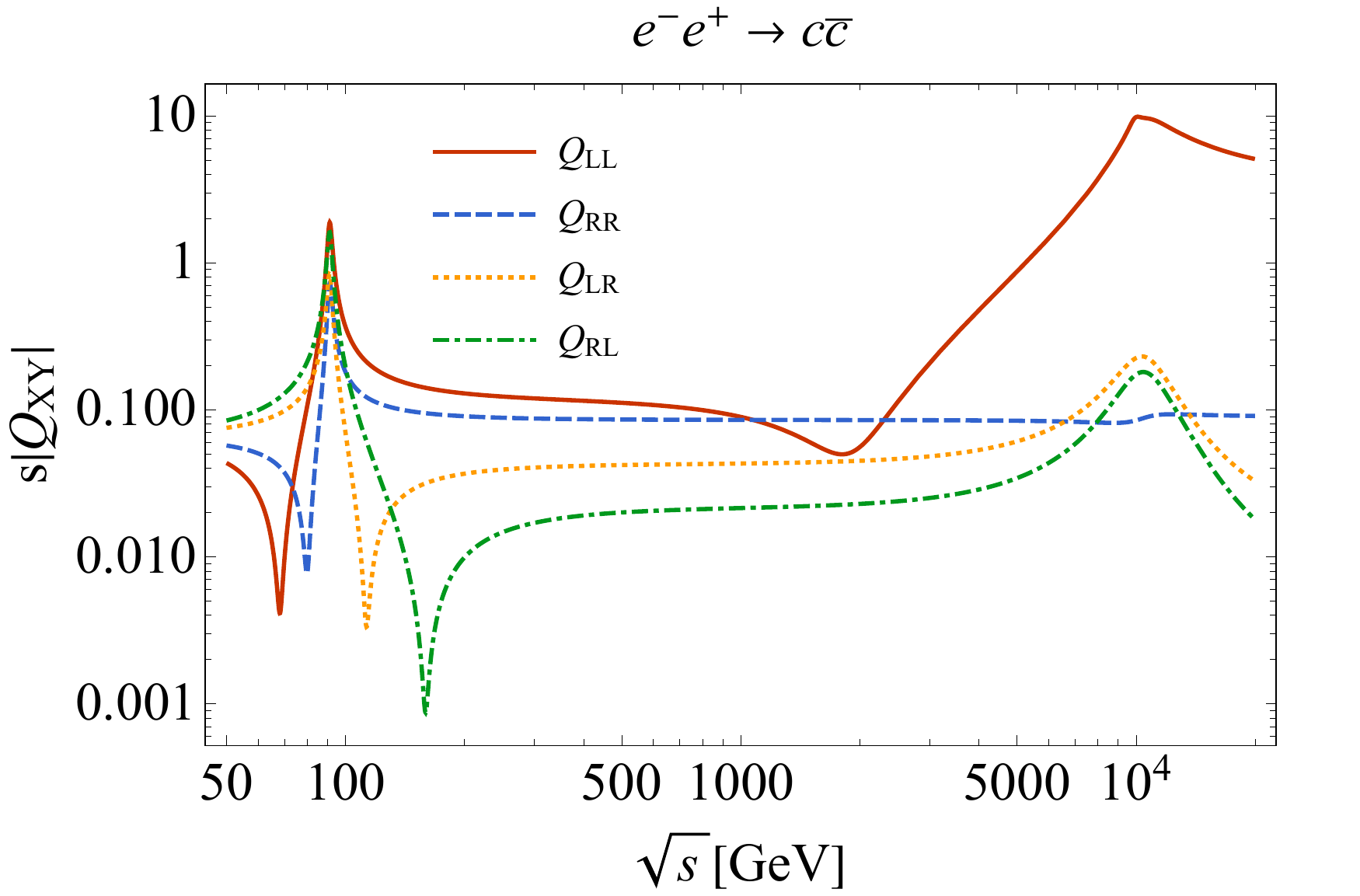}\\[1em]
\includegraphics[bb=0 0 504 346,height=5cm]{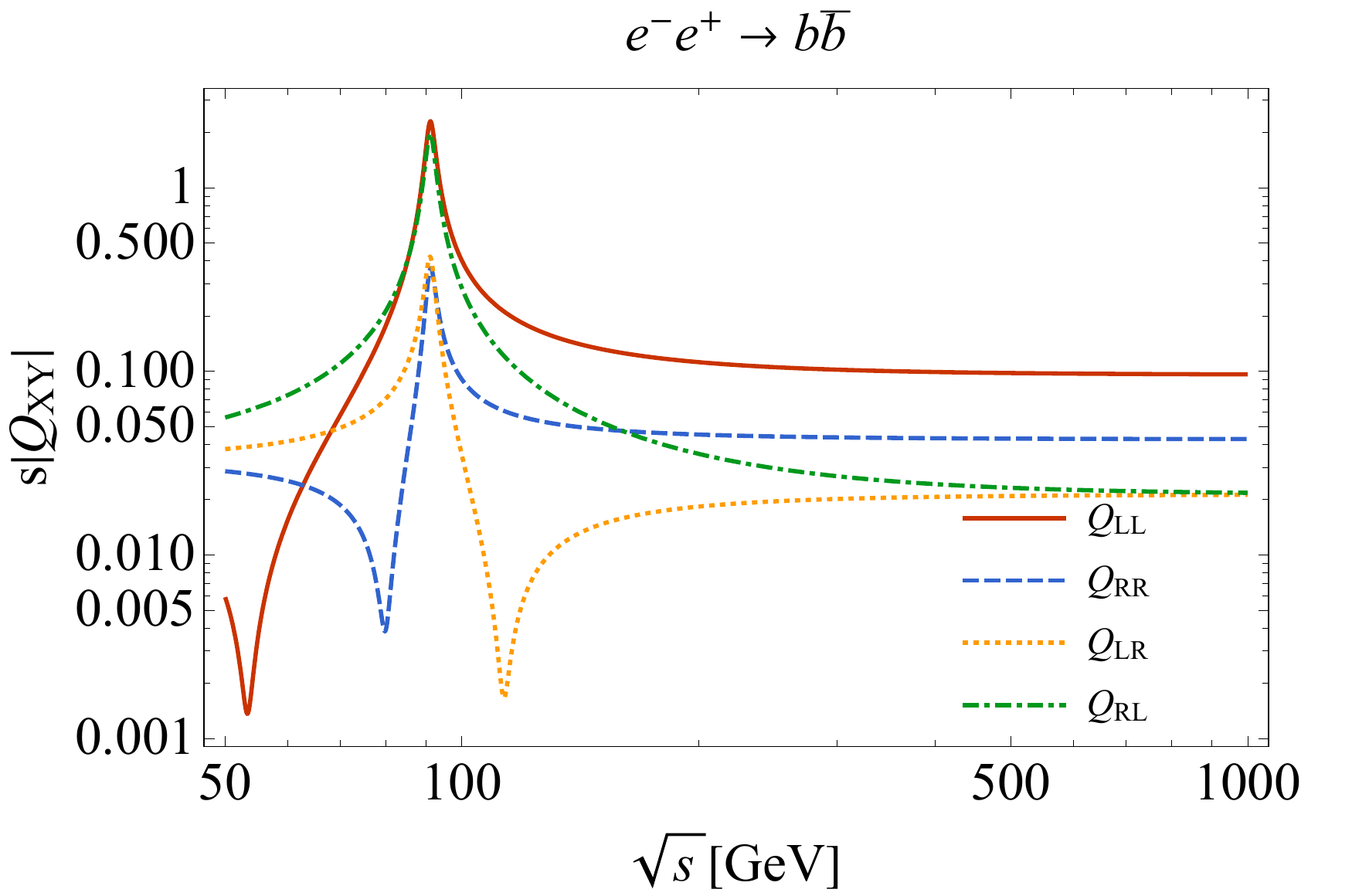}
\includegraphics[bb=0 0 504 334,height=5cm]{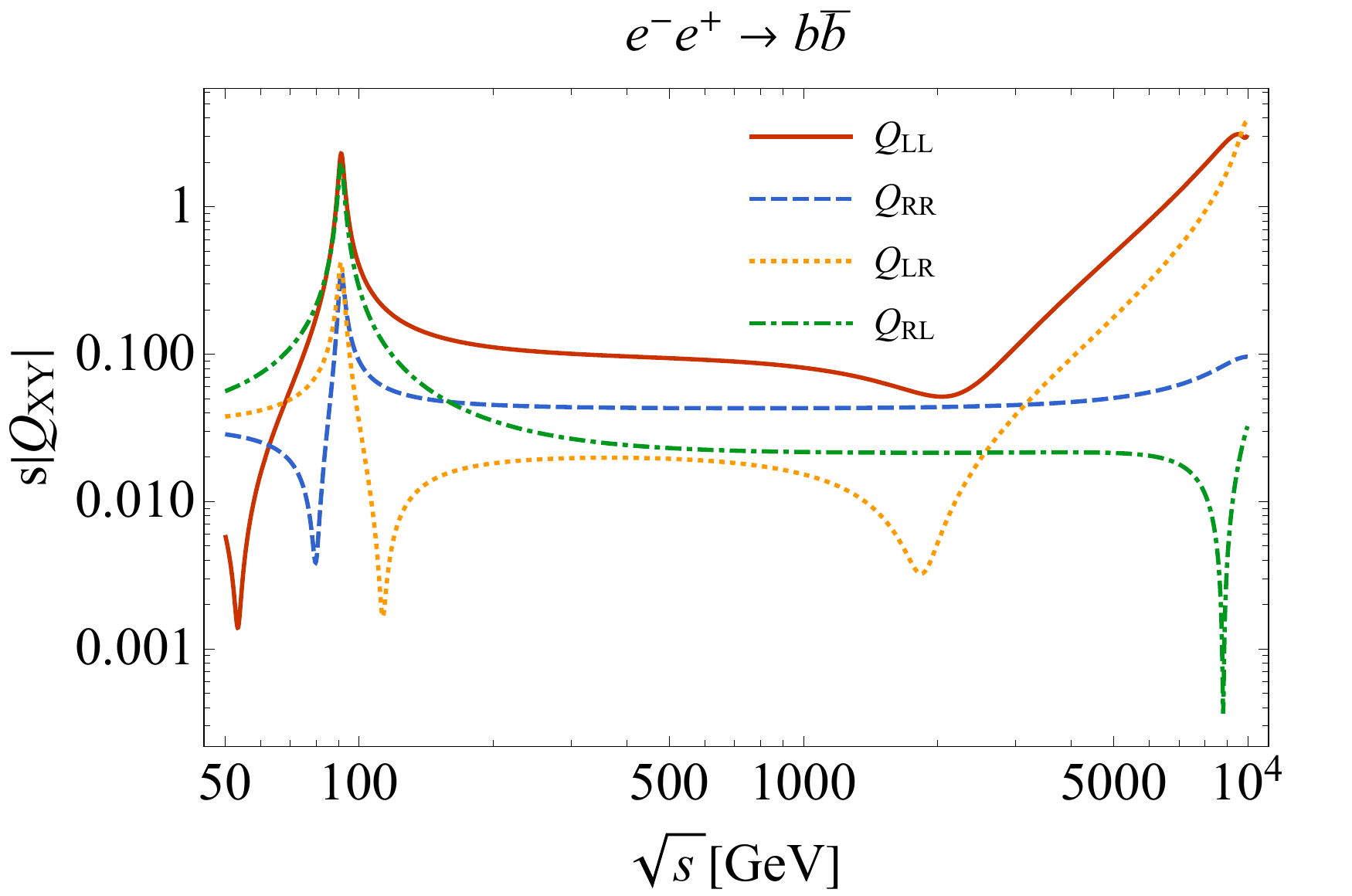}\\[1em]
\includegraphics[bb=0 0 504 344,height=5cm]{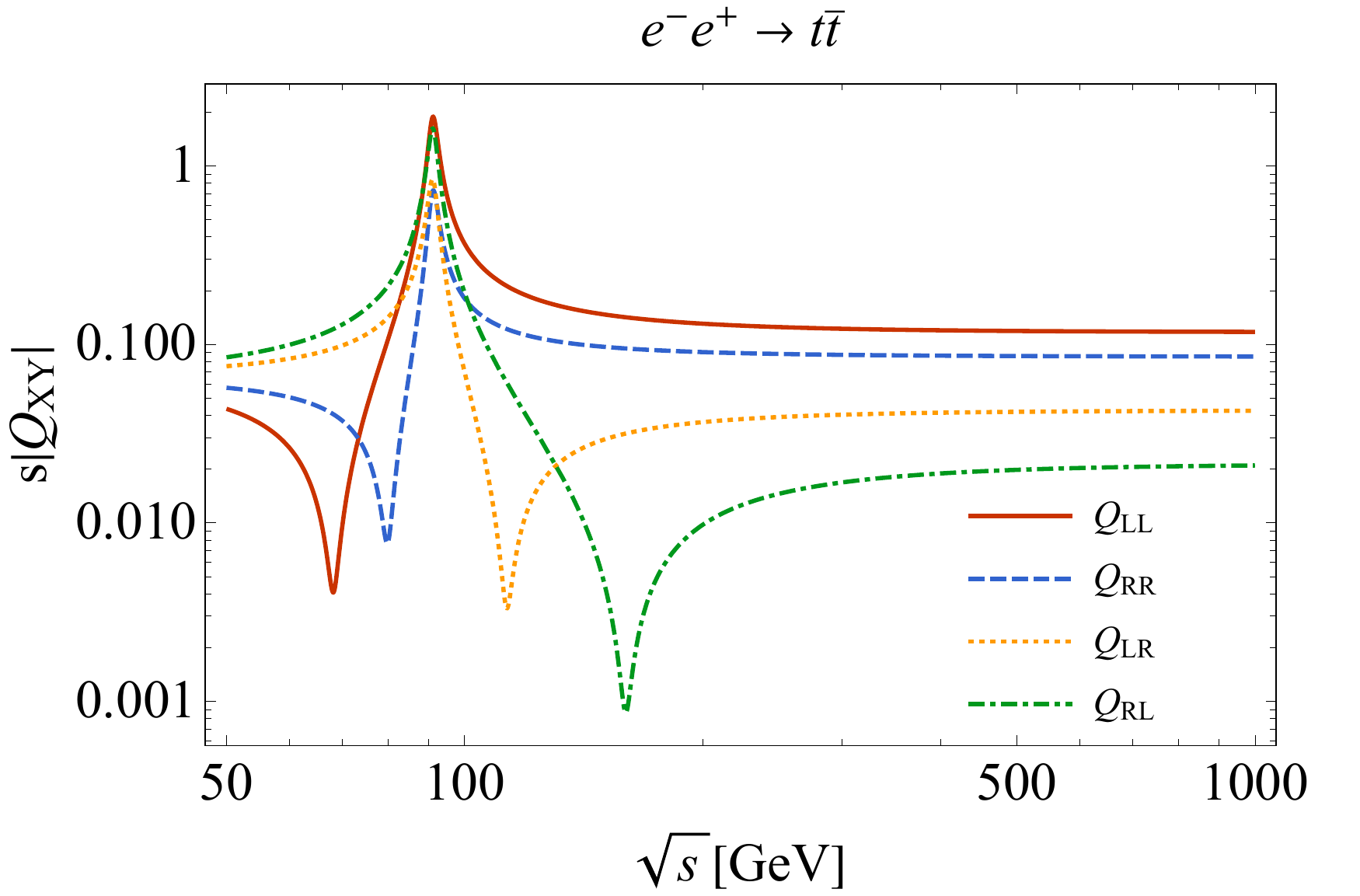}
\includegraphics[bb=0 0 504 332,height=5cm]{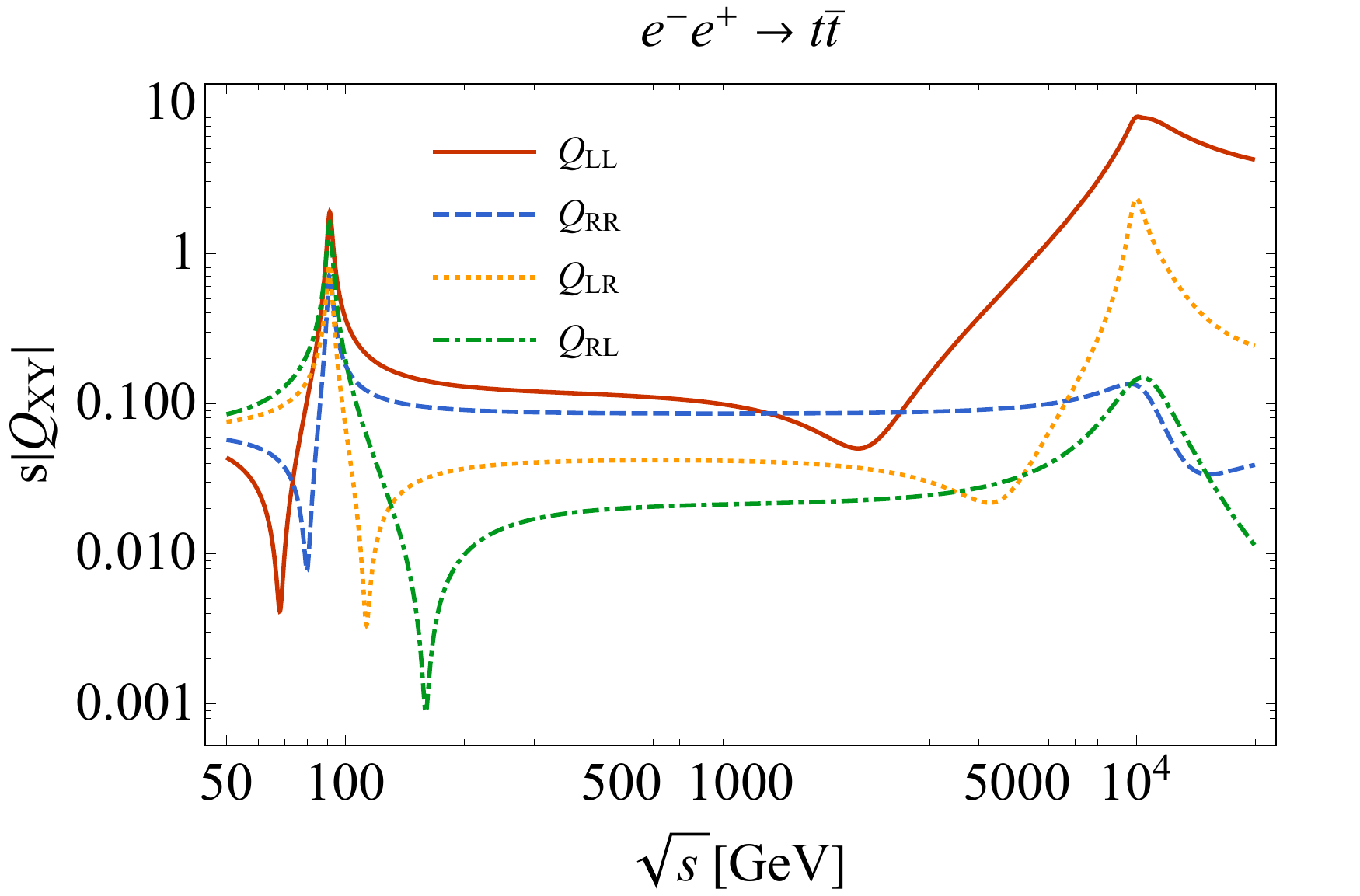}\\[1em]
 \caption{\small
 The amplitude $s|Q_{e_X f_Y}|(e^-e^+\to f\bar{f})$
 $(X,Y=L, R\, ; f\bar{f}=\mu^-\mu^+,c\bar{c},b\bar{b},t\bar{t})$ vs $\sqrt{s}\,$[GeV] for
 the SM (left side figures)
 and  the GHU (B) (right side figures) in
 Table~\ref{Table:Mass-Width-Vector-Bosons} 
 are shown.
In each figure $Q_{e_X f_Y}$ is denoted as $Q_{X Y}$.
 The energy ranges $\sqrt{s}$ in the left and right side figures are
 $\sqrt{s}=[50,1000] \,$GeV and $[50,2\times 10^4] \,$GeV, respectively.
 } 
 \label{Figure:sQ_ef-s_dependence}
\end{center}
\end{figure}

In GHU 
\begin{align}
Q_{e_{X} f_{Y}} &= Q_{e_{X} f_{Y}}^\SM + Q_{e_{X} f_{Y}}^{Z'} ~, \cr
\noalign{\kern 5pt}
Q_{e_{X} f_{Y}}^{Z'} & \simeq \sum_{V = Z^{(1)}, \gamma^{(1)}, Z_R^{(1)}}
 \frac{g_{V e}^X g_{V f}^Y g_{w}^2} {(s-m_V^2) + i m_V \Gamma_V}  ~,
\end{align}
where we have retained contributions from first KK modes in $Q_{e_{X} f_{Y}}^{Z'}$.
For $\sqrt{s} \lesssim 200\,$GeV, $Q_{e_{X} f_{Y}} \sim Q_{e_{X} f_{Y}}^\SM $ to good approximation.
In Figure~\ref{Figure:sQ_ef-s_dependence} the $\sqrt{s}$-dependence of $s |Q_{e_{X} f_{Y}}|$ is
plotted.  $Q_{e_{X} f_{Y}}$ has a peak around $\sqrt{s}\simeq m_{Z'}\simeq 10 \,$TeV.
The dominant component is $Q_{e_{L} f_{L}}$, which develops significant deviation from the SM.
$Q_{e_{L} f_{L}}$ has a dip around $\sqrt{s}\simeq 1.7 \,$TeV.
For $f = b, t$, an additional dip is seen in the  2--5 TeV region for $Q_{e_{L} f_{R}}$.

We stress that due to the interference effects among $\gamma$, $Z$
and $Z'$ bosons, the GHU prediction for the total cross section shown in 
Figures~\ref{Figure:sigma-ef-Peff-theta} and \ref{Figure:sigma-ef-mKK-theta}
deviates from that in the SM  even well below the masses of $Z'$ bosons.
Also, from Figure~\ref{Figure:sQ_ef-s_dependence}, the behavior of 
the various components of the scattering amplitudes $Q_{e_X f_Y}$ is different so that
by using the polarized electron-positron beams, one can investigate physics at
10 TeV region in more detail than with unpolarized beams.

Let us look at differential cross sections.  In Figure~\ref{Figure:dsigma-ef-theta=010}, 
$d\sigma^{f\bar{f}} /d\cos\theta$ are shown for $f\bar{f}= \mu^-\mu^+, c\bar{c}, b\bar{b}$,
at  $\sqrt{s}=250\,$GeV and  for $f\bar{f}= t\bar{t}$ at $\sqrt{s}=500 \,$GeV.
Differential cross sections in the forward region are  larger than 
those of the backward region $(\cos\theta=[0,1])$ regardless of the polarization.
The deviation from the SM are seen in the forward region with less statistical errors.
The differential cross sections of  the 100\% left- and right-handed polarized initial electron are
given by the formulas in Eq.~(\ref{Eq:dsigma_LR-RL-mf=0}).
In the SM the $Z$ couplings are different for left-handed and right-handed fermions which
leads to $Q_{e_{L} f_{L}} \not= Q_{e_{L} f_{R}}$ and $Q_{e_{R} f_{R}} \not= Q_{e_{R} f_{L}}$ and
therefore  forward-backward asymmetry.   

In GHU   coupling constants of the left-handed fermions to
$Z'$ bosons are, in most cases, much larger than those of the right-handed ones.
The magnitude of the left-handed fermion couplings is rather large so that
the amount of the deviation in $d\sigma^{f\bar{f}}/d\cos\theta$ from the SM becomes
large for either left-handed polarized or unpolarized electron beams, whereas the deviation
becomes small for right-handed electron beams.
$\Delta_{d\sigma}^{f\bar{f}}(P_{e^-},P_{e^+}, \cos \theta)$ in (\ref{Eq:Delta_dsigma})
is plotted in the right column of Figure~\ref{Figure:dsigma-ef-theta=010}.
The deviation can be clearly seen in $e^- e^+$ collisions at $\sqrt{s}=250 \,$GeV 
with 250$\,$fb$^{-1}$ data  for $f\bar{f}=\mu^-\mu^+,c\bar{c},b\bar{b}$ and
at $\sqrt{s}=500 \,$GeV with 500$\,$fb$^{-1}$ data for $f\bar{f}=t\bar{t}$.

\begin{figure}[thb]
\begin{center}
\includegraphics[bb=0 0 504 341,height=5cm]{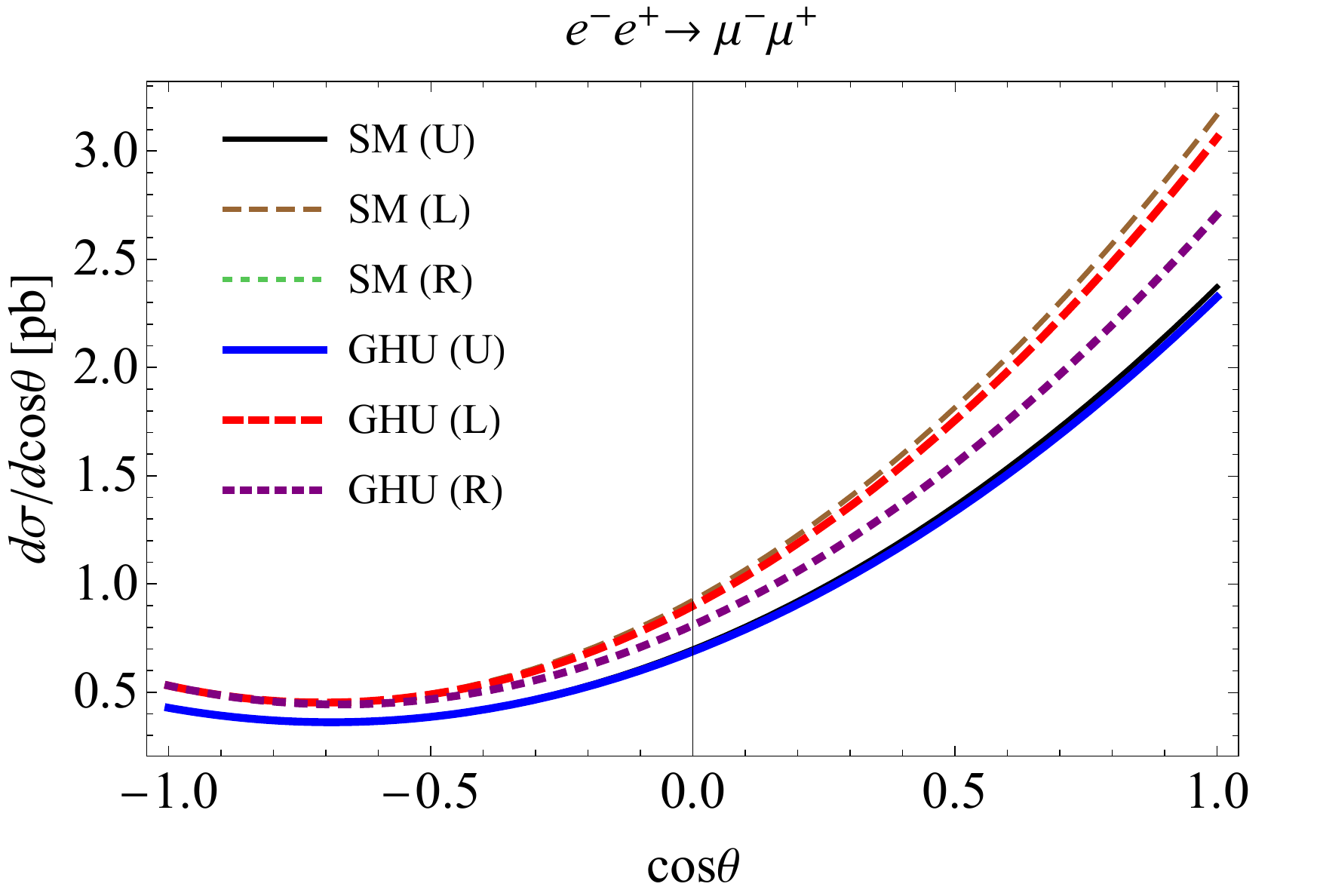}
\includegraphics[bb=0 0 504 327,height=5.cm]{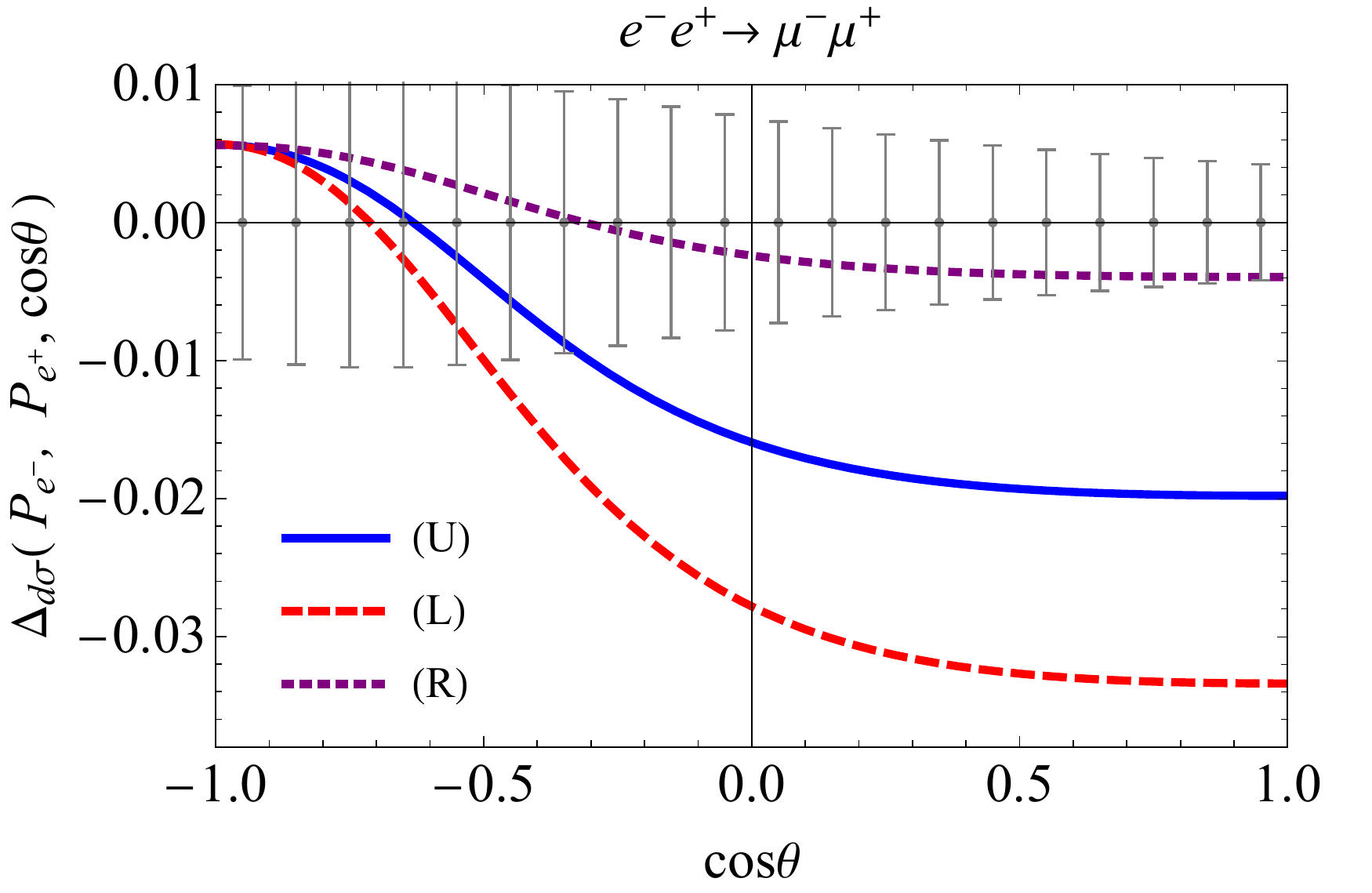}\\[0.5em]
\includegraphics[bb=0 0 504 341,height=5cm]{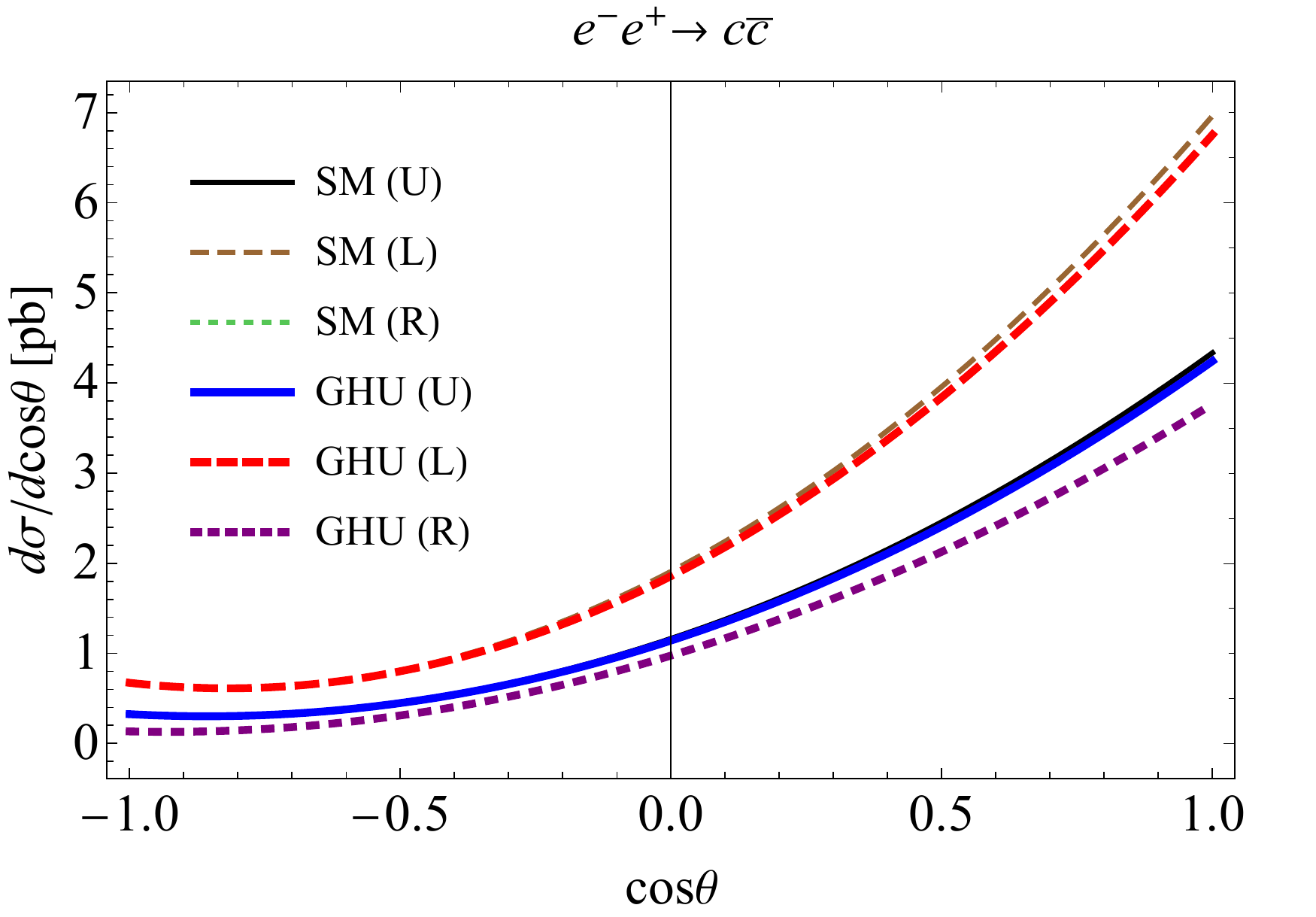}
\includegraphics[bb=0 0 504 327,height=5.cm]{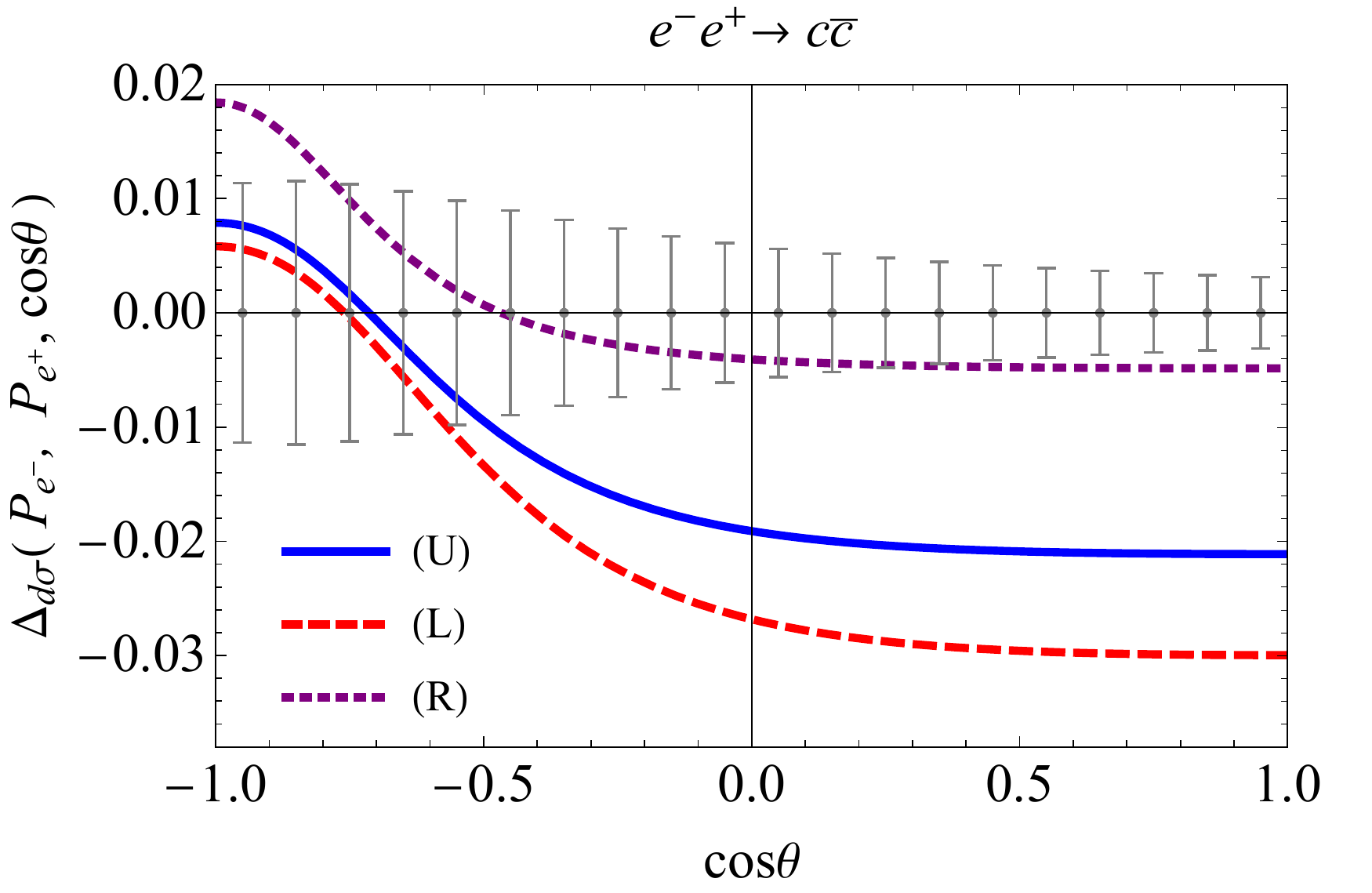}\\[0.5em]
\includegraphics[bb=0 0 504 341,height=5cm]{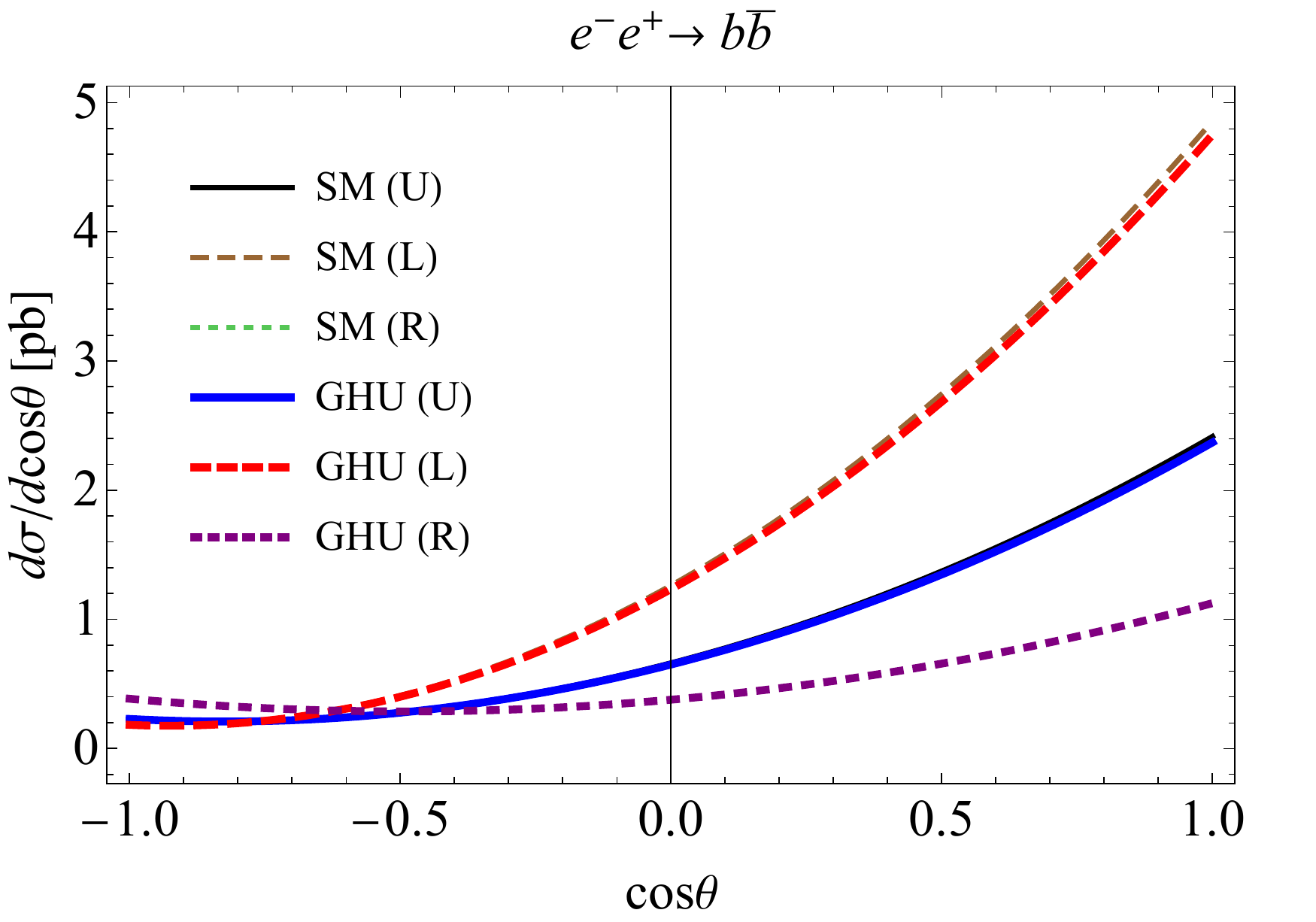}
\includegraphics[bb=0 0 504 327,height=5.cm]{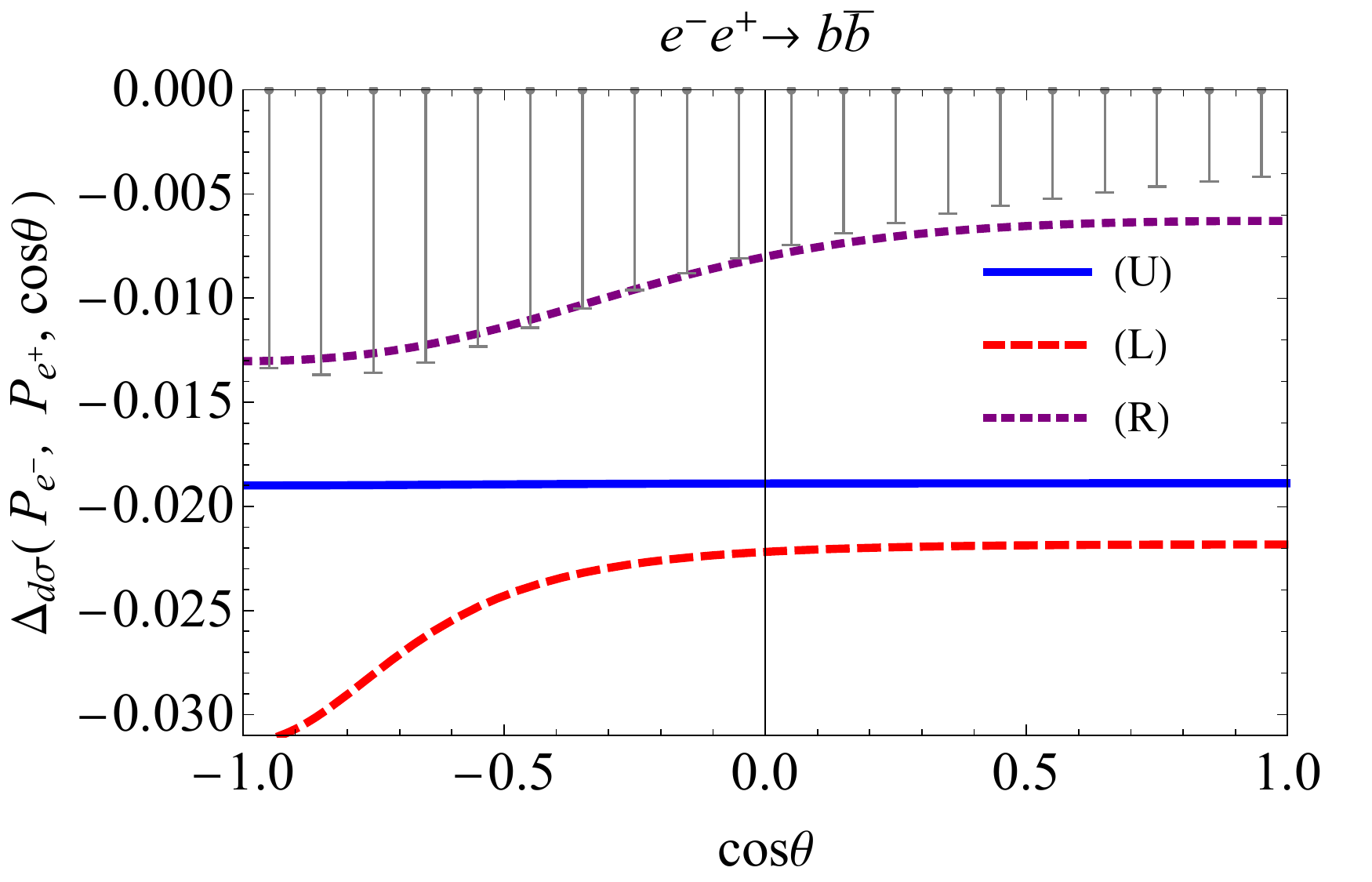}\\[0.5em]
\includegraphics[bb=0 0 504 341,height=5cm]{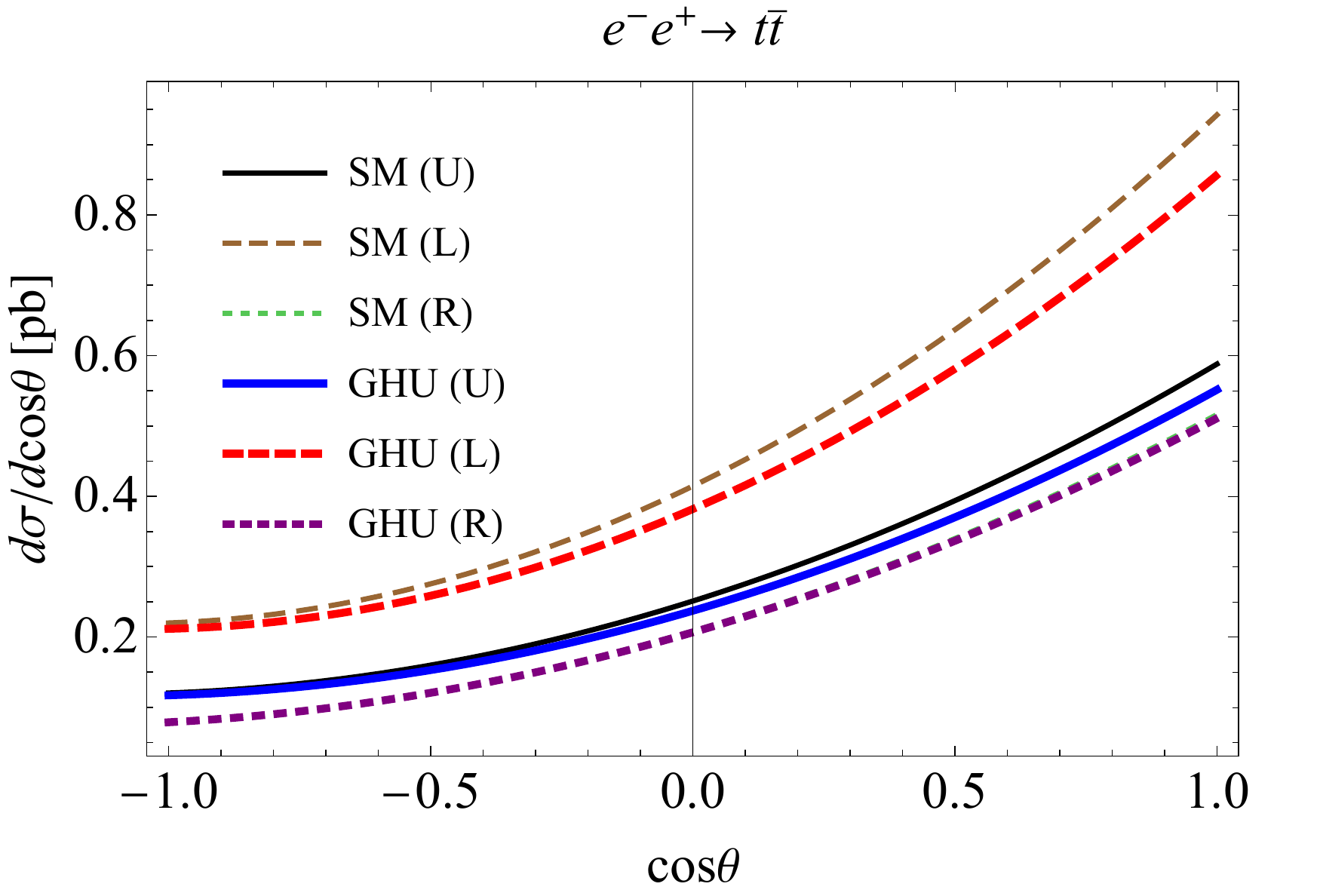}
\includegraphics[bb=0 0 504 327,height=5.cm]{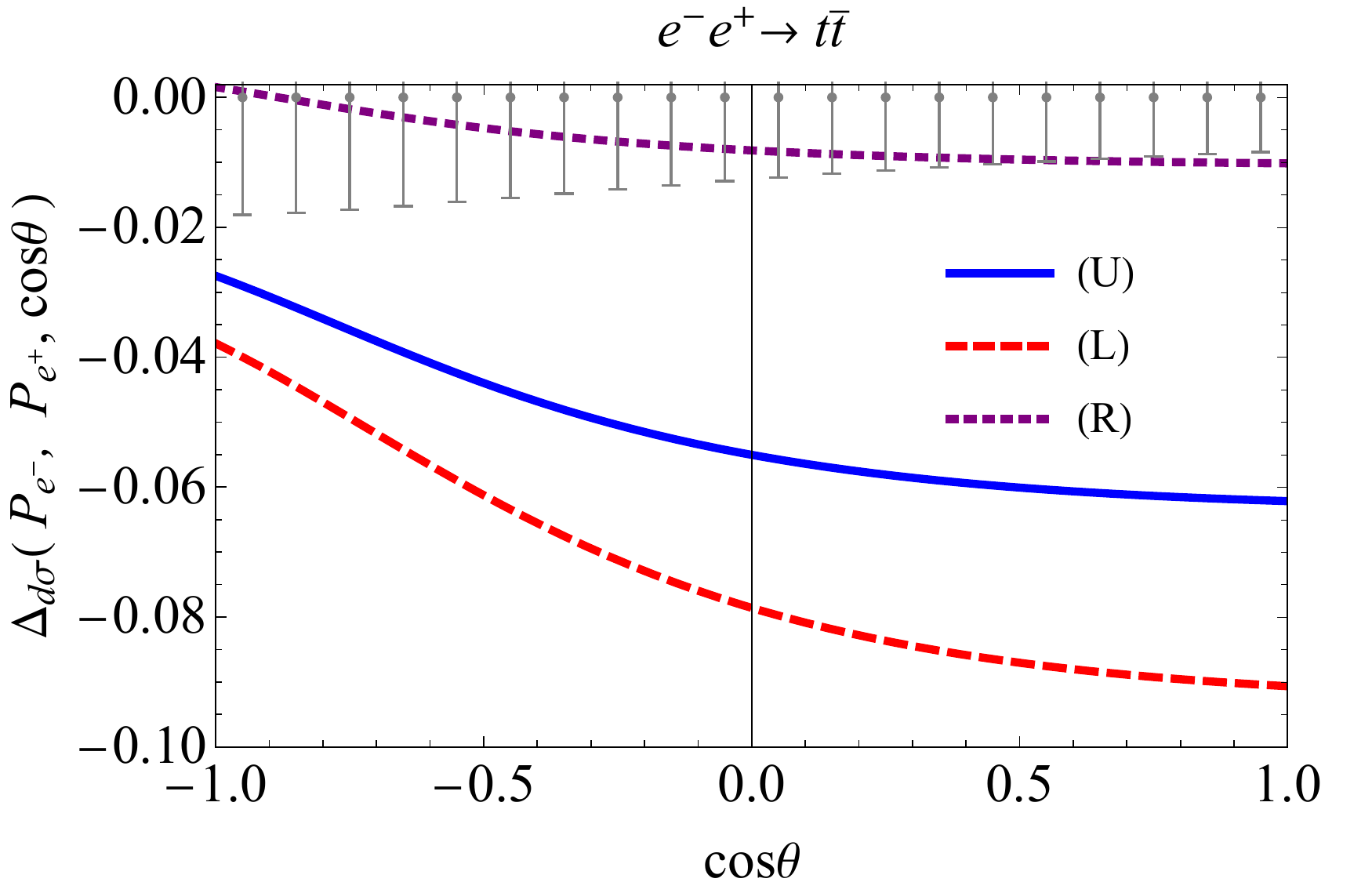}\\
 \caption{\small
 Differential cross sections  $d\sigma^{f\bar{f}}/d\cos\theta$
 $(f\bar{f}=\mu^-\mu^+,c\bar{c},b\bar{b},t\bar{t})$ are shown.
 The left side figures show the $\theta$ dependence of  $d\sigma^{f\bar{f}}/d\cos\theta$
in the SM and the GHU (B) in  Table~\ref{Table:Mass-Width-Vector-Bosons}
 with three sets $(P_{e^-},P_{e^+})=(0,0) (U), (-0.8,+0.3) (L), (+0.8,-0.3) (R)$.  
$\sqrt{s}=250 \,$GeV for $f\bar{f}=\mu^-\mu^+,c\bar{c},b\bar{b}$,  and
 $\sqrt{s}=500 \,$GeV for $f\bar{f}=t\bar{t}$.
 The right side figures show the $\theta$ dependence of 
$\Delta_{d\sigma}^{f\bar{f}}(P_{e^-},P_{e^+}, \cos\theta )$ in (\ref{Eq:Delta_dsigma}).
The error bars  represent   statistical errors in the SM 
at $\sqrt{s}=250 \,$GeV with 250$\,$fb$^{-1}$ data  for $f\bar{f}=\mu^-\mu^+,c\bar{c},b\bar{b}$
and  at $\sqrt{s}=500\,$GeV with 500$\,$fb$^{-1}$ data  for $f\bar{f}=t\bar{t}$.
Each bin is given by $\cos\theta=[k-0.05,k+0.05]$ ($k=-0.95,-0.85,\cdots,0.95$).
 } 
 \label{Figure:dsigma-ef-theta=010}
\end{center}
\end{figure}

\subsection{Forward-backward asymmetry}

The forward-backward asymmetry $A_{FB}^{f\bar{f}}$ is shown in  Figure~\ref{Figure:AFB-ef-theta=010}.
From Eq.~(\ref{Eq:A_FB}), 
$A_{FB}^{f\bar{f}}(P_{e^-},P_{e^+})$ with $(P_{e^-},P_{e^+})=(0,0),(-1,0),(+1,0)$
are given
by 
\begin{align}
A_{FB}^{f\bar{f}}(0, 0) ~ &\simeq
\frac{3}{4}\frac{
\{|Q_{e_{R} f_{R}}|^{2}+|Q_{e_{L} f_{L}}|^{2} \}
-\{|Q_{e_{R} f_{L}}|^{2} +|Q_{e_{L} f_{R}}|^{2} \}
}{
\{|Q_{e_{R} f_{R}}|^{2} +|Q_{e_{L} f_{L}}|^{2} \}
+\{|Q_{e_{R} f_{L}}|^{2} +|Q_{e_{L} f_{R}}|^{2} \} } ~, \cr 
A_{FB}^{f\bar{f}}(-1, 0) &\simeq
 \frac{3}{4}\frac{|Q_{e_{L} f_{L}}|^{2}-|Q_{e_{L} f_{R}}|^{2}}
 {|Q_{e_{L} f_{L}}|^{2}+|Q_{e_{L} f_{R}}|^{2}} ~, \cr
A_{FB}^{f\bar{f}}(1, 0) ~ &\simeq
 \frac{3}{4}\frac{|Q_{e_{R} f_{R}}|^{2}-|Q_{e_{R} f_{L}}|^{2}}
 {|Q_{e_{R} f_{R}}|^{2}+|Q_{e_{R} f_{L}}|^{2}}
\end{align}
for $\sqrt{s}\gg m_f$.
In the SM, the forward-backward asymmetry $A_{FB}^{f\bar{f}}$
becomes constant for $\sqrt{s}\gg m_Z$.
For $f\bar{f}=\mu^-\mu^+$, for instance, 
$A_{FB}^{\mu^-\mu^+}(P_{e^-},P_{e^+})\simeq 3/4$  at $Z$-pole $\sqrt{s}=m_Z$   since
$|Q_{e_L\mu_L}|\gg |Q_{e_R\mu_L}|,|Q_{e_L\mu_R}|,|Q_{e_R\mu_L}|$, 
and  $A_{FB}^{\mu^-\mu^+}(P_{e^-},P_{e^+})$ approaches constant for
$\sqrt{s}\gg m_Z$. 

In the GHU (B) in Table~\ref{Table:Mass-Width-Vector-Bosons}, due to the
interference effects between $Z$ and $Z'$ bosons,
$|Q_{e_L\mu_L}|$ can be smaller than $|Q_{e_L\mu_R}|$ in some energy
region (around $\sqrt{s} \sim 1.7\,$TeV).
Consequently $A_{FB}^{f\bar{f}}$ can become negative even for $\sqrt{s}\gg m_Z$
as shown in Figure~\ref{Figure:AFB-ef-theta=010}.
Deviation from the SM starts to show up around $\sqrt{s} = 250\,$GeV.  As shown
in the middle and right columns in Figure~\ref{Figure:AFB-ef-theta=010},  the amount of the deviation
$\Delta_{FB}^{f\bar{f}}(P_{e^-},P_{e^+}=0)$ in Eq.~(\ref{Eq:Delta_A_FB}) becomes
significant for $P_{e^-} \sim -1$ even at $\sqrt{s} = 250\,$GeV.

\begin{figure}[thb]
\begin{center}
\includegraphics[bb=0 0 504 336,height=3.3cm]{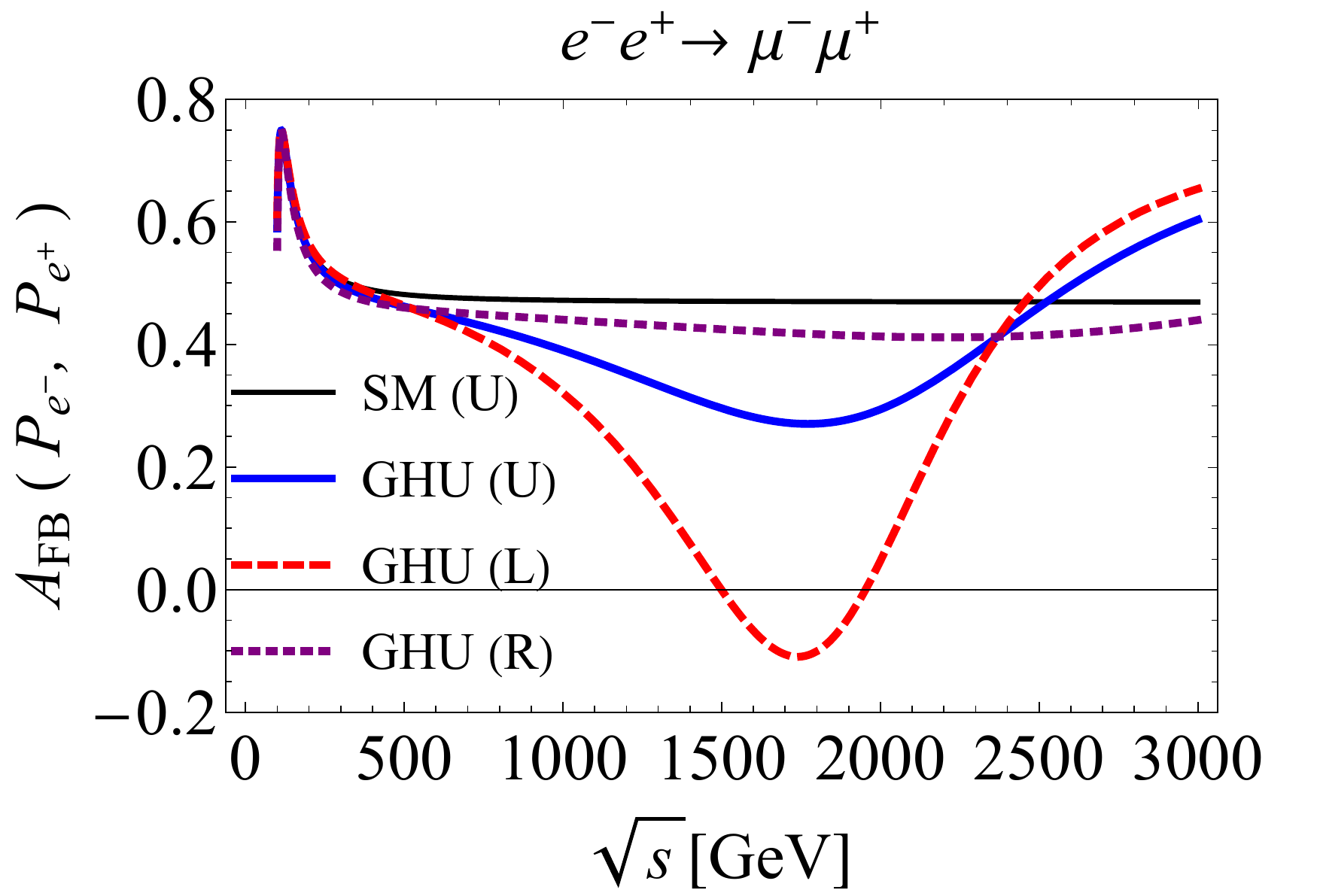}
\includegraphics[bb=0 0 504 320,height=3.3cm]{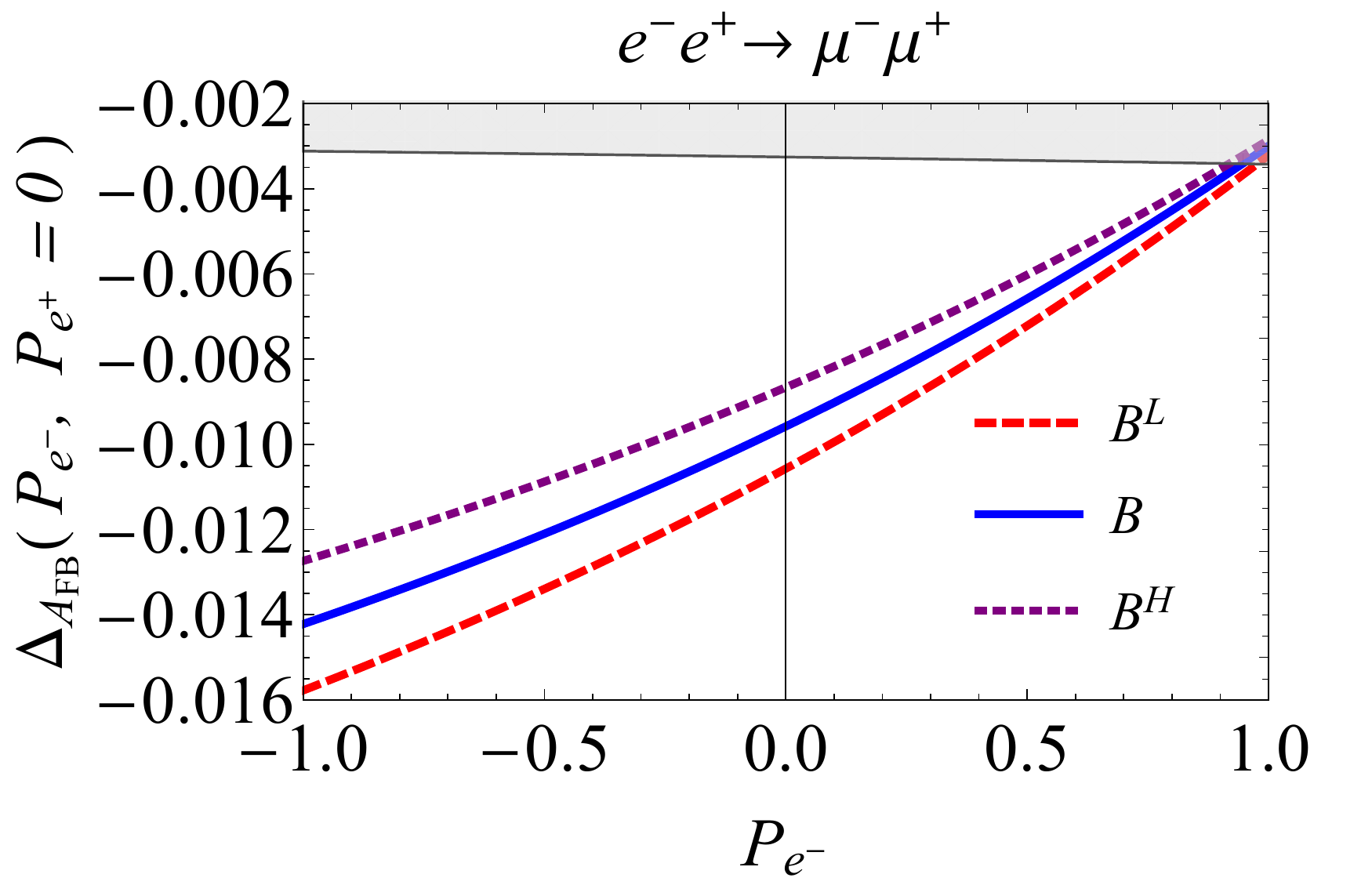}
\includegraphics[bb=0 0 504 320,height=3.3cm]{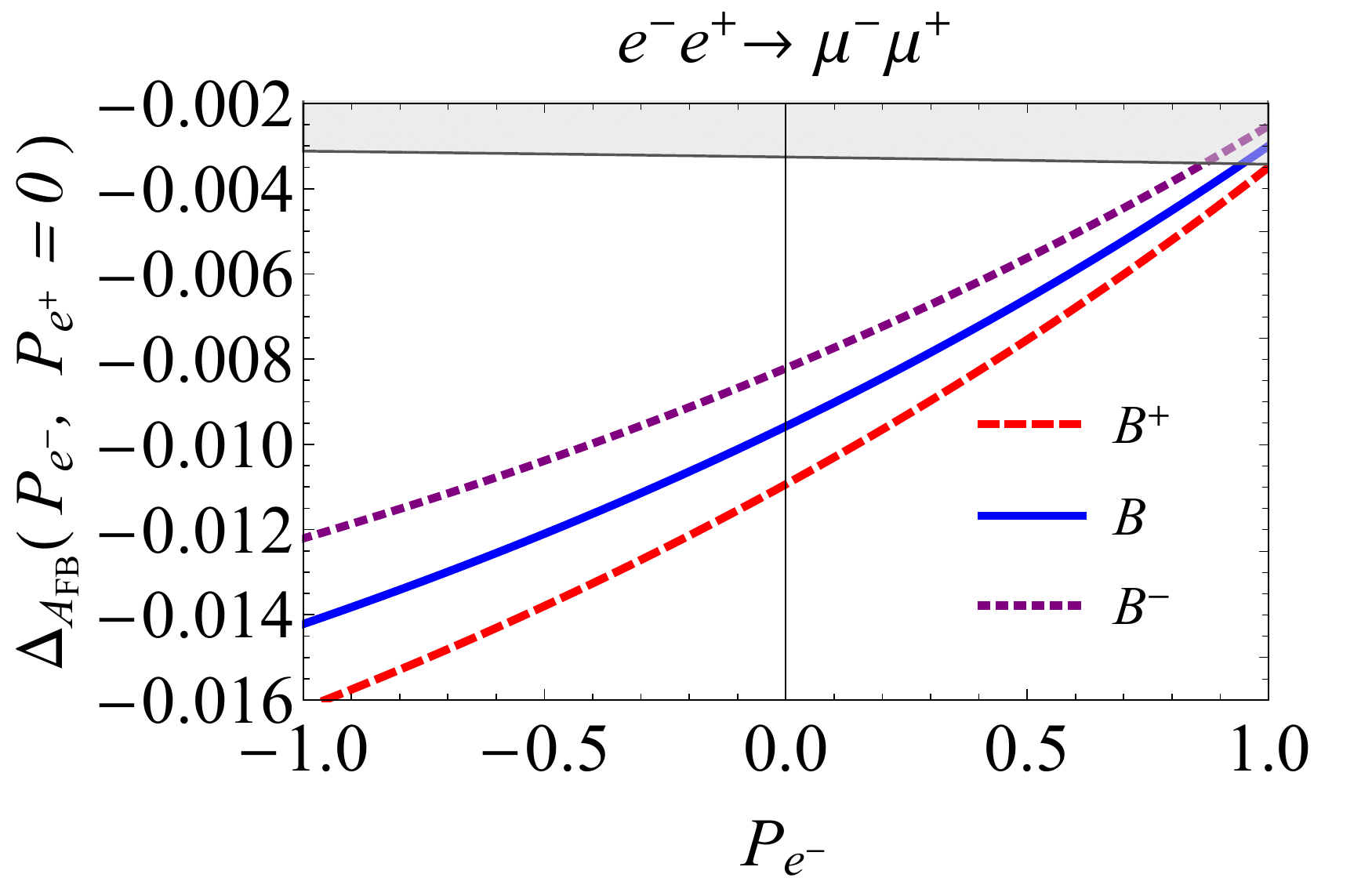}\\[0.5em]
\includegraphics[bb=0 0 504 336,height=3.3cm]{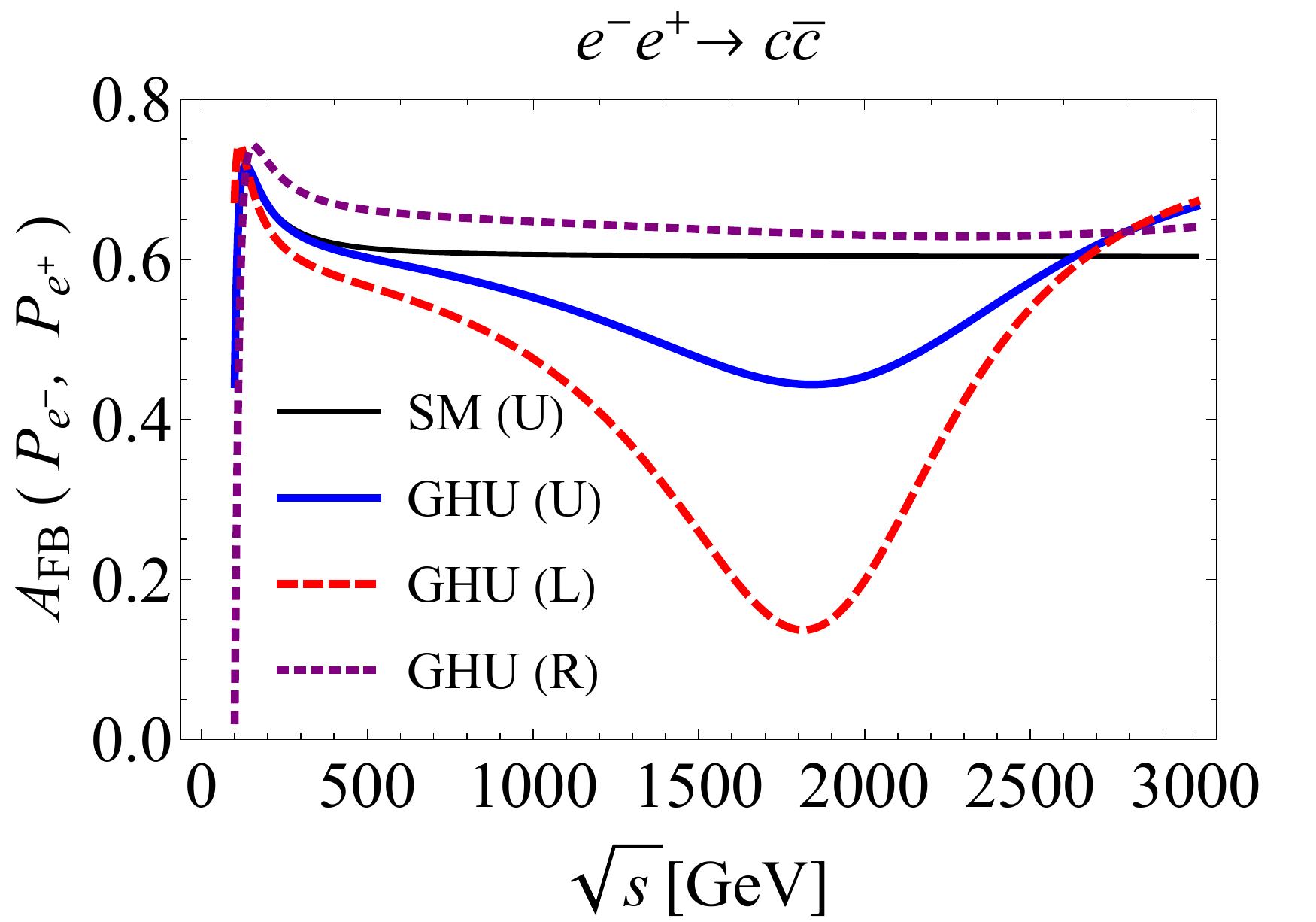}
\includegraphics[bb=0 0 504 320,height=3.3cm]{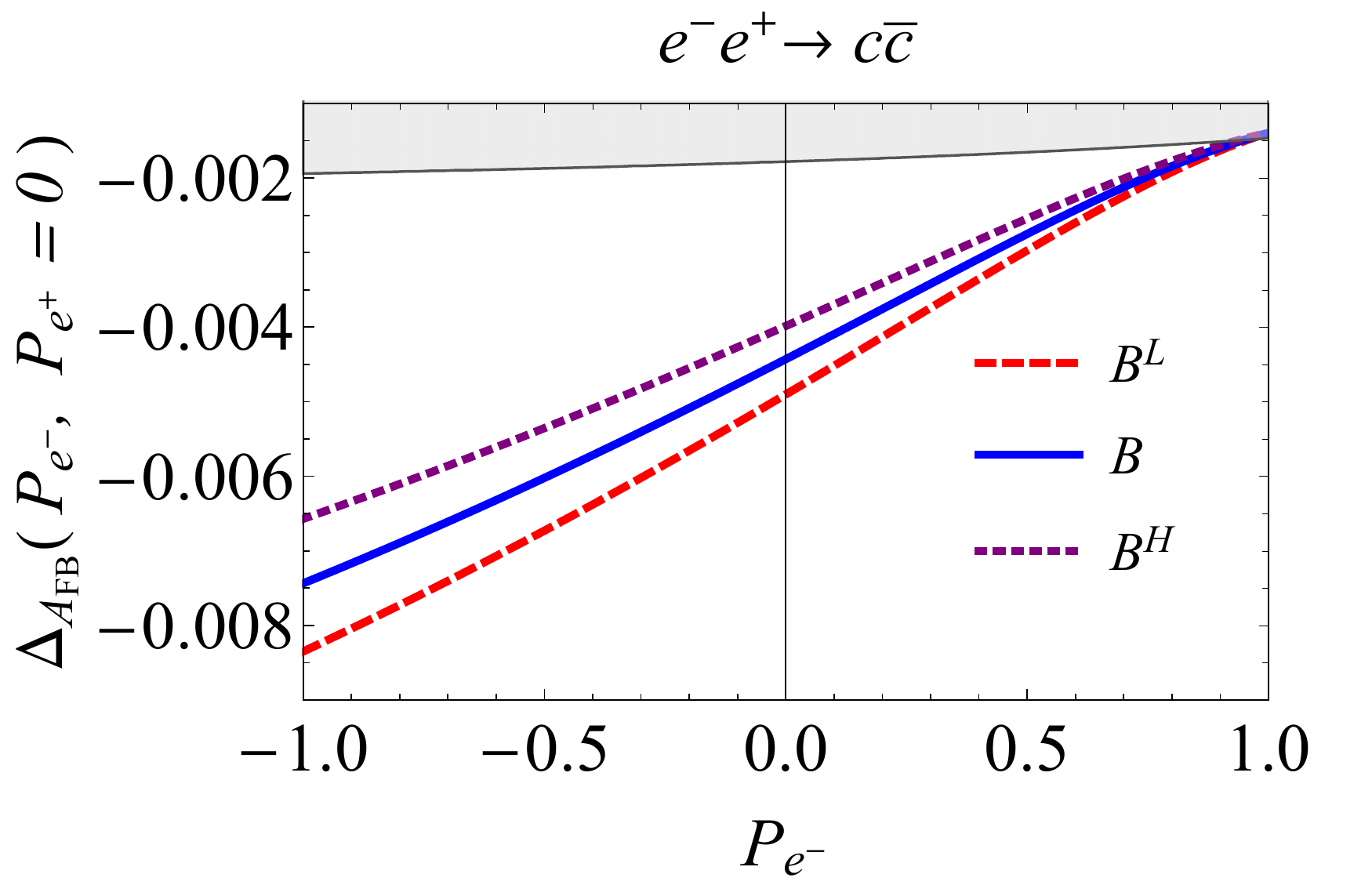}
\includegraphics[bb=0 0 504 320,height=3.3cm]{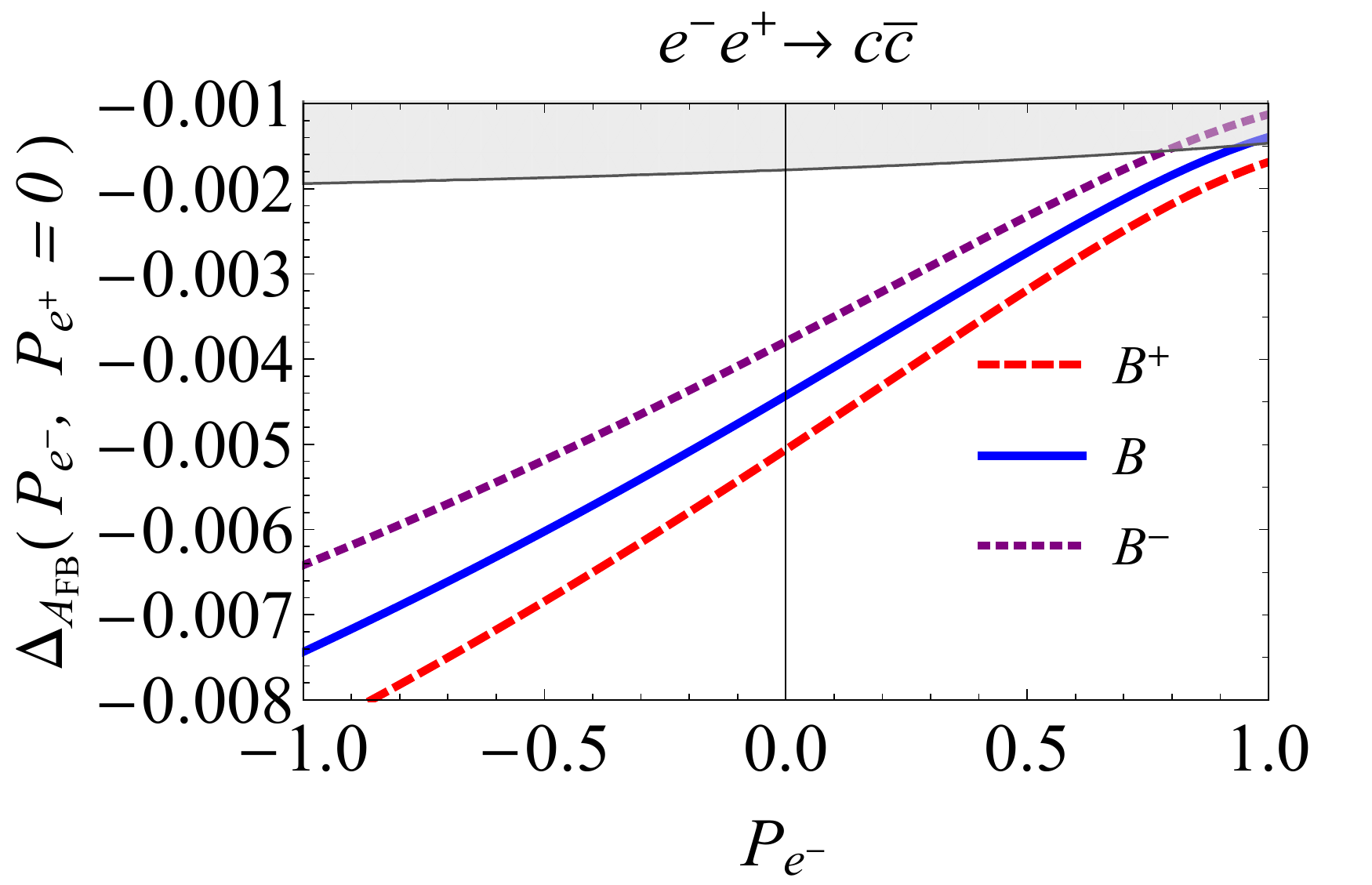}\\[0.5em]
\includegraphics[bb=0 0 504 336,height=3.3cm]{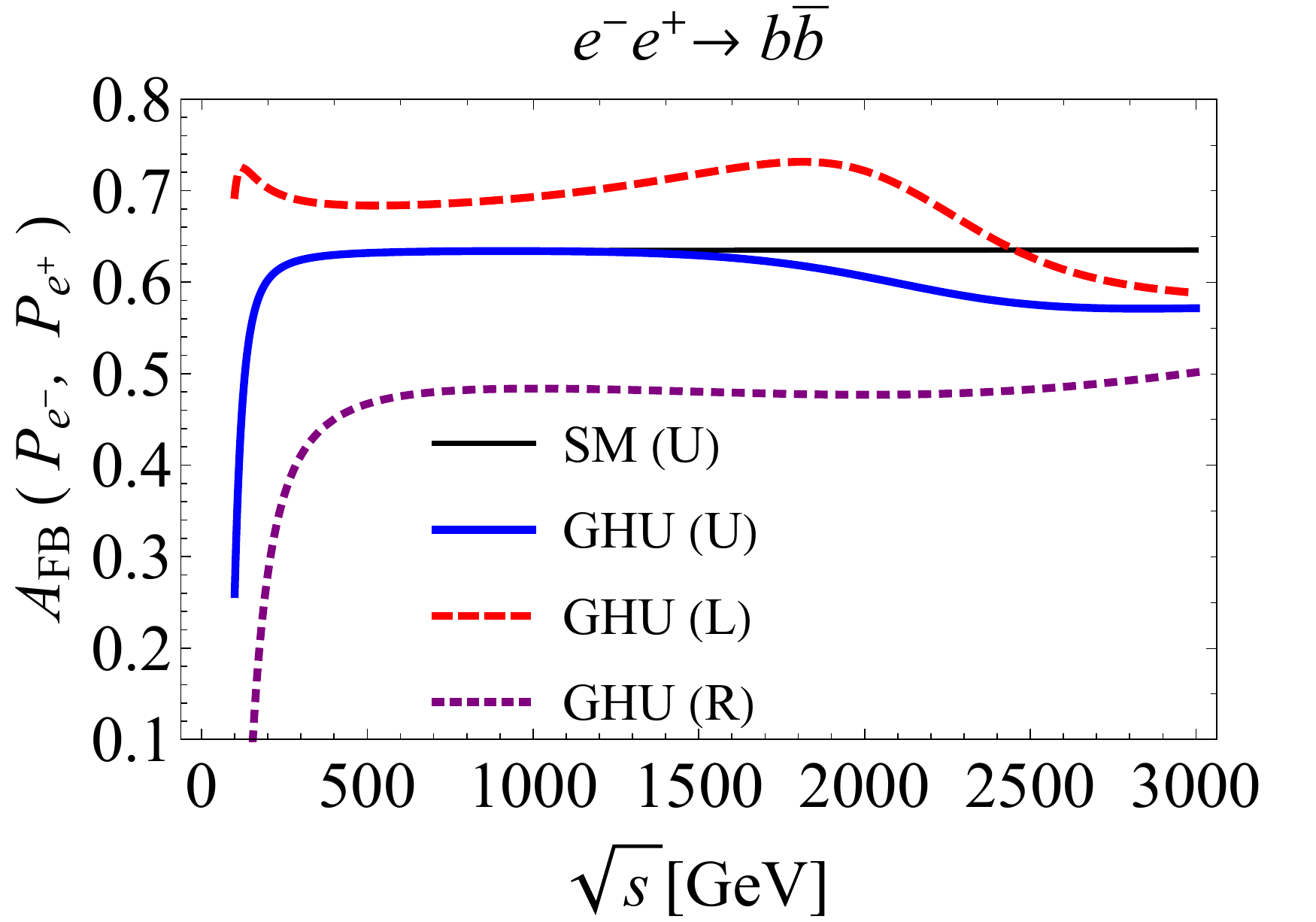}
\includegraphics[bb=0 0 504 320,height=3.3cm]{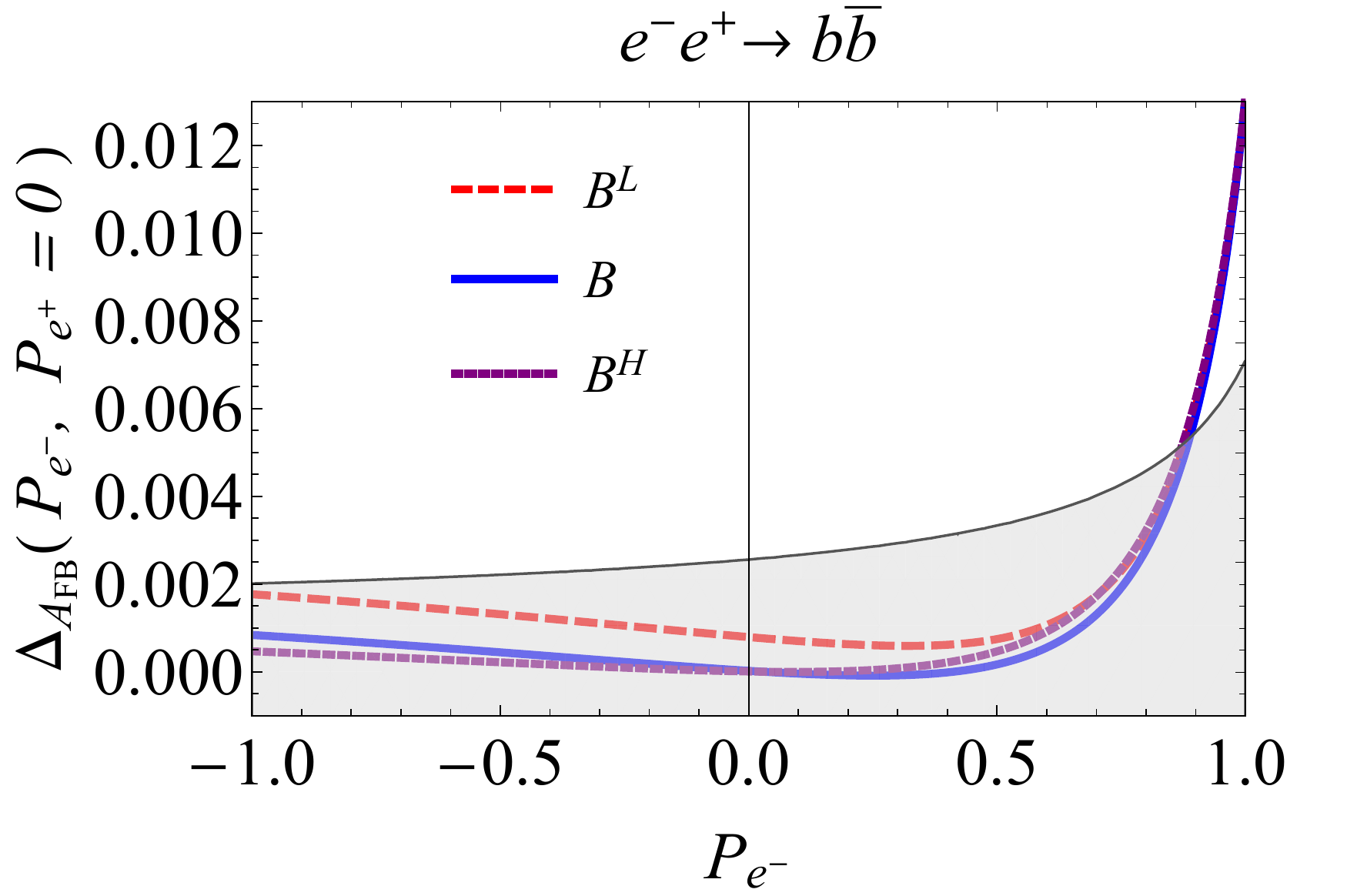}
\includegraphics[bb=0 0 504 320,height=3.3cm]{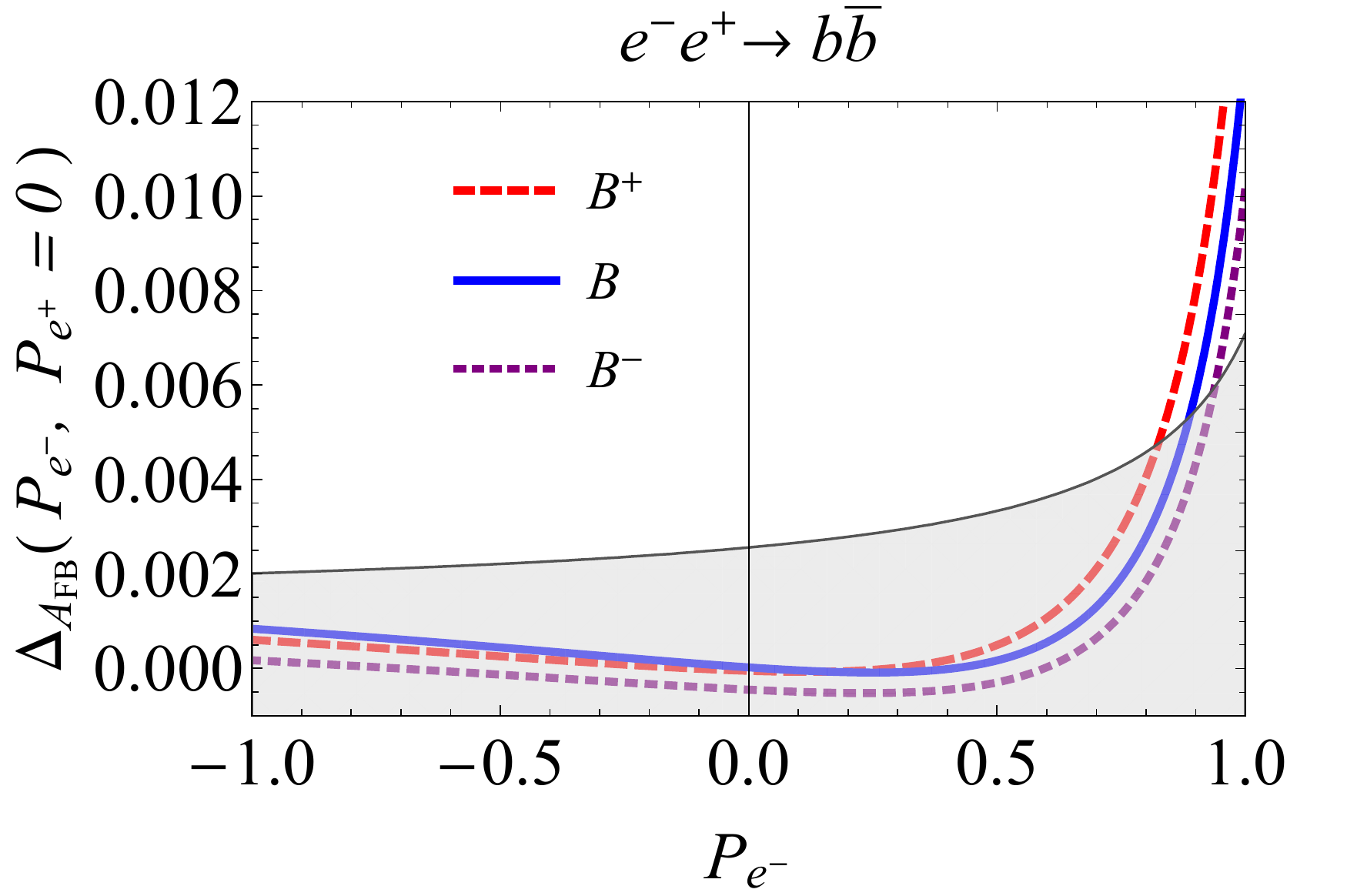}\\[0.5em]
\includegraphics[bb=0 0 504 336,height=3.3cm]{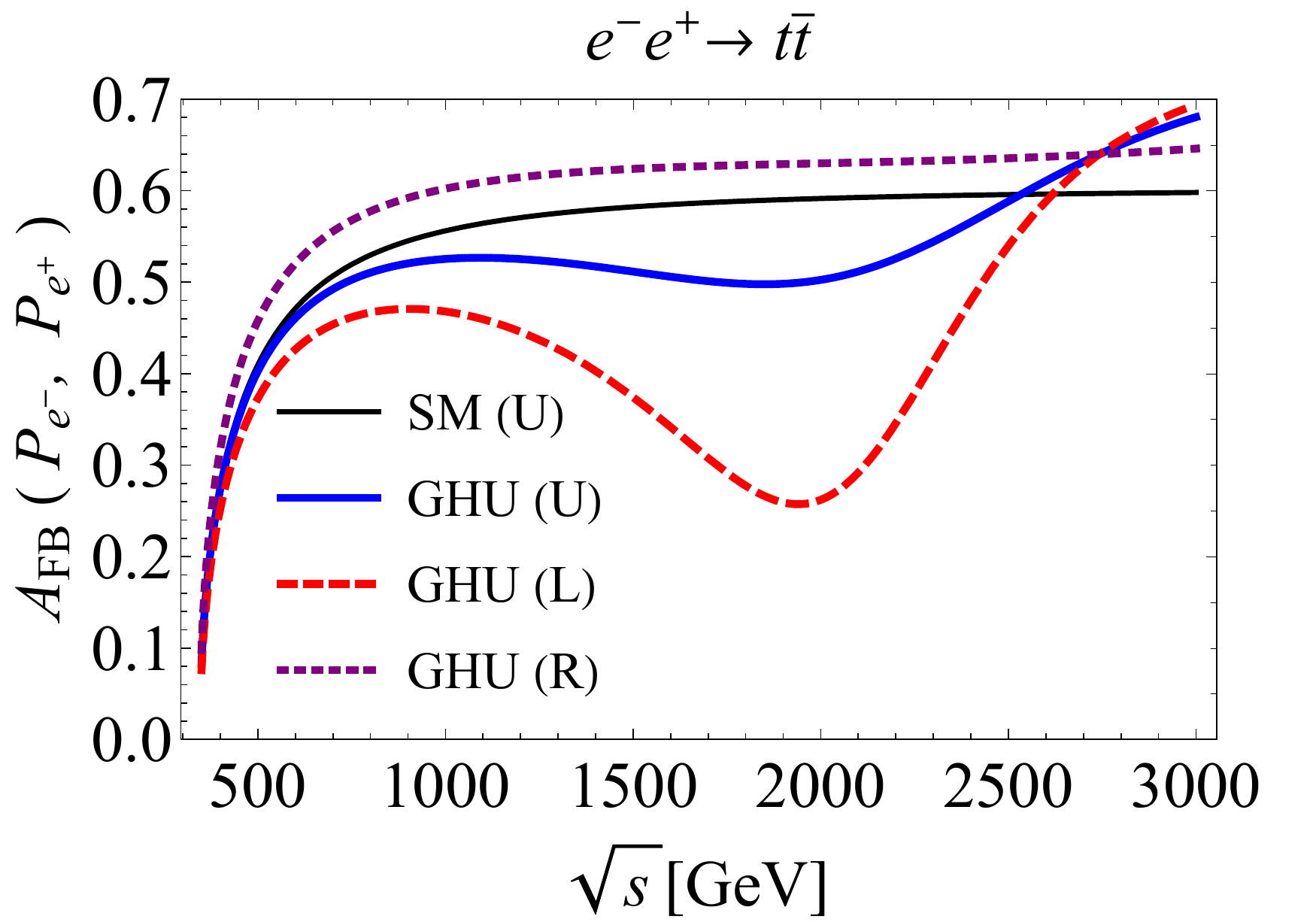}
\includegraphics[bb=0 0 504 320,height=3.3cm]{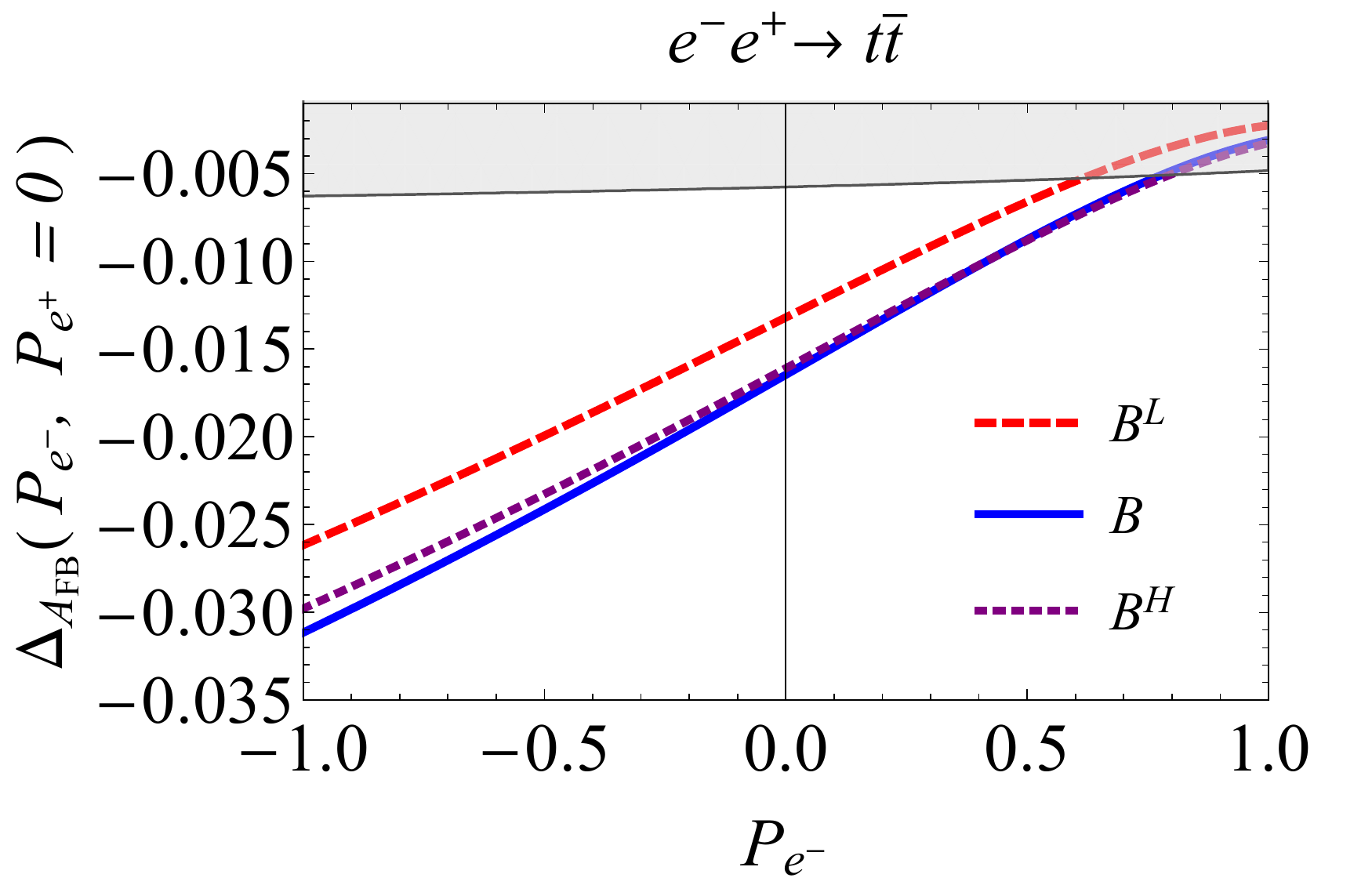}
\includegraphics[bb=0 0 504 320,height=3.3cm]{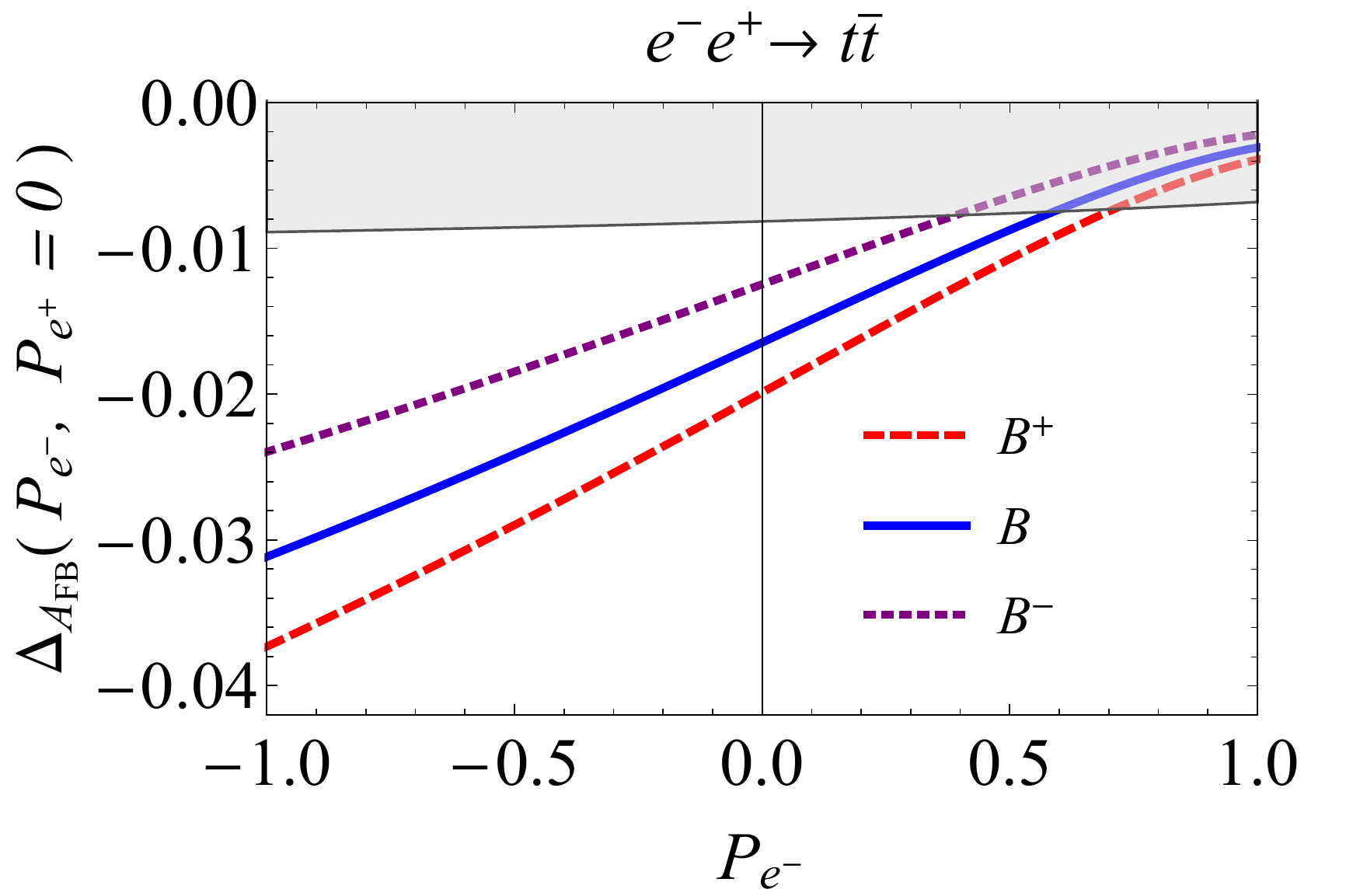}\\
 \caption{\small
 Forward-backward asymmetries $A_{FB}^{f\bar{f}}$
 $(f\bar{f}=\mu^-\mu^+,c\bar{c},b\bar{b},t\bar{t})$ are shown.
 The left side figures show the $\sqrt{s}$ dependence of  $A_{FB}^{f\bar{f}}$ 
 in the SM and the GHU  (B) in  Table~\ref{Table:Mass-Width-Vector-Bosons}.
 Three cases of polarization of electron  and positron beams 
 $(P_{e^-},P_{e^+})=(0,0)(U), (-0.8,+0.3) (L), (+0.8,-0.3)(R)$ are depicted for GHU.
The energy range $\sqrt{s}$ is  $[80,3000] \,$GeV for $f\bar{f}=\mu^-\mu^+,c\bar{c},b\bar{b}$ 
and  $[350,3000] \,$GeV  for $f\bar{f}=t\bar{t}$.
The central figures show the electron polarization $P_{e^-}$
 dependence of  the deviation from the SM
 $\Delta_{A_{FB}}^{f\bar{f}}(P_{e^-},P_{e^+}=0)$ in Eq.~(\ref{Eq:Delta_A_FB})
  for the GHU  (B$^{\rm L}$), (B), (B$^{\rm H}$) in  Table~\ref{Table:Mass-Width-Vector-Bosons}.
 The right side figures show the electron polarization $P_{e^-}$
 dependence of  the deviation
  for  the GHU (B$^+$), (B), (B$^-$) in  Table~\ref{Table:Mass-Width-Vector-Bosons}.
 The gray band in the central and right side figures represent
the statistical error in the SM  at $\sqrt{s}=250 \,$GeV with 250$\,$fb$^{-1}$ data
 for $f\bar{f}=\mu^-\mu^+,c\bar{c},b\bar{b}$  and
 at $\sqrt{s}=500 \,$GeV with 500$\,$fb$^{-1}$ data  for $f\bar{f}=t\bar{t}$.
 } 
 \label{Figure:AFB-ef-theta=010}
\end{center}
\end{figure}

\subsection{Left-right asymmetry}

The integrated left-right asymmetry in $e^-e^+\to f\bar{f}$ 
$(f\bar{f}=\mu^-\mu^+,c\bar{c},b\bar{b},t\bar{t})$,
$A_{LR}^{f\bar{f}}$, is shown in Figure~\ref{Figure:ALR-ef-theta=010}.
The integrated left-right asymmetry
$A_{LR}^{{f\bar{f}}}$ in Eq.~(\ref{Eq:A_LR-mf=0})  is given by 
\begin{align}
 A_{LR}^{{f\bar{f}}}
 &\simeq 
\frac{
[|Q_{e_{L} f_{L}}|^{2} + |Q_{e_{L} f_{R}}|^{2}]
 - [|Q_{e_{R} f_{R}}|^{2} + | Q_{e_{R} f_{L}}|^{2}] 
}{
[|Q_{e_{L} f_{L}}|^{2} + |Q_{e_{L} f_{R}}|^{2}]
 + [|Q_{e_{R} f_{R}}|^{2} + | Q_{e_{R} f_{L}}|^{2}] 
}
\end{align}
for $m_f \ll \sqrt{s}$.  In the center-of-mass energy region of interest
$|Q_{e_Lf_L}| \gg |Q_{e_Lf_R}|$ and $|Q_{e_Rf_R}| \gg |Q_{e_Rf_L}|$ are
satisfied so that 
\begin{align}
 A_{LR}^{{f\bar{f}}}
 &\simeq 
\frac{|Q_{e_{L} f_{L}}|^{2} - |Q_{e_{R} f_{R}}|^{2}}
 {|Q_{e_{L} f_{L}}|^{2}+ |Q_{e_{R} f_{R}}|^{2}} ~.
\end{align}
In the GHU (B) in Table~\ref{Table:Mass-Width-Vector-Bosons}, due to the
interference effects between $Z$ and $Z'$ bosons,
$|Q_{e_L\mu_L}|$ becomes smaller than $|Q_{e_R\mu_R}|$ {in the region around 
$\sqrt{s} = 1 \sim 2\,$TeV 
as shown in Figure~\ref{Figure:sQ_ef-s_dependence}.   
Consequently $A_{LR}^{f\bar{f}}$ can be negative even for $\sqrt{s}\gg m_Z$.

\begin{figure}[thb]
\begin{center}
\includegraphics[bb=0 0 504 332,height=5cm]{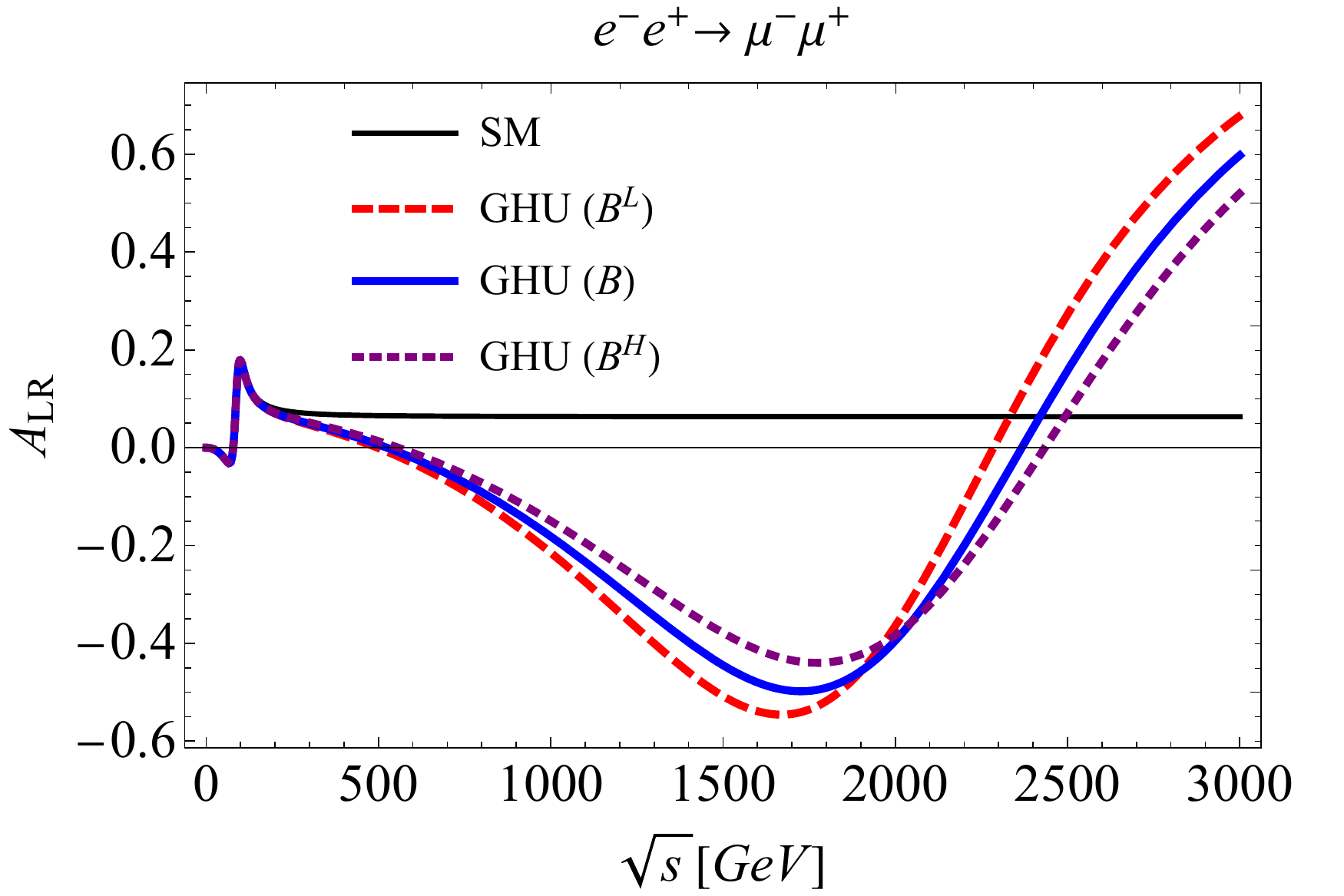}
\includegraphics[bb=0 0 504 332,height=5cm]{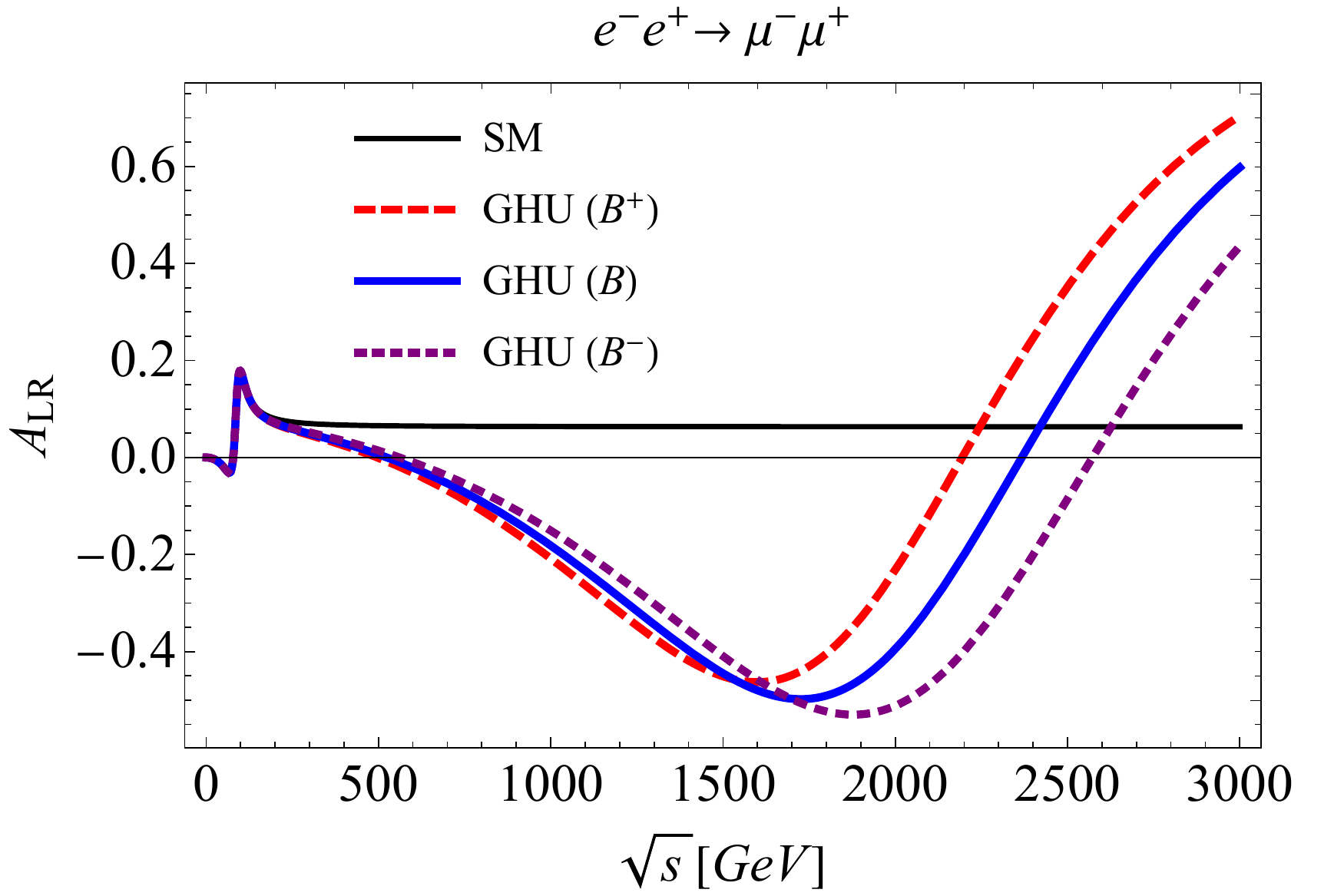}\\[0.5em]
\includegraphics[bb=0 0 504 332,height=5cm]{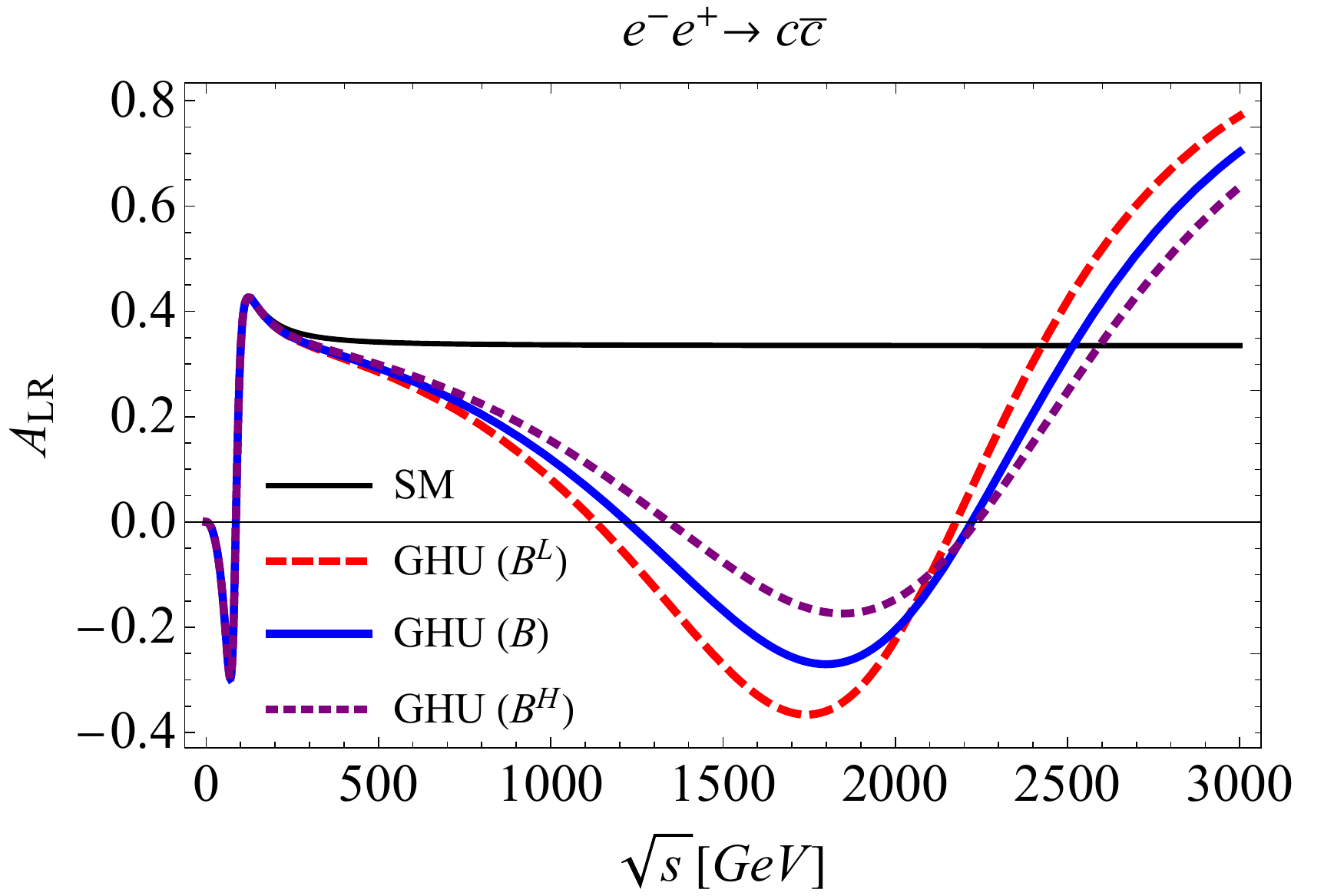}
\includegraphics[bb=0 0 504 332,height=5cm]{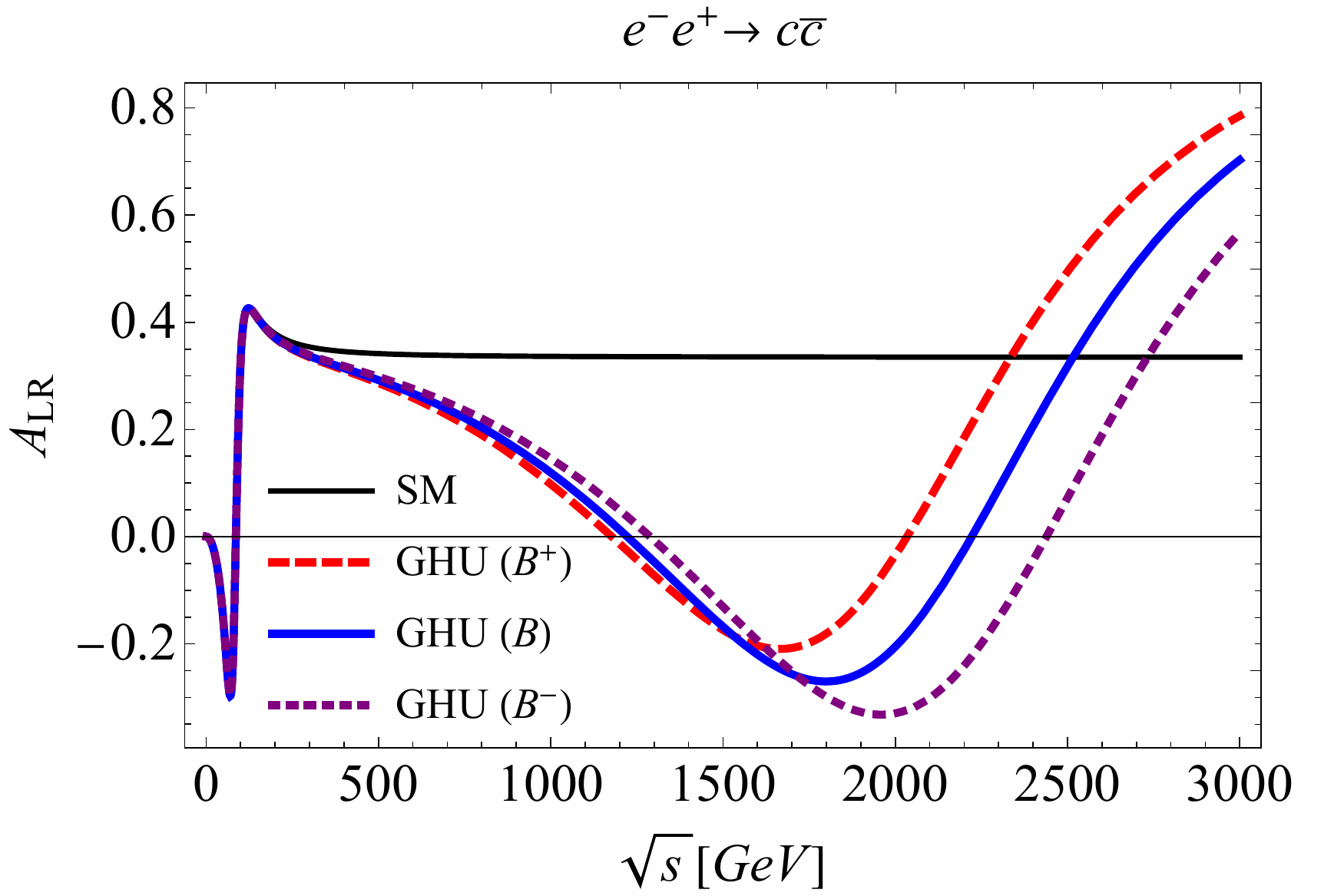}\\[0.5em]
\includegraphics[bb=0 0 504 332,height=5cm]{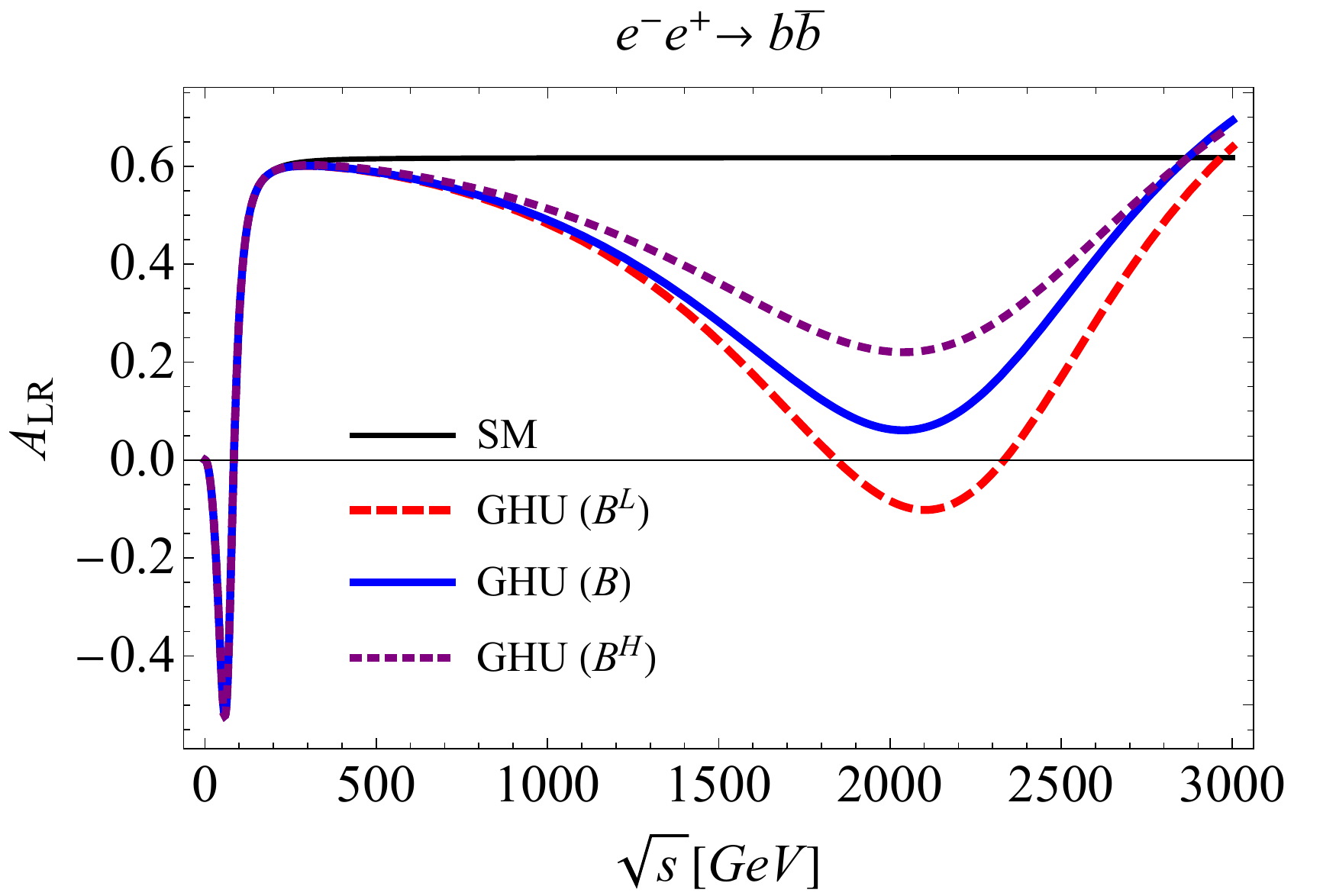}
\includegraphics[bb=0 0 504 332,height=5cm]{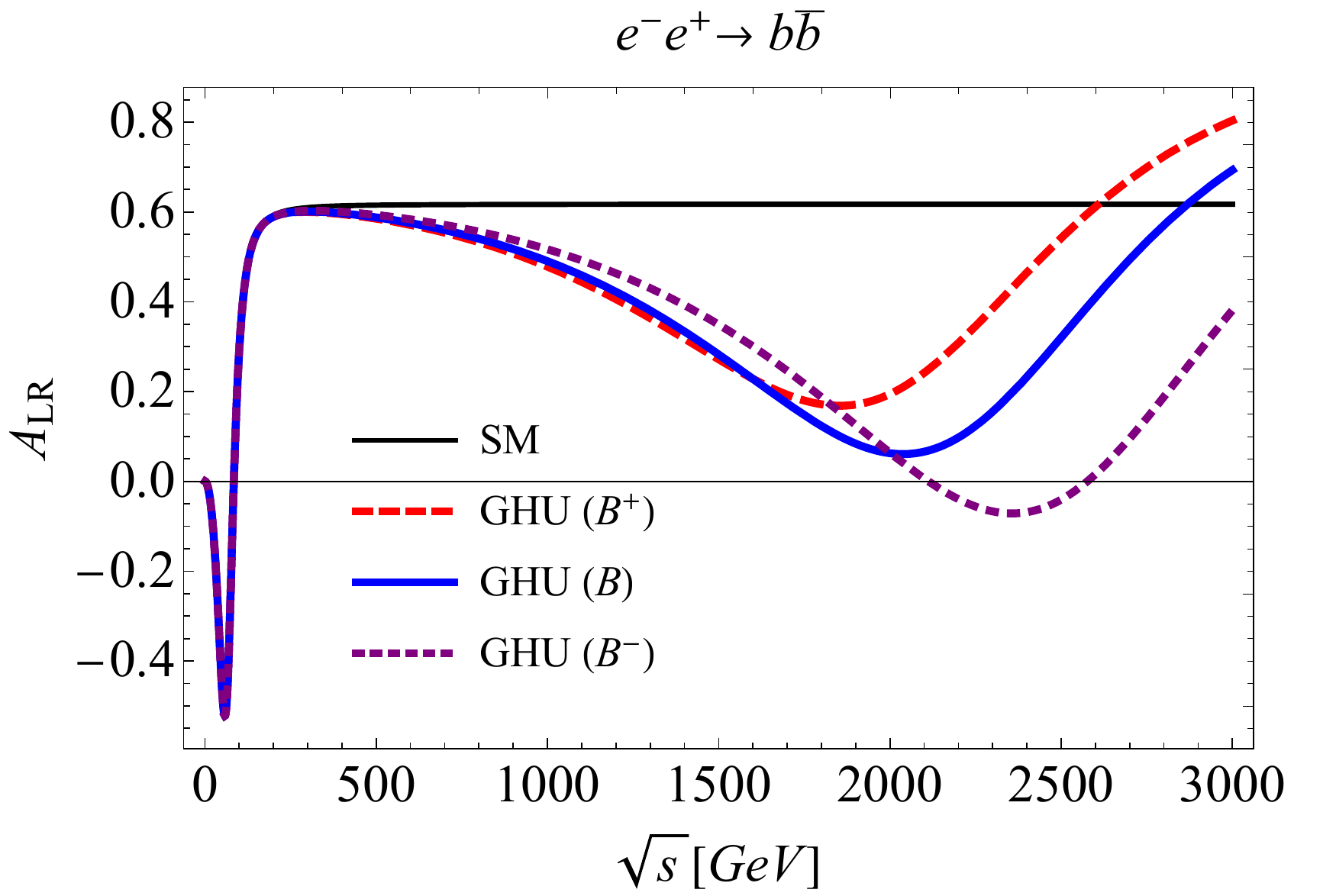}\\[0.5em]
\includegraphics[bb=0 0 504 332,height=5cm]{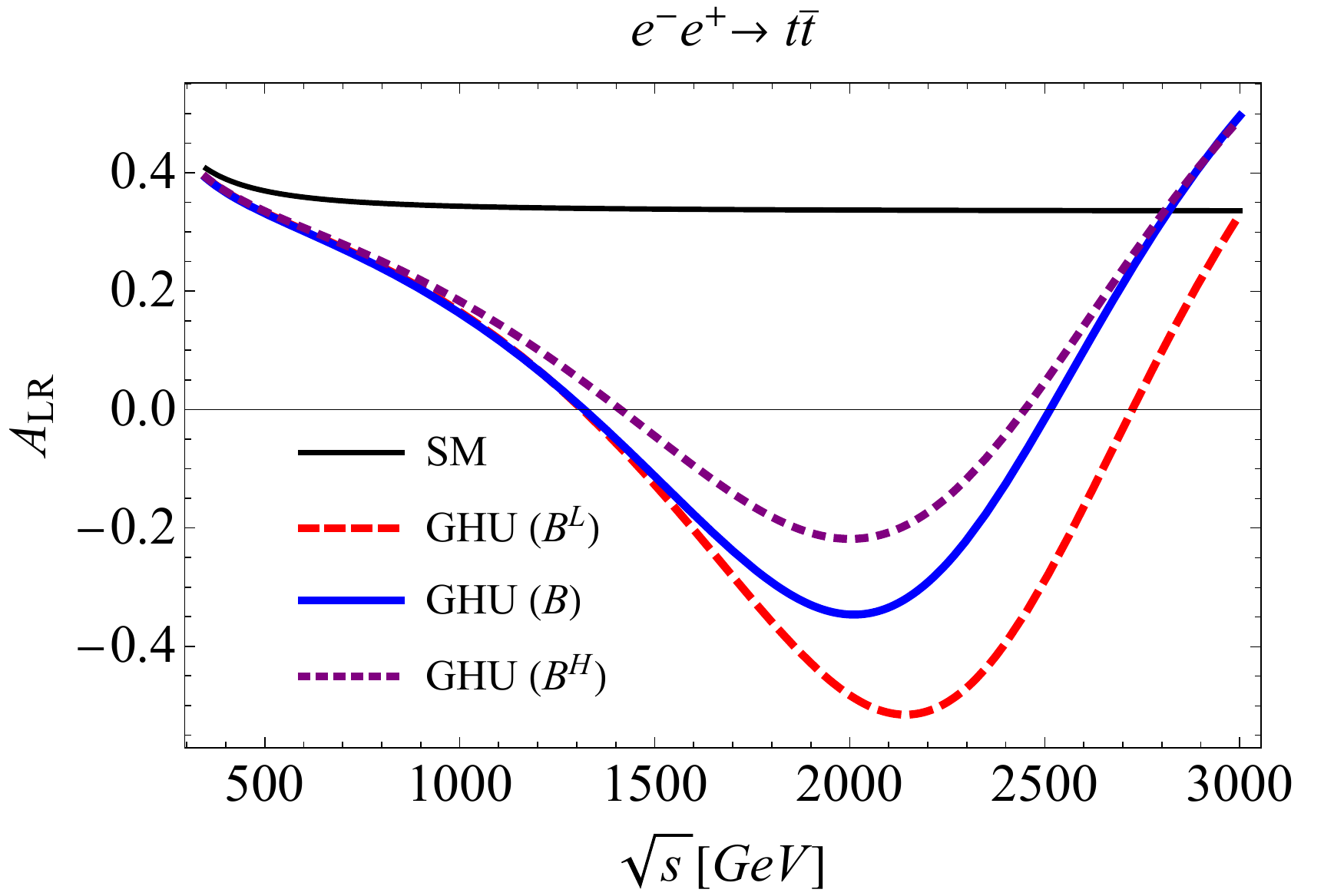}
\includegraphics[bb=0 0 504 332,height=5cm]{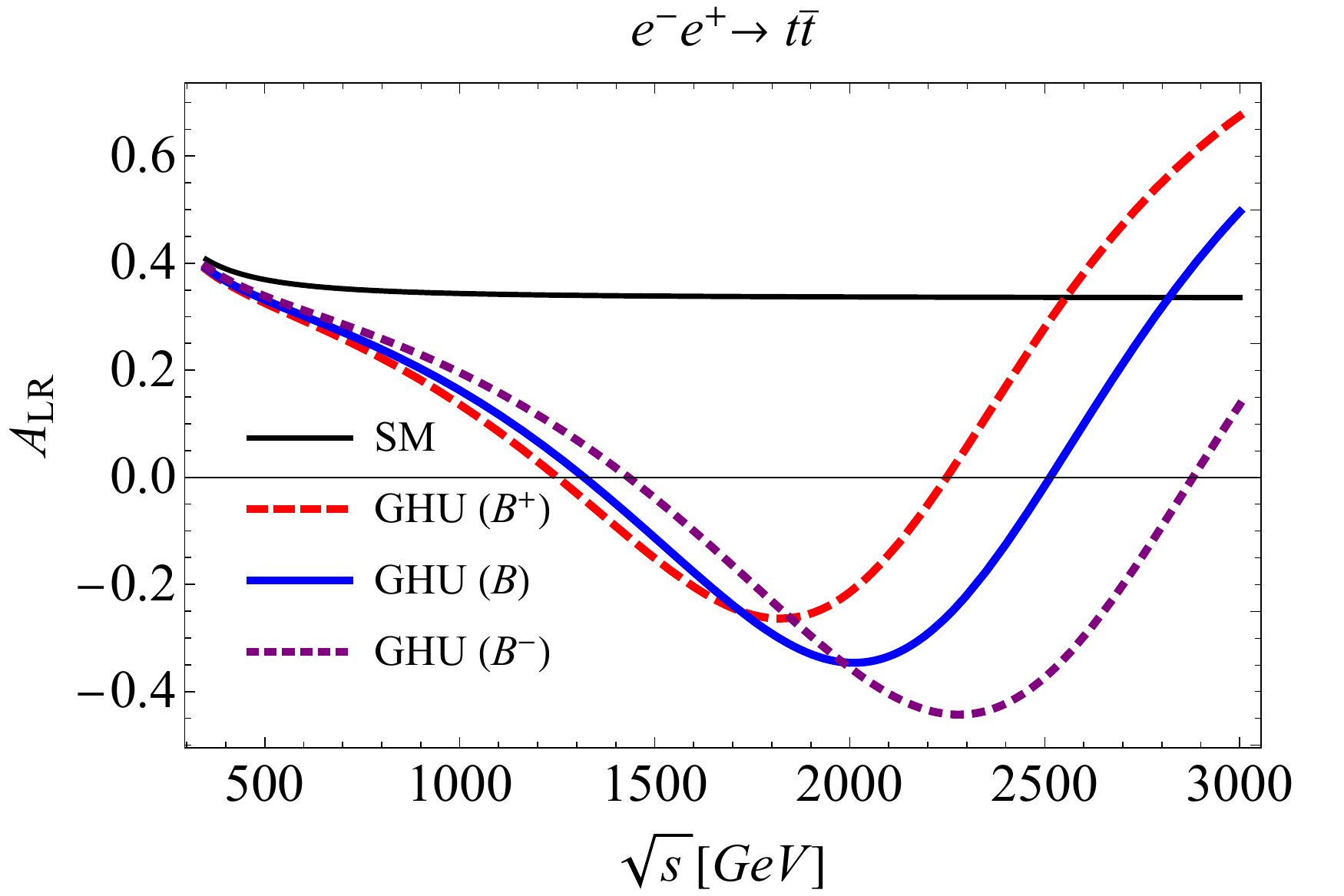}\\
 \caption{\small
 Left-right asymmetries $A_{LR}^{f\bar{f}}$
 $(f\bar{f}=\mu^-\mu^+,c\bar{c},b\bar{b},t\bar{t})$ are shown.
 The left and right side figures show the $\sqrt{s}$ dependence of the
 left-right asymmetry $A_{LR}^{f\bar{f}}$
 for the SM and the GHU
 (B$^{\rm L}$), (B), (B$^{\rm H}$)
 and for the SM and the GHU
 (B$^{+}$), (B), (B$^{-}$)
 in Table~\ref{Table:Mass-Width-Vector-Bosons}.
 The energy ranges $\sqrt{s}$ in the above figures are
 $[80,3000] \,$GeV and $[350,3000] \,$GeV
 for $f\bar{f}=\mu^-\mu^+,c\bar{c},b\bar{b}$ and 
 for $f\bar{f}=t\bar{t}$, respectively.
 } 
 \label{Figure:ALR-ef-theta=010}
\end{center}
\end{figure}

The differential left-right asymmetry of $e^-e^+\to f\bar{f}$
$(f\bar{f}=\mu^-\mu^+,c\bar{c},b\bar{b},t\bar{t})$,
$A_{LR}^{f\bar{f}}(\cos\theta)$, is given by Eq.~(\ref{Eq:A_LR-cos-mf=0}), 
and is displayed  in Figure~\ref{Figure:ALR-cos-ef-theta=010}.
In most of  center-of-mass energy region of interest,   relations
$|Q_{e_Lf_L}|\gg |Q_{e_Lf_R}|$ and $|Q_{e_Rf_R}|\gg |Q_{e_Rf_L}|$ are
satisfied so that in the forward region $\cos \theta>0$, 
the differential left-right asymmetry is approximately 
\begin{align}
A_{LR}^{f\bar{f}}(\cos\theta)
 &\simeq \frac{|Q_{e_{L} f_{L}}|^{2}-|Q_{e_{R} f_{R}}|^{2}}
{|Q_{e_{L} f_{L}}|^{2}+|Q_{e_{R} f_{R}}|^{2}}.
\end{align}

\begin{figure}[thb]
\begin{center}
\includegraphics[bb=0 0 504 332,height=5cm]{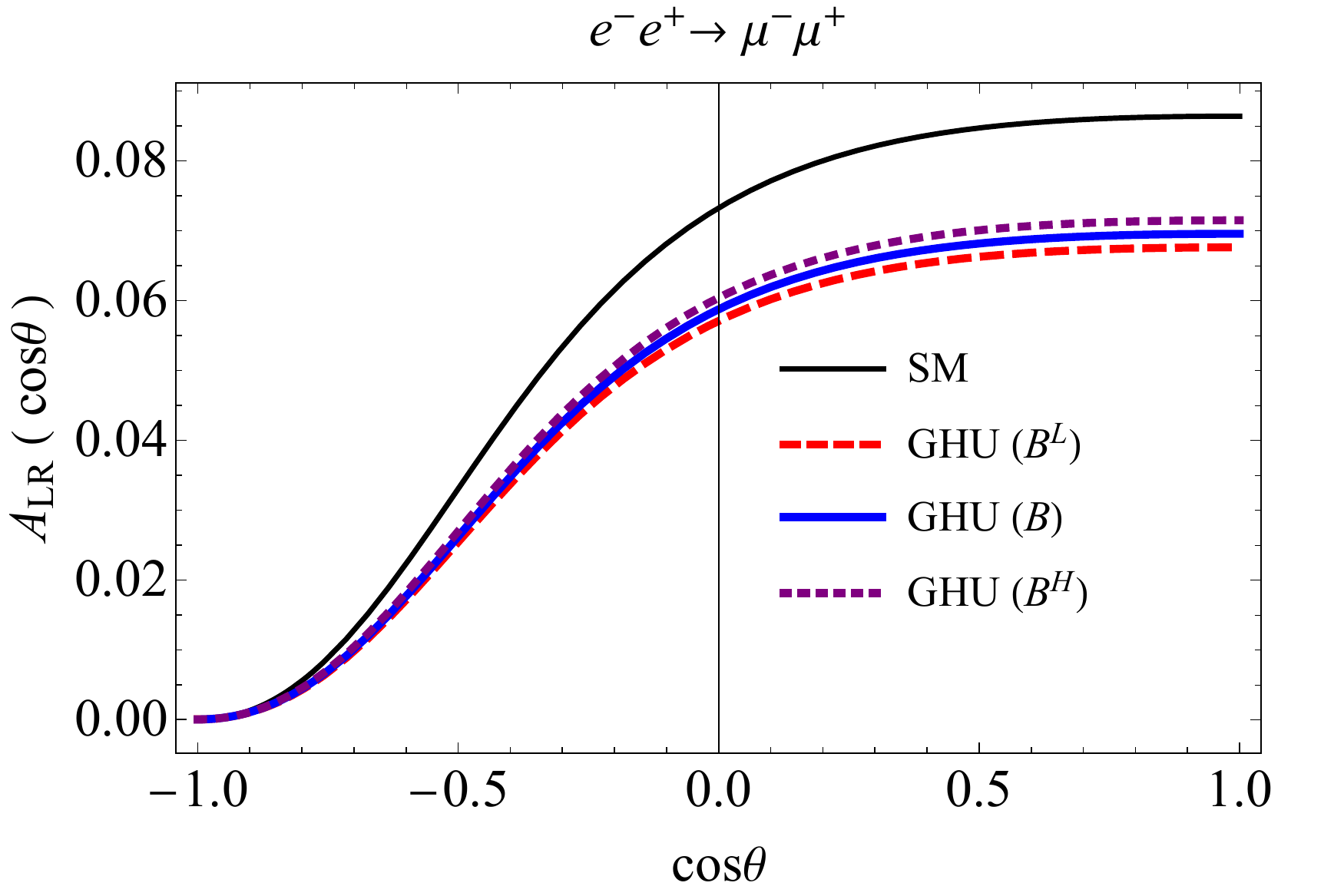}
\includegraphics[bb=0 0 504 332,height=5cm]{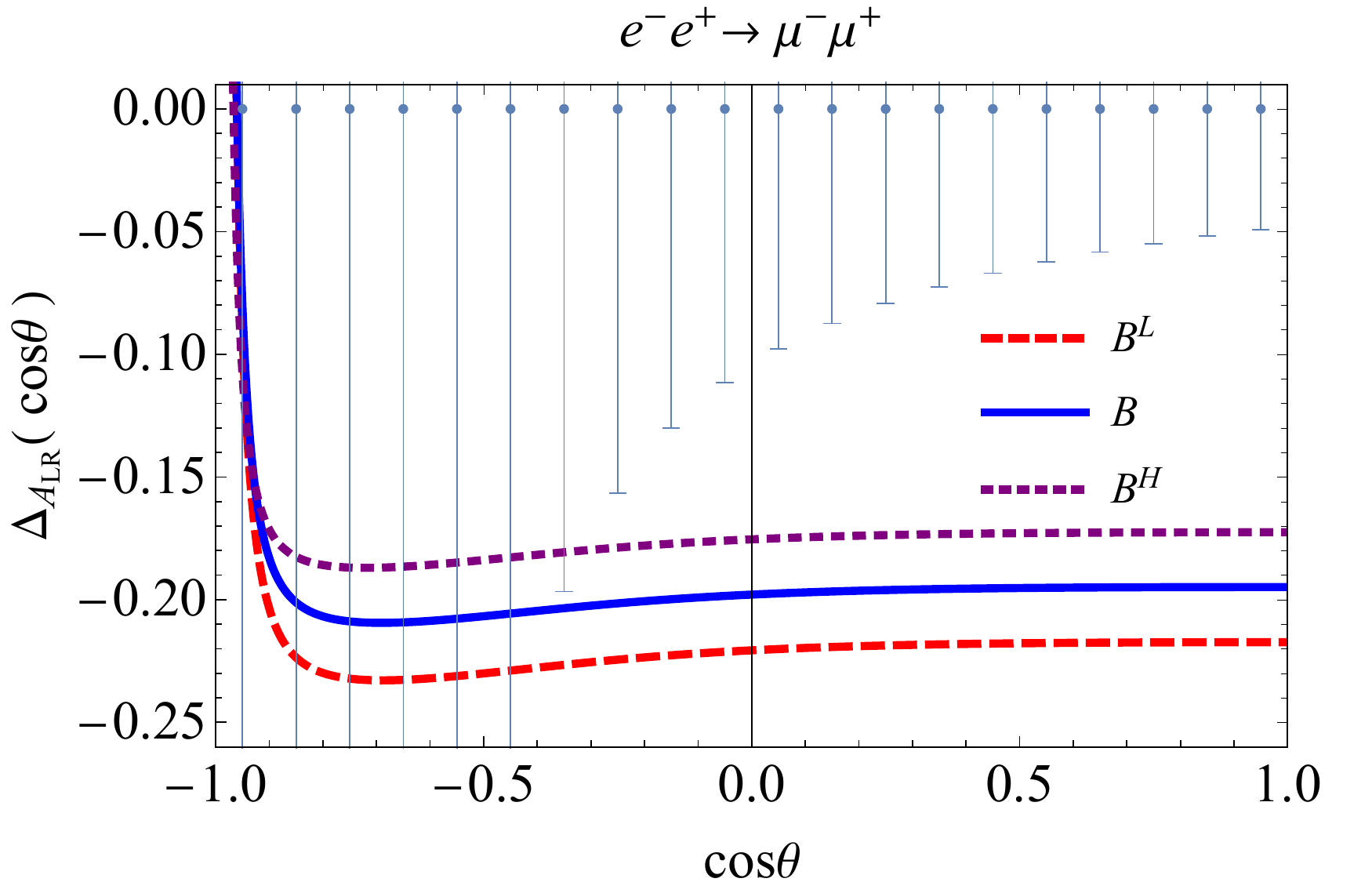}\\[1em]
\includegraphics[bb=0 0 504 332,height=5cm]{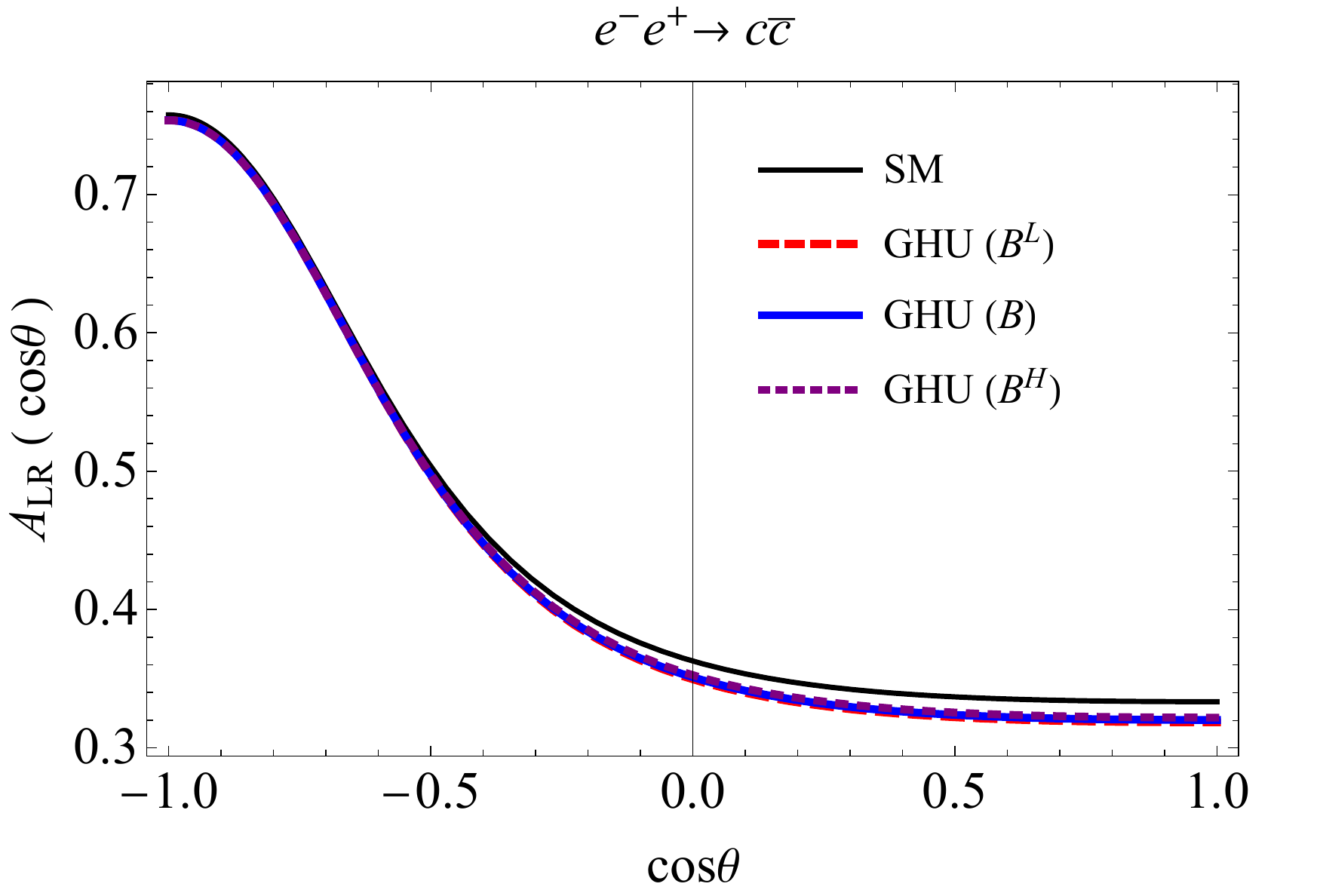}
\includegraphics[bb=0 0 504 332,height=5cm]{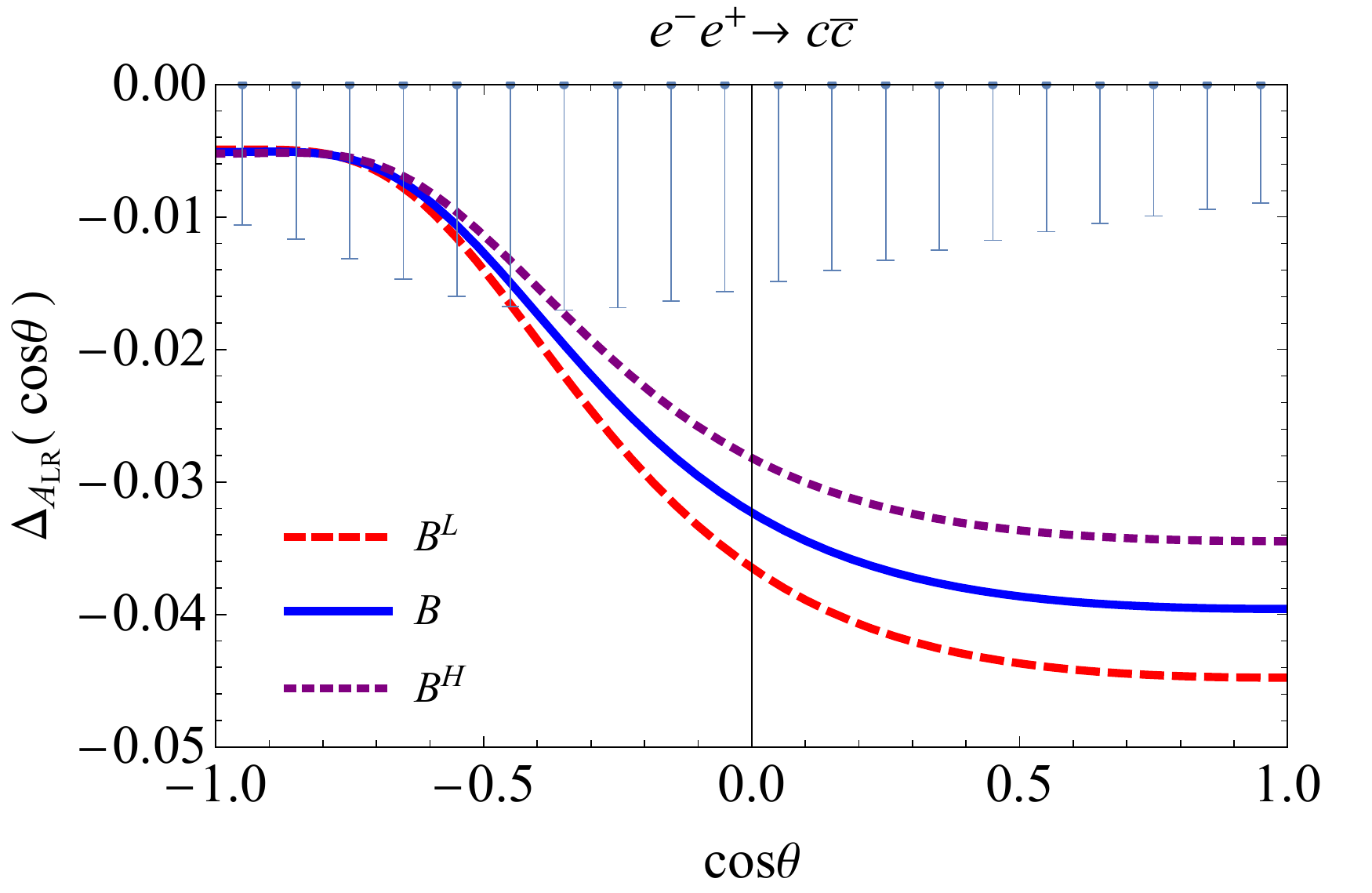}\\[1em]
\includegraphics[bb=0 0 504 332,height=5cm]{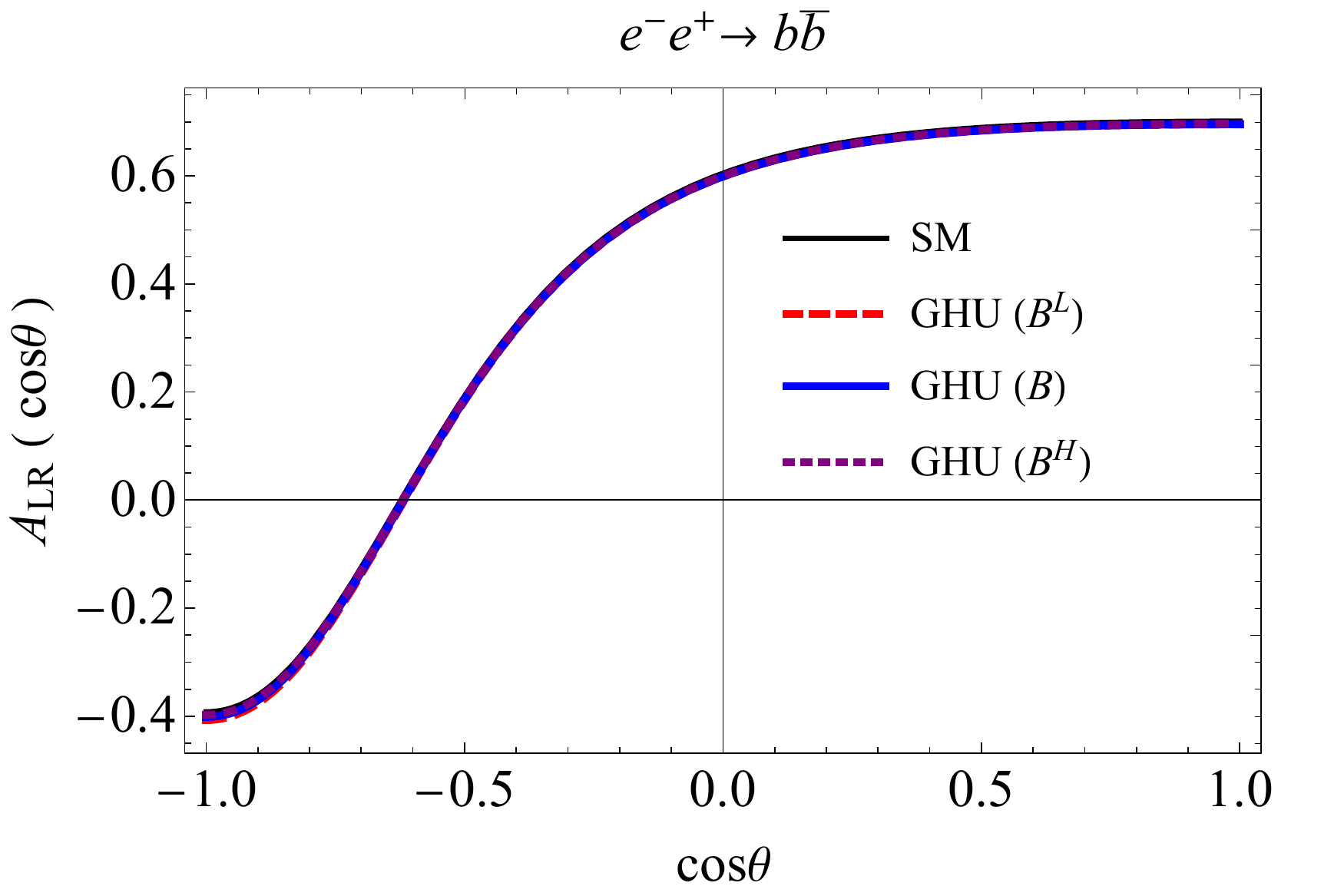}
\includegraphics[bb=0 0 504 332,height=5cm]{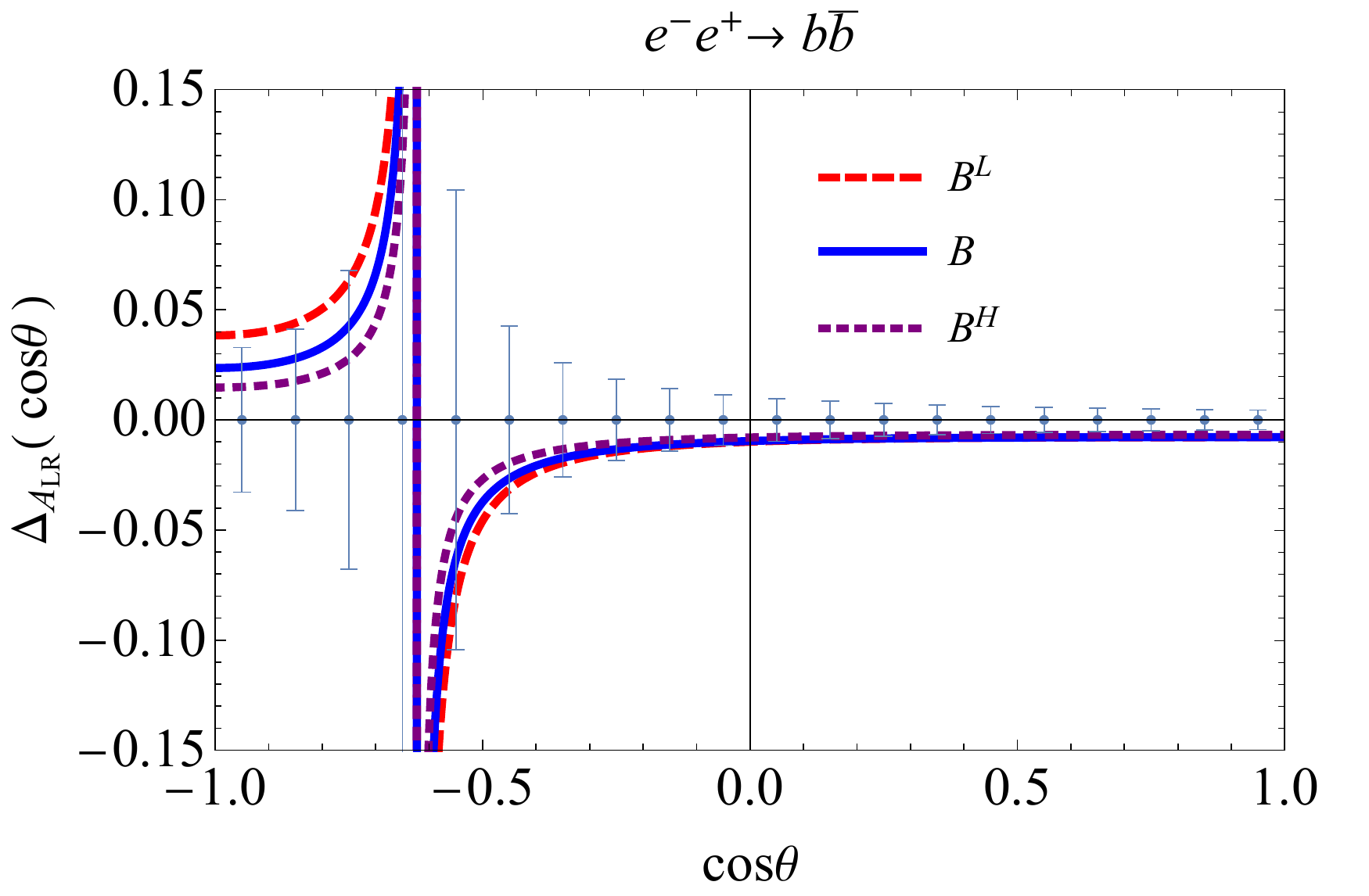}\\[1em]
\includegraphics[bb=0 0 504 332,height=5cm]{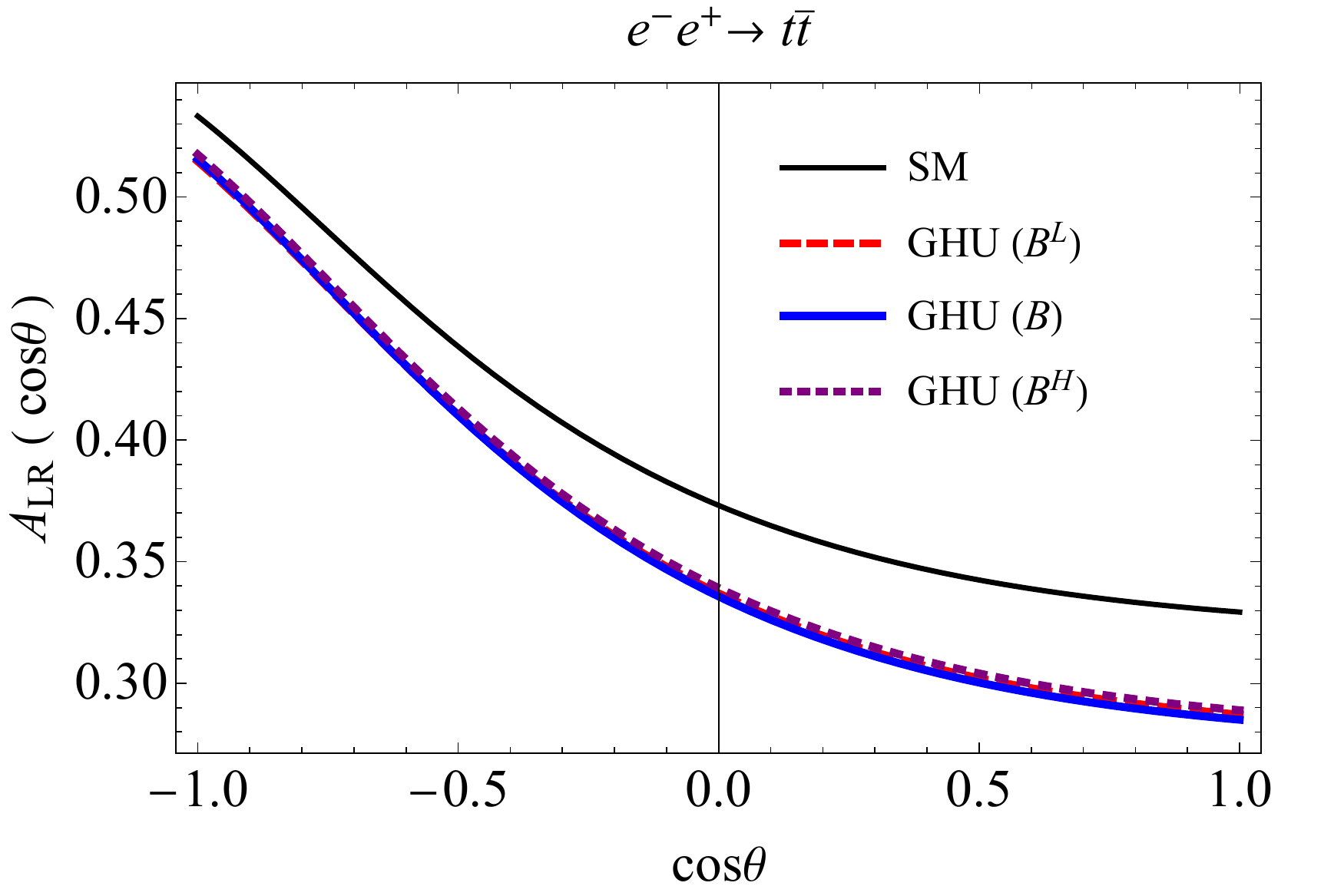}
\includegraphics[bb=0 0 504 332,height=5cm]{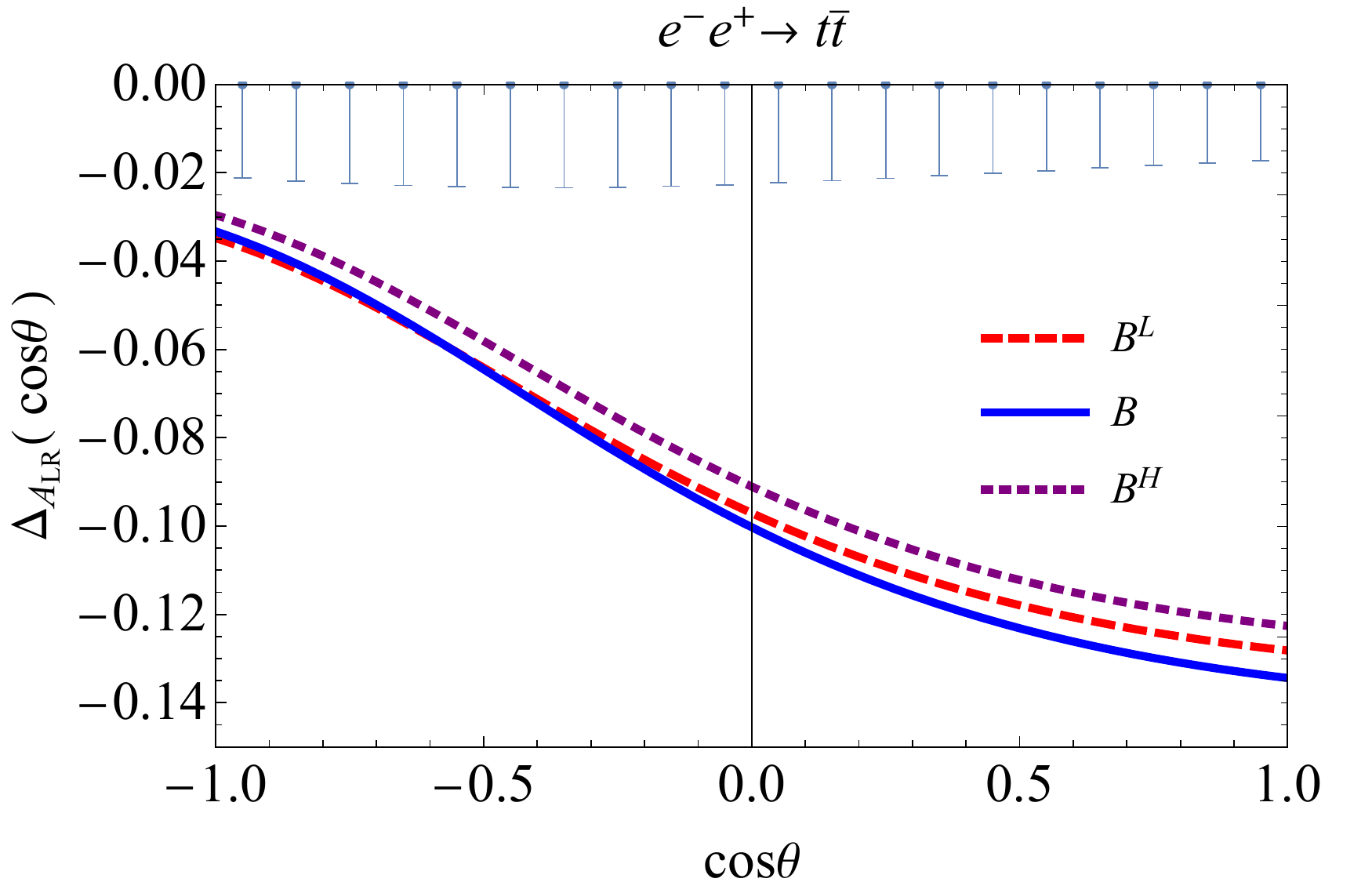}\\
 \caption{\small
Differential left-right asymmetries
 $A_{LR}^{f\bar{f}}(\cos\theta)$
 $(f\bar{f}=\mu^-\mu^+,c\bar{c},b\bar{b},t\bar{t})$ are shown.
 The left side figures show the $\theta$ dependence of  $A_{LR}^{f\bar{f}}(\cos\theta)$
 for the SM and the GHU
 (B$^{\rm L}$), (B), (B$^{\rm H}$) in  Table~\ref{Table:Mass-Width-Vector-Bosons}.
 The right side figures show the $\theta$ dependence of
the deviation of the differential left-right asymmetry from the SM, 
 $\Delta_{A_{LR}}^{f\bar{f}}(\cos\theta)$
 in Eq.~(\ref{Eq:Delta_A_LR})  for  the GHU  (B$^{\rm L}$), (B), (B$^{\rm H}$).
 The error bars in the right side figures represent
 the statistical error in Eq.~(\ref{Eq:Error-A_LR}) 
 at $\sqrt{s}=250 \,$GeV with 250$\,$fb$^{-1}$ data and
 $(P_{e^-},P_{e^+})=(-0.8,+0.3), (+0.8,-0.3)$
 for $f\bar{f}=\mu^-\mu^+,c\bar{c},b\bar{b}$ and
 at $\sqrt{s}=500 \,$GeV with 500$\,$fb$^{-1}$ data  for $f\bar{f}=t\bar{t}$.
 } 
 \label{Figure:ALR-cos-ef-theta=010}
\end{center}
\end{figure}

\subsection{Left-right forward-backward asymmetry}

The left-right forward-backward asymmetry $A_{LR,FB}^{f\bar{f}}(\cos\theta)$ is given by
in Eq.~(\ref{Eq:A_LRFB-mf=0}).  It is shown in Figure~\ref{Figure:ALRFB-ef-theta=010}.
For $|Q_{e_Lf_L}|\gg |Q_{e_Lf_R}|$, $|Q_{e_Rf_R}|\gg|Q_{e_Rf_L}|$ and
$m_f\ll \sqrt{s}$,
the left-right forward-backward asymmetry can be written in terms of  the integrated left-right asymmetry
$A_{LR}^{f\bar{f}}$ by
\begin{align}
A_{LR,FB}^{f\bar{f}}(\cos\theta)&\simeq 
 \frac{2 \cos\theta}{1+\cos^2\theta}
\frac{|Q_{e_{L}f_{L}}|^{2}-|Q_{e_{R}f_{R}}|^{2}}
 {|Q_{e_{L}f_{L}}|^{2}+|Q_{e_{R}f_{R}}|^{2}}
\simeq \frac{2\cos\theta}{1+\cos^2\theta}
A_{LR}^{f\bar{f}} ~.
\end{align}

\begin{figure}[thb]
\begin{center}
\includegraphics[bb=0 0 504 327,height=5cm]{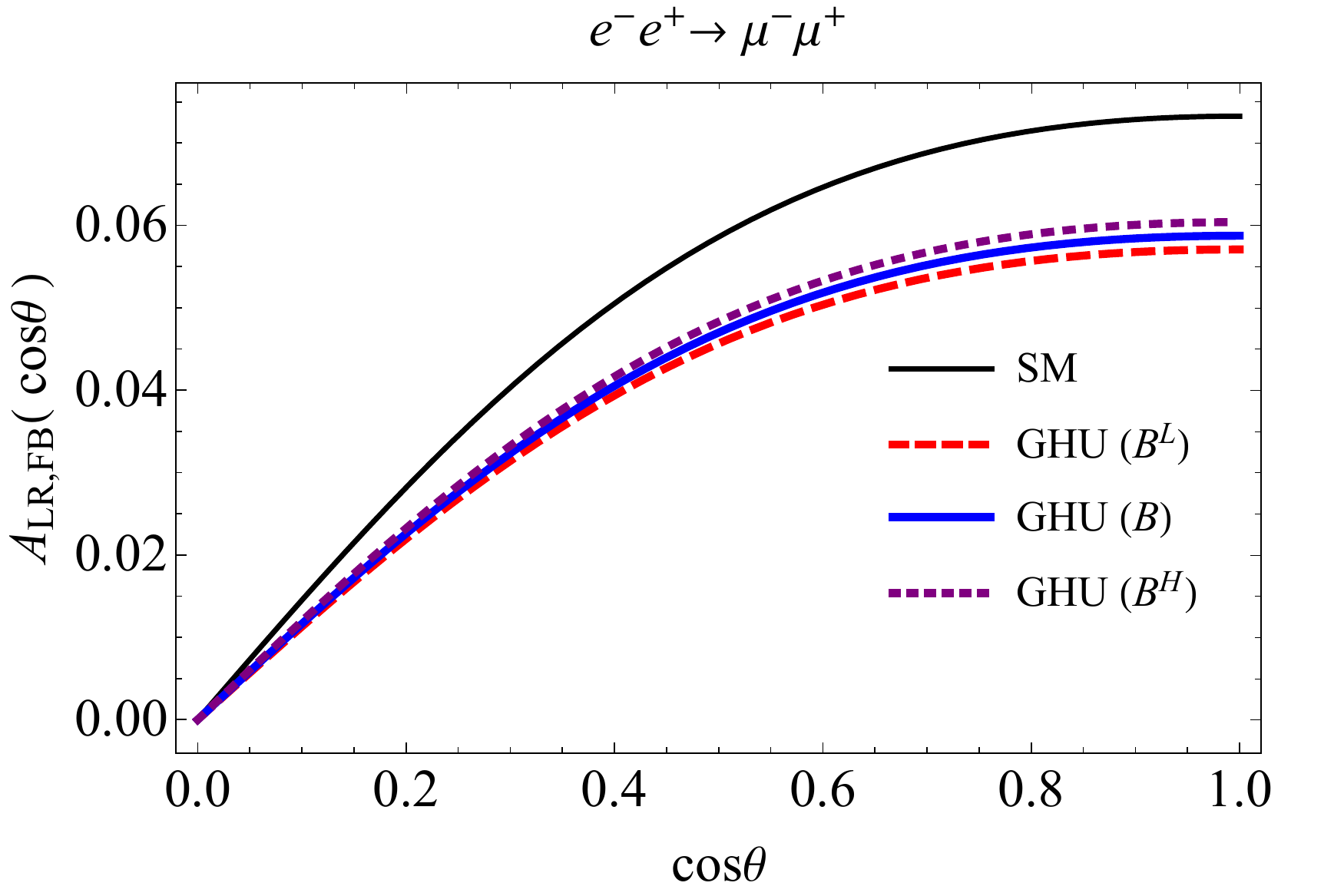}
\includegraphics[bb=0 0 504 325,height=5cm]{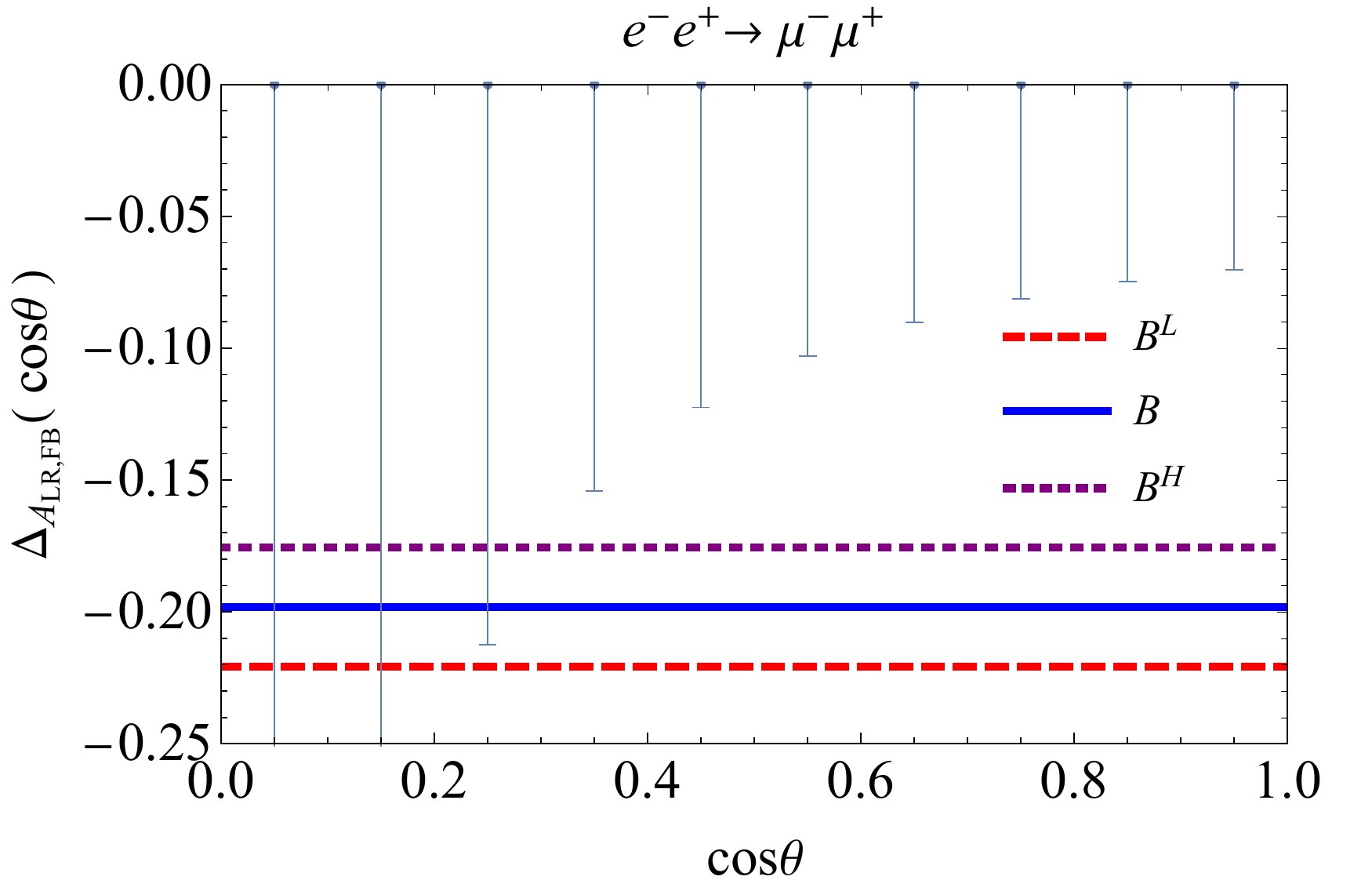}\\[1em]
\includegraphics[bb=0 0 504 327,height=5cm]{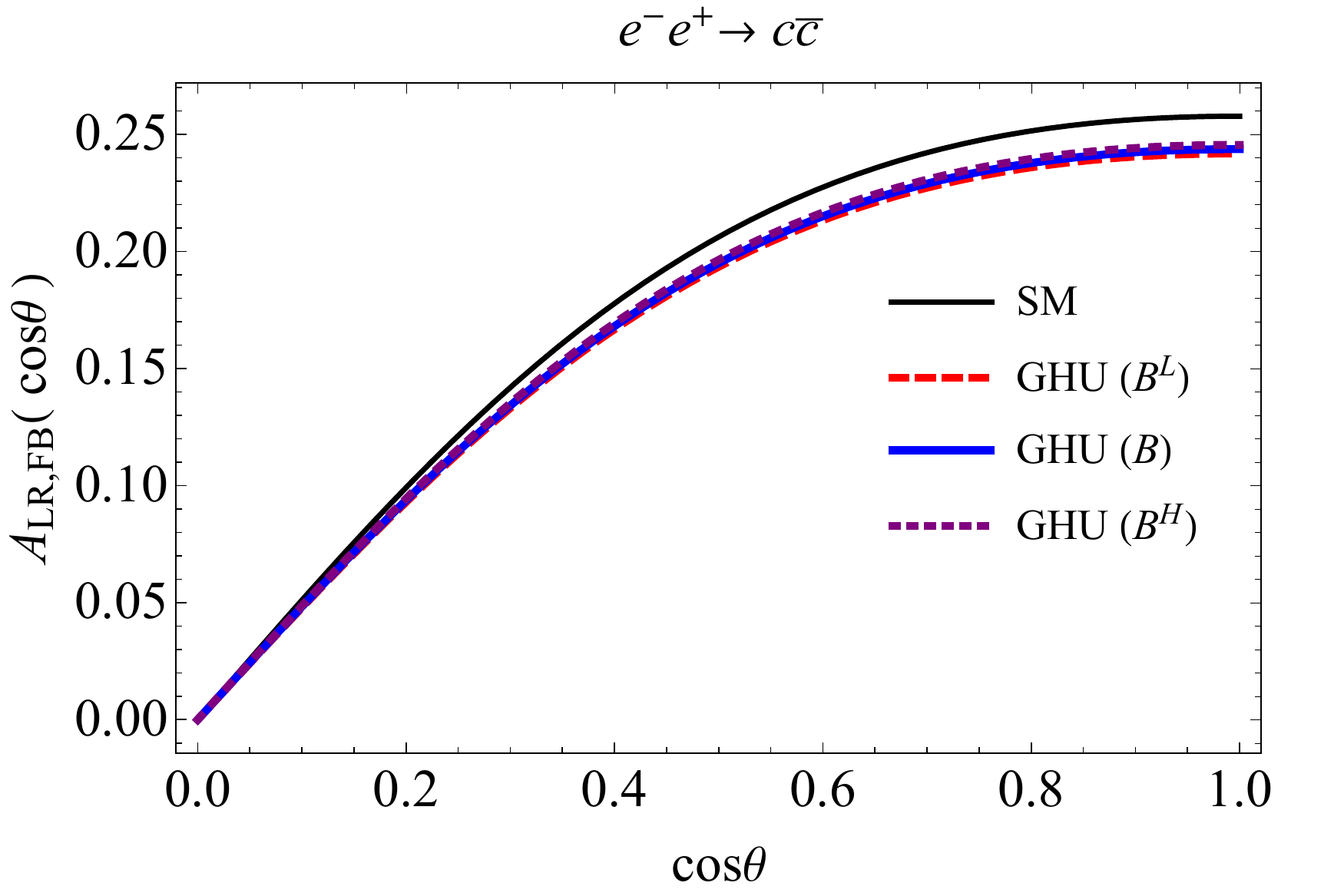}
\includegraphics[bb=0 0 504 325,height=5cm]{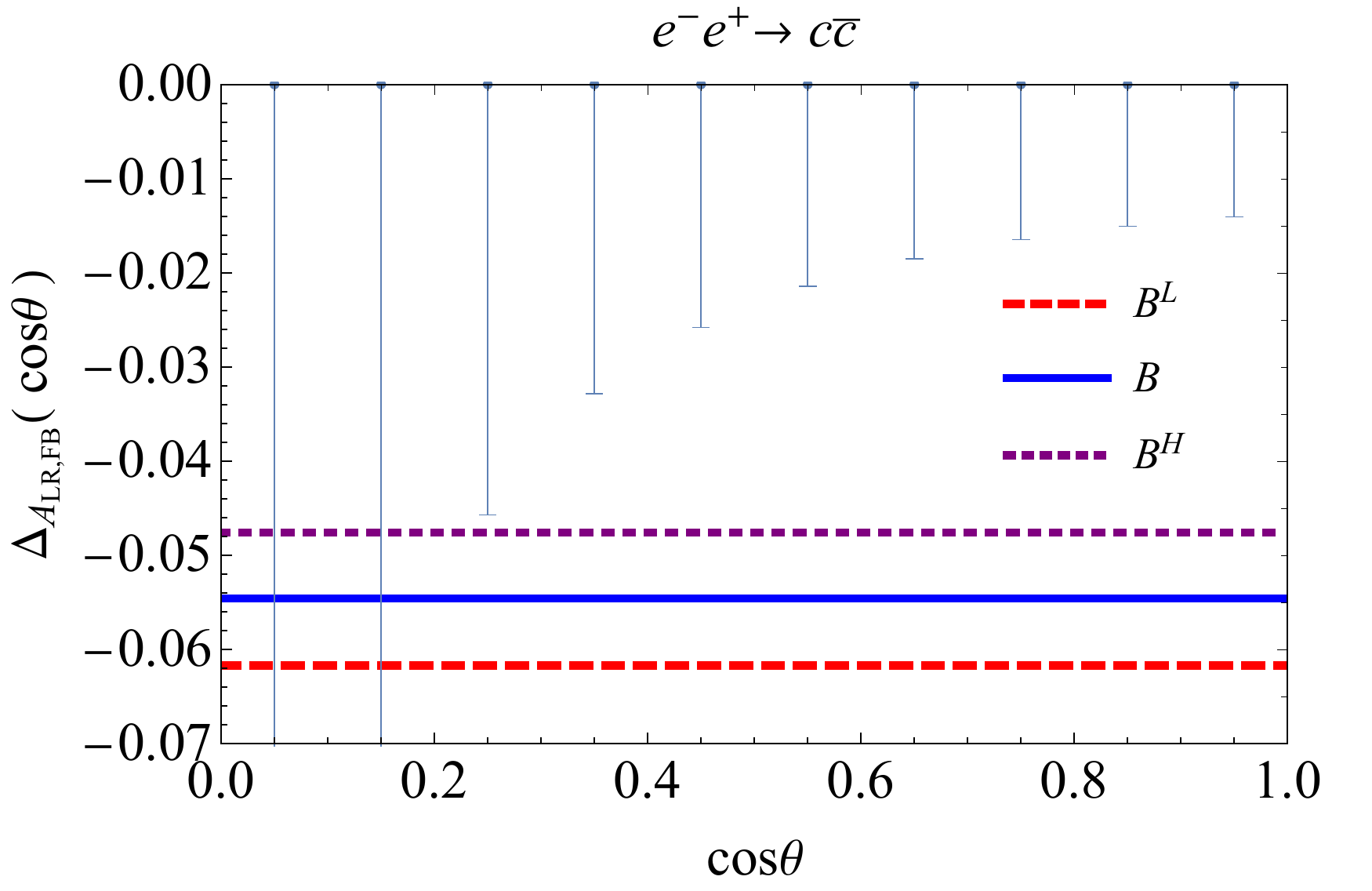}\\[1em]
\includegraphics[bb=0 0 504 327,height=5cm]{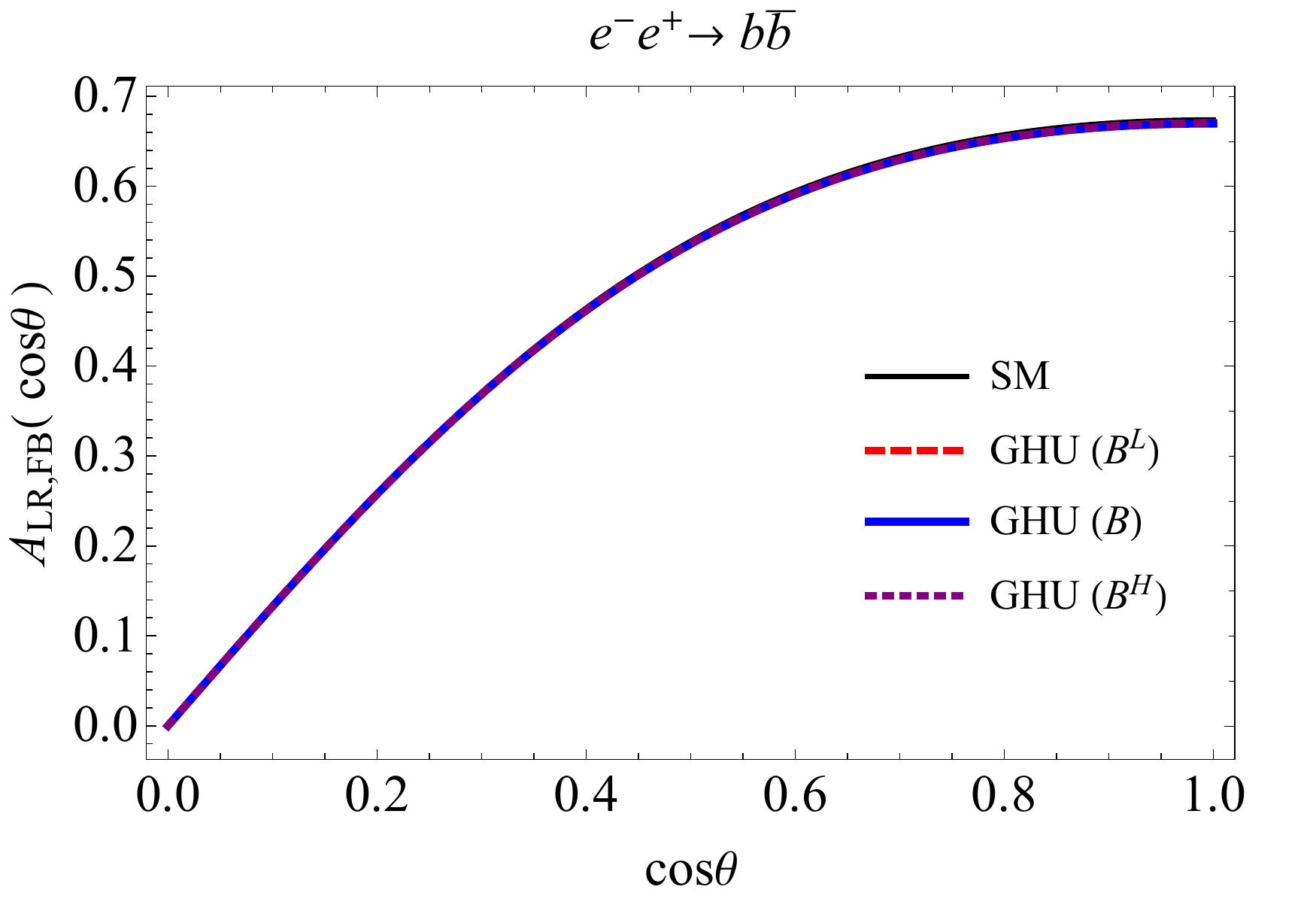}
\includegraphics[bb=0 0 504 325,height=5cm]{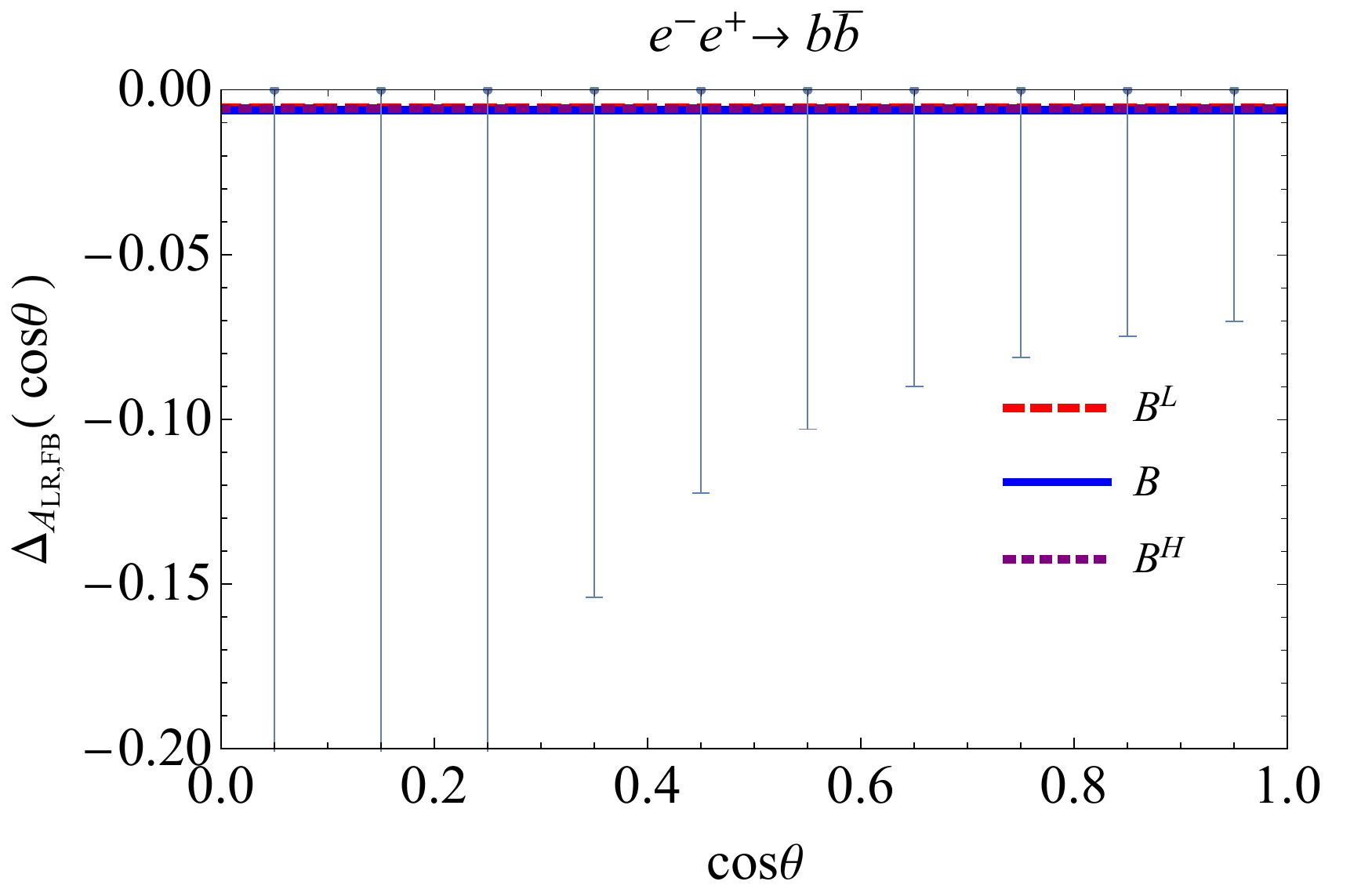}\\[1em]
\includegraphics[bb=0 0 504 327,height=5cm]{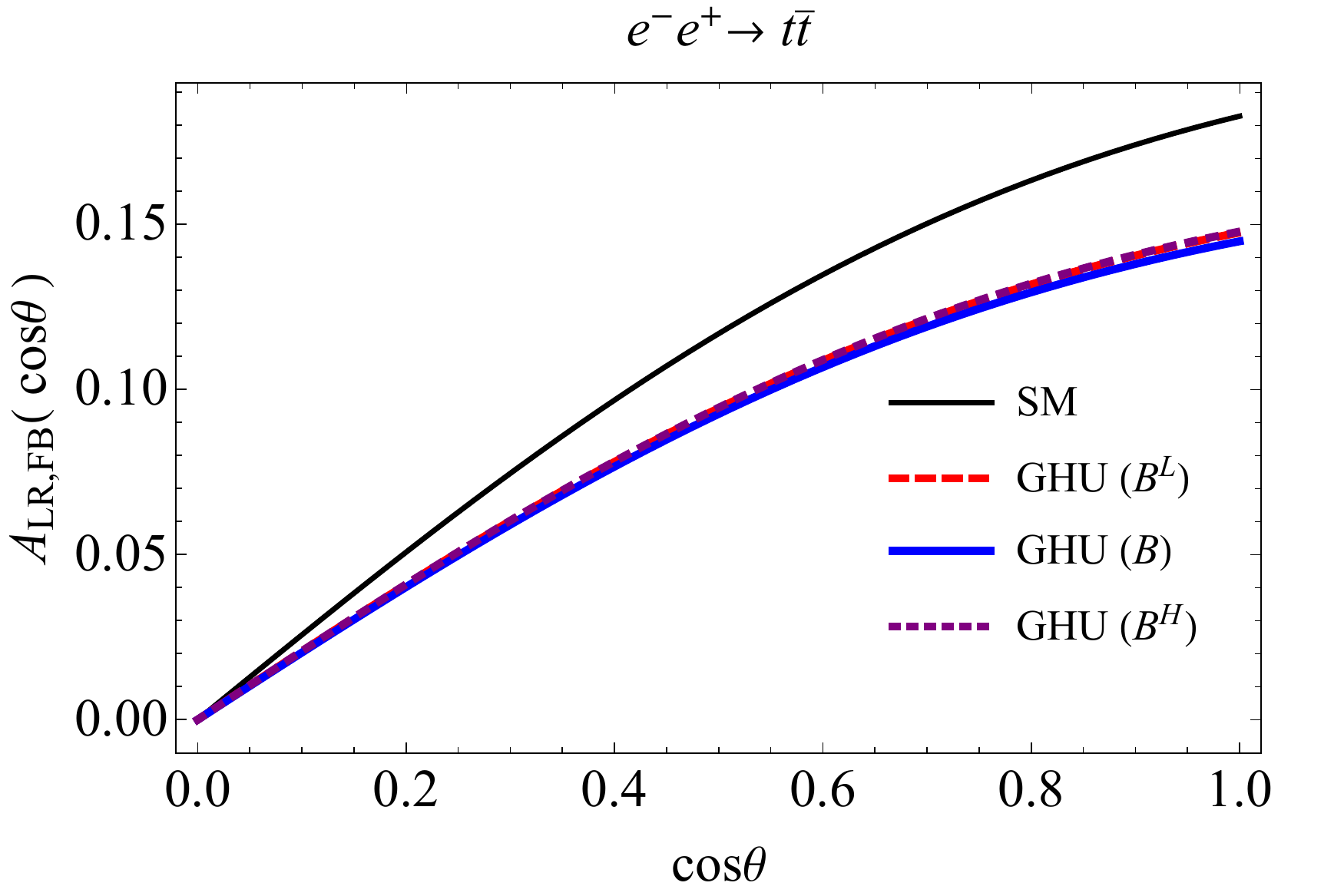}
\includegraphics[bb=0 0 504 325,height=5cm]{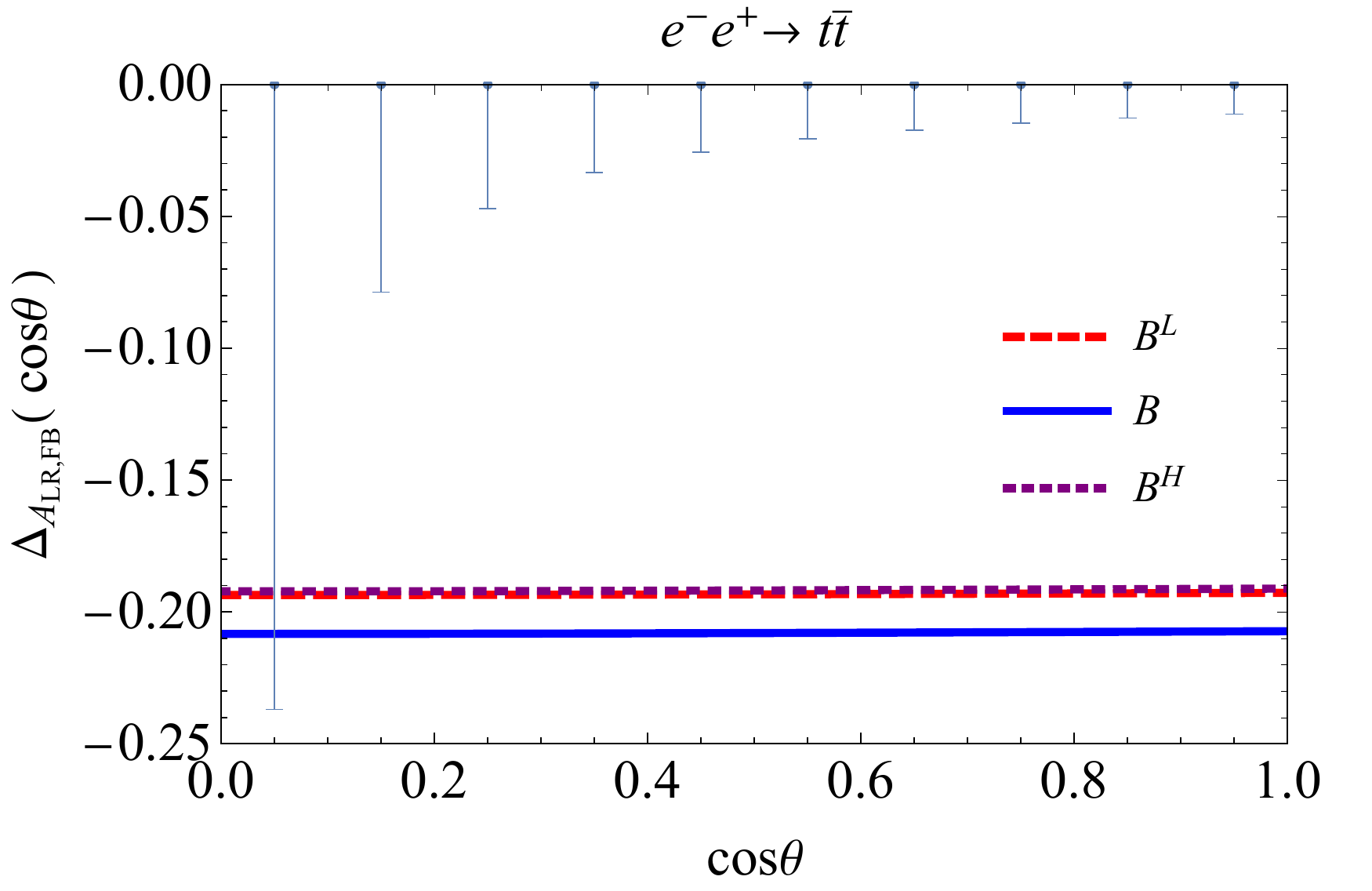}\\[1em]
\vskip -.5cm
 \caption{\small
Left-right forward-backward asymmetries
 $A_{LR,FB}^{f\bar{f}}(\cos\theta)$
 $(f\bar{f}=\mu^-\mu^+,c\bar{c},b\bar{b},t\bar{t})$ are shown.
 The left side figures show the $\sqrt{s}$ dependence of  $A_{LR,FB}^{f\bar{f}}(\cos\theta)$
 for the SM and the GHU
 (B$^{\rm L}$), (B), (B$^{\rm H}$) in  Table~\ref{Table:Mass-Width-Vector-Bosons}.
 The right figure shows the $\cos\theta$  dependence of  the deviation of
 the left-right asymmetry from the SM, 
$\Delta_{A_{LR, FB}}(\cos\theta)$
 in Eq.~(\ref{Eq:Delta-A_LRFB}) for  the GHU (B$^{\rm L}$), (B), (B$^{\rm H}$) in
 Table~\ref{Table:Mass-Width-Vector-Bosons}. 
 The error bars in the right side figures stand for
 the statistical error in Eq.~(\ref{Eq:Error-A_LR}) 
 at $\sqrt{s}=250 \,$GeV with 250$\,$fb$^{-1}$ data and
 $(P_{e^-},P_{e^+})=(-0.8,+0.3), (+0.8,-0.3)$
 for $f\bar{f}=\mu^-\mu^+,c\bar{c},b\bar{b}$
 and
 at $\sqrt{s}=500\,$GeV with 500$\,$fb$^{-1}$ data
 for $f\bar{f}=t\bar{t}$.
 } 
 \label{Figure:ALRFB-ef-theta=010}
\end{center}
\end{figure}

\section{Summary and discussions}
\label{Sec:Summary}

In the present paper, we evaluated  total and differential cross sections,
forward-backward asymmetries, differential and integrated left-right asymmetries, 
and  left-right forward-backward asymmetries in the process  $e^-e^+\to f\bar{f}$
($f\bar{f}= \mu^-\mu^+, c\bar{c}, b\bar{b}, t\bar{t})$ in the GUT inspired GHU model. 
We showed that  significant deviations in  total cross sections from those in the SM 
can be detected even in the early stage of the ILC experiment at 250$\,$GeV
with $L_{\rm int}=250 \,$fb$^{-1}$ data. 
By examining the dependence on the polarization of electrons and positrons, 
the two GHU models, the A- and B-models, can be distinguished  up to $m_{\rm KK}\simeq 15 \,$TeV. 
In  differential cross sections and  forward-backward asymmetry
for $f\bar{f}=\mu^-\mu^+$ deviations from the SM are observed
with the polarization $(P_{e^-},P_{e^+})=(-0.8,+0.3)$ and $m_{\rm KK} \sim 10 \,$TeV.
Deviations from the SM in the differential left-right asymmetry and the left-right forward-backward asymmetry 
for $f\bar{f}=\mu^-\mu^+$ are also observed with $m_{\rm KK} \sim 10 \,$TeV.
In these analyses we have checked that contributions from the second KK modes are 
negligible compared to those from the first KK modes in the energy region $\sqrt{s} \leq 1 \,$TeV. 

The main reason for having  these large effects lies in the fact that couplings of
leptons and quarks to $Z'$  bosons exhibit large parity violation.  In the GUT inspired
GHU (the B-model) left-handed leptons and light quarks have much larger couplings
than right-handed ones as shown in Tables 
\ref{Table:Couplings-Zprime_thetaH=010-mKK=13}--\ref{Table:Couplings-Zprime_thetaH=009-mKK=13}.
The magnitudes of those left-handed couplings are much larger than those of the $Z$ couplings.
This is a special feature in GHU models formulated in the Randall-Sundrum (RS) warped space.
KK gauge bosons in the RS space, including $Z'$  bosons in our case, are localized 
near the IR brane at $z=z_L$.   
In GHU  both left- and right-handed components of each lepton or quark are 
in one gauge multiplet, and each lepton or quark acquires a  mass mainly through
the Hosotani mechanism.  It implies in the RS space that  if the left-handed 
component is localized near the IR brane as in the B-model, then the right-handed
component is localized near the UV brane, and the left-handed component has
a large coupling to  $Z'$  bosons as the overlap of wave functions becomes large.

There have been many GHU models formulated in flat space, particularly 
on $M^4 \times (S^1/Z_2)$\cite{Adachi:2010cc,Kubo:2001zc,Csaki:2002ur,Scrucca:2003ra,Khojali:2017azj}.
In flat space $Z'$  bosons are symmetrically distributed around the midpoint in the fifth dimension.
In most cases leptons and quarks have uniform wave functions in the fifth dimension
so that there arises no large parity violation in the $Z'$ couplings.
In some models on $M^4 \times (S^1/Z_2)$ additional  kink-mass terms are introduced to
make left-handed and right-handed components are localized near
one brane or the other brane.  
Even in this case no large parity violation emerges in the $Z'$ couplings 
as $Z'$  bosons are symmetrically distributed in the fifth dimension.

In the composite Higgs model composite vector bosons play the role of 
$Z'$ bosons\cite{Bellazzini:2012tv}.   It has been argued that the composite
Higgs model is AdS dual of five-dimensional gauge theory\cite{Agashe:2004rs}.
In this picture $Z'$ bosons correspond to KK gauge bosons as in GHU.
In most of the composite Higgs models leptons and quarks except for the top quark 
are supposed to be localized near the UV brane so that they do not couple to $Z'$ 
bosons very much.   Except for the $e^- e^+ \go t \bar t$ process 
one does not expect significant deviations   from the SM due to $Z'$ bosons.

In the present paper, we focused on the analysis of the $s$-channel scattering
processes $e^-e^+\to f\bar{f}$ $(f\bar{f}\not=e^-e^+)$ mediated by
neutral vector bosons $Z'$ in the GHU B-model.
For $e^-e^+\to e^-e^+$, there is a contribution  not only from 
the $s$-channel scattering process  but also from the $t$-channel scattering process.
The formulas for several observables in the
scattering process  $e^-e^+\to f\bar{f}$   in Sec.~\ref{Sec:Observables}
need to be modified for  $e^-e^+\to e^-e^+$.
It has been pointed out in Ref.~\cite{Richard:2018zhl}
that for the scattering process  $e^-e^+\to e^-e^+$, 
deviations from the SM in the GHU A-model can be detected even
in the early stage of the ILC experiment at 250 GeV, and therefore
we  expect  similar  deviations from the SM in the GHU B-model as well.
We plan to give  a detailed analysis of
the $e^-e^+\to e^-e^+$ scattering process in GHU  in near future.

The scenario of gauge-Higgs unification (GHU) leads to distinct signals in 
electron-positron collision experiments.  Clear deviations from the SM should be observed
in the early stage of ILC 250$\,$GeV experiments.  In particular, GHU predicts
strong dependence on the polarization of electron and positron beams, with which one 
can explore physics at the KK mass scale of 15$\,$TeV.

\section*{Acknowledgments}

This work was supported in part 
by European Regional Development Fund-Project Engineering Applications of 
Microworld Physics (No.\ CZ.02.1.01/0.0/0.0/16\_019/0000766) (Y.O.), 
by the National Natural Science Foundation of China (Grant Nos.~11775092, 
11675061, 11521064, 11435003 and 11947213) (S.F.), 
by the International Postdoctoral Exchange Fellowship Program (IPEFP)
(S.F.),  
and by Japan Society for the Promotion of Science, 
Grants-in-Aid  for Scientific Research, No.\ 19K03873 (Y.H.)
and Nos.\  18H05543 and 19K23440 (N.Y.).

\appendix

\section{Formulas of total and partial decay widths}
\label{Sec:Decay-width}

We summarize  formulas of total and partial decay widths of 
a vector boson in a tree-level approximation.
The total decay width of a vector boson $\Gamma_{V'}$ is the sum of  
partial decay widths for all possible final states: 
\begin{align}
\Gamma_{V'} &=
\sum_{\sum_am_{\chi_a}<m_{V'}}\Gamma \Big(V' \to \prod_a \chi_a \Big) ~, 
\end{align}
where $\Gamma(V'\to \prod_a \chi_a)$ represents the partial
decay width of $V'$ to the final state $\prod_a \chi_a$. 
$m_{V'}$ and $m_{\chi_a}$ are masses of $V'$ and $\chi_a$, respectively. 

In general, the partial decay width of $V'$ to two particles
$\chi_1\chi_2$ is given by 
\begin{align}
\Gamma \left(V' \to \chi_1\chi_2\right) &= \frac{1}{16\pi m_{V'}}
\sqrt{\lambda\left(1,\frac{{m_{\chi_1}}}{{m_{V'}}},\frac{{m_{\chi_2}}}{{m_{V'}}}\right)}\
 \vert {\cal M}_{\chi_1\chi_2}\vert^2 ~,  \cr
\noalign{\kern 5pt}
\lambda(A,B,C) &= A^4+B^4+C^4-2(A^2B^2+B^2C^2+C^2A^2) ~,
\end{align}
where $m_{\chi_i}$ $(i=1,2)$ is the mass of the particle $\chi_j$
and ${\cal M}_{\chi_1\chi_2}$ stands for the amplitude for $V'\to\chi_1\chi_2$.
For  fermion final states $\chi_1\chi_2=f_1f_2$
\begin{align}
\vert \mathcal{M}_{f_1f_2}\vert^2 
=\frac{2}{3} N_cm_{V'}^2 \left\{(g_L^2+g_R^2)\left[1-\frac{m_{f_1}^2+m_{f_2}^2}{2 m_{V'}^2}-\frac{(m_{f_1}^2-m_{f_2}^2)^2}{2 m_{V'}^4}\right]+6g_L g_R\frac{m_{f_1} m_{f_2}}{m_{V'}^2} \right\},
\end{align}
where $g_{L/R}$ is the left- (right-)handed coupling constant of $V'$ to $f_1$ and $f_2$,
and $N_c$ is a color factor in the $SU(N_c)$ gauge group.

For $\chi_1\chi_2=V_1V_2$ where $V_1, V_2$ are gauge bosons
\begin{align}
\vert \mathcal{M}_{V_1V_2}\vert^2 
=\frac{1}{12} \frac{m_{V'}^6}{m_{f_1}^2 m_{f_2}^2}g_{V' V_1 V_2}^2 \Bigg\{& \left(1+\frac{m_{V_1}^4}{m_{V'}^4}+\frac{m_{V_2}^4}{m_{V'}^4}+10
\frac{m_{V_1}^2 m_{V'}^2 + m_{V_2}^2 m_{V'}^2 + m_{V_1}^2 m_{V_2}^2}{m_{V'}^4}\right)
\nonumber\\
\times&\left(1-\frac{(m_{V_1}+m_{V_2})^2}{m_{V'}^2}\right)\left(1-\frac{(m_{V_1}-m_{V_2})^2}{m_{V'}^2}\right)\Bigg\} ~.
\end{align}
Here $m_{V_i}$ $(i=1,2)$ is the mass of the gauge boson $V_i$, and
$g_{V' V_1 V_2}$ is the coupling constant of $V'$ to $V_1$ and $V_2$.
For $\chi_1\chi_2=VH$ where $V$ and $H$ are a gauge boson and scalar boson 
\begin{align}
\vert \mathcal{M}_{VH}\vert^2 
=\frac{2}{3} g_{V' VH}^2\Bigg\{\frac{(m_{V'}^2+m_{V}^2-m_{H}^2)^2}{8m_{V'}^2m_{V}^2}+1\Bigg\},
\end{align}
where $m_{V}$ and $m_H$ are the mass of the gauge boson $V$ and the
scalar $H$, respectively, and
$g_{V' VH}$ is the coupling constant of $V'$ to $V$ and $H$.
Normalization of $g_{V' V_1 V_2}$ and $g_{V' VH}$ is given in Ref.~\cite{Funatsu:2016uvi}.

\bibliographystyle{utphys} 
\bibliography{../../arxiv/reference}

\end{document}